\documentclass[oneside,reqno,11pt,a4paper]{amsart}
\usepackage[T1]{fontenc}
\usepackage[margin=2.5cm]{geometry}
\usepackage{amsmath,amsfonts,amssymb,amscd,amsthm}
\usepackage[usenames,dvipsnames,svgnames]{xcolor}
\usepackage[utf8]{inputenc}
\usepackage{graphicx}
\usepackage{grffile}
\graphicspath{{figures/}}
\usepackage[labelfont=rm,width=0.9\textwidth]{caption}
\usepackage[labelformat=simple]{subcaption}

\usepackage[style=english]{csquotes}
\usepackage[natbib=true,firstinits=true,sorting=none,style=numeric-comp]{biblatex}
\usepackage{makecell}
\usepackage{hyperref}
\usepackage{acronym}
\usepackage{listings}
\usepackage{textgreek}
\usepackage{url}
\usepackage{color}
\usepackage{framed}
\usepackage{fancyvrb}
\usepackage{newtxtext}
\usepackage{newtxmath} 
\usepackage{mathtools}
\usepackage{microtype} 
\usepackage[final]{pdfpages}
\usepackage{tikz}
\usepackage{booktabs}
\usepackage{multirow}

\usepackage{algorithm}
\usepackage{algpseudocode}

\usepackage{changes}
\setaddedmarkup{\textcolor{black}{#1}}

\hypersetup{
    breaklinks=true,
    bookmarksopen=true,
    pdftitle={Modelling coupled surface-bulk active viscous flows in animal cells with unfitted finite elements}, 
    pdfauthor={Eric Neiva, Hervé Turlier}, 
    colorlinks=true,                       
    linkcolor=black,                       
    citecolor=blue,                        
    filecolor=black,                       
    urlcolor=blue                          
}

\definecolor{bg}{rgb}{0.93,0.93,0.93}

\acrodef{ale}[ALE]{Arbitrary Lagrangian-Eulerian}
\acrodef{pde}[PDE]{partial differential equation}
\acrodef{fe}[FE]{finite element}
\acrodef{fem}[FEM]{finite element method}
\acrodef{xfem}[XFEM]{extended finite element method}
\acrodef{dof}[DOF]{degree of freedom}
\acrodefplural{dof}[DOFs]{degrees of freedom}
\acrodef{agfe}[AgFE]{aggregated FE}
\acrodef{agfem}[AgFEM]{aggregated FEM}
\acrodef{cg}[CG]{continuous Galerkin}
\acrodef{dg}[DG]{discontinuous Galerkin}
\acrodef{bc}[BC]{boundary condition}
\acrodef{gp}[GP]{ghost penalty}
\acrodef{tdc}[TDC]{tangential differential calculus}

\definecolor{shadecolor}{gray}{.92}
\definecolor{incolor}{rgb}{0,0,.7}
\definecolor{outcolor}{rgb}{.65,0,0}
\definecolor{syntaxcolor}{rgb}{.65,0,0}

\definecolor{FigBlue}{RGB}{0,0,255}
\definecolor{FigCyan}{RGB}{0,255,255}
\definecolor{FigPurple}{RGB}{222,135,170}
\definecolor{FigYellow}{RGB}{255,221,85}
\definecolor{FigRed}{RGB}{255,0,0}

\begin{document}

\title[Unfitted FE modelling of surface-bulk 
viscous flows in animal cells]{Unfitted finite 
element modelling of \\ surface-bulk viscous 
flows in animal cells}

\author[E. Neiva]{Eric Neiva$^{\text{a},\text{b},*}$}

\author[H. Turlier]{Hervé Turlier$^{\text{a},\text{c}}$}

\thanks{\null\\
$^{\text{a}}$ Center for Interdisciplinary Research in Biology (CIRB), Collège de France, Université PSL, CNRS, INSERM, Paris, France.\\
$^{\text{b}}$ \emph{Present address:} Departament de Mecànica de Fluids (MF), Escola Politècnica Superior d'Enginyeria de Vilanova i la Geltrú (EPSEVG), Universitat Politècnica de Catalunya · BarcelonaTech (UPC), 08800 Vilanova i la Geltrú, Spain.\\
$^{\text{c}}$ \emph{Present address:} Center for Integrative Biology (CBI), University of Toulouse, CNRS, Toulouse, France.\\
$^*$ Corresponding author.\\
E-mails: {\tt eric.neiva@upc.edu} (EN), 
{\tt herve.turlier@cnrs.fr} (HT)
}

\date{\today}

\begin{abstract}

This work presents a novel unfitted finite element framework to simulate coupled surface-bulk problems in time-dependent domains, focusing on fluid-fluid interactions in animal cells between the actomyosin cortex and the cytoplasm. The cortex, a thin layer beneath the plasma membrane, provides structural integrity and drives shape changes by generating surface contractile forces akin to tension. Cortical contractions generate Marangoni-like surface flows and induce intracellular cytoplasmic flows that are essential for processes such as cell division, migration, and polarization, particularly in large animal cells. Despite its importance, the spatiotemporal regulation of cortex-cytoplasm interactions remains poorly understood and computational modelling can be very challenging because surface-bulk dynamics often lead to large cell deformations. To address these challenges, we propose a sharp-interface framework that uniquely combines the trace finite element method for surface flows with the aggregated finite element method for bulk flows. This approach enables accurate and stable simulations on fixed Cartesian grids without remeshing. The model also incorporates mechanochemical feedback through the surface transport of a molecular regulator of active tension. We solve the resulting mixed-dimensional system on a fixed Cartesian grid using a level-set-based method to track the evolving surface. Numerical experiments validate the accuracy and stability of the method, capturing phenomena such as self-organised pattern formation, curvature-driven relaxation, and cell cleavage. This novel framework offers a powerful and extendable tool for investigating increasingly complex morphogenetic processes in animal cells.

\end{abstract}

\maketitle


\section{Introduction}
\label{sec:intro}


Morphogenesis is the biological process by which a cell, tissue or organism takes shape and develops its distinctive form~\cite{lecuit2007cell}. Researchers in this field seek to understand the underlying genetic, cellular, mechanical, and environmental factors that drive the formation of complex biological structures. The ultimate goal is to unravel the fundamental principles that govern the development of living organisms. Understanding morphogenesis has significant far-reaching implications for various fields, including regenerative medicine~\cite{sasai2013next}, cancer research~\cite{yamada2007modeling} or environmental and ecological studies~\cite{wichard2015green}.

\begin{figure}[h!]
  \centering
  \begin{subfigure}{0.56\textwidth}
    \centering
    \includegraphics[width=0.9\textwidth]{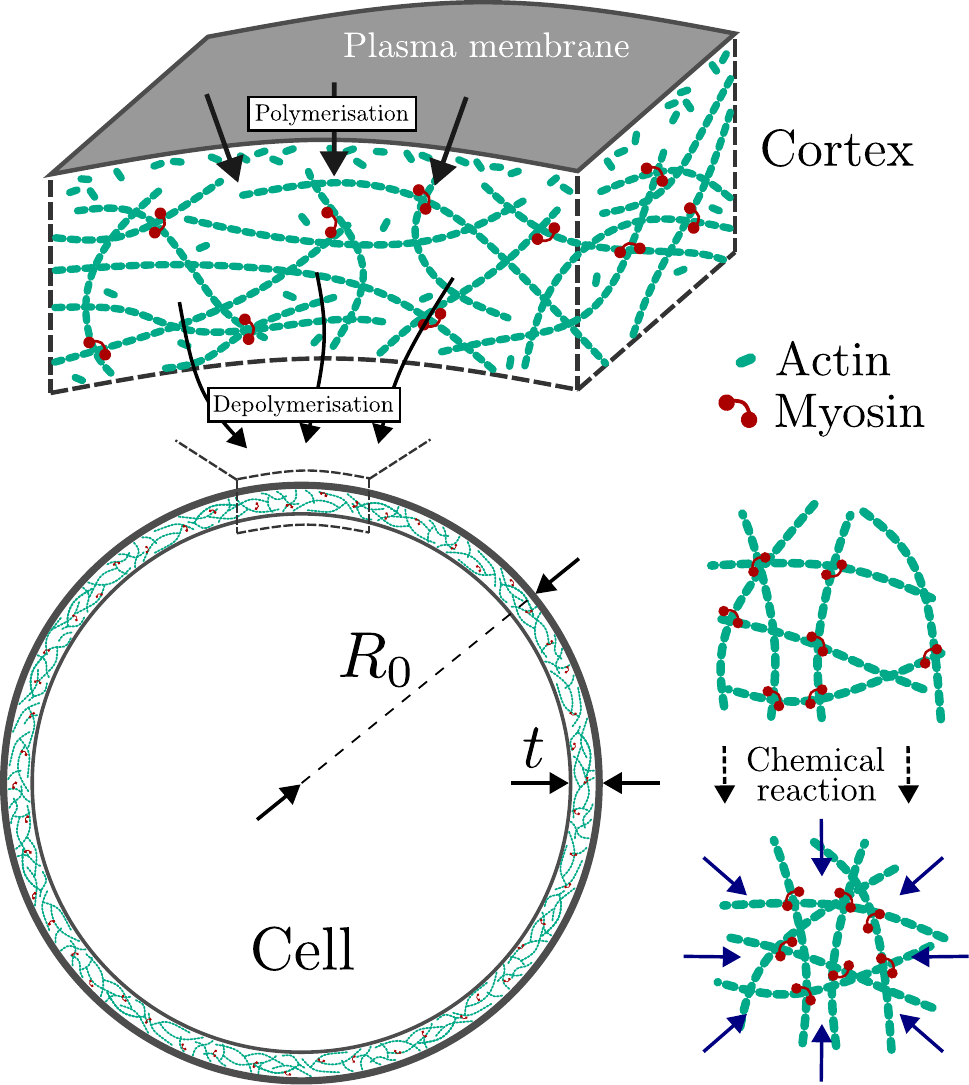}
    \caption{}\label{fig:animal-cells-a}
  \end{subfigure}
  \begin{subfigure}{0.42\textwidth}
    \centering
    \includegraphics[width=0.9\textwidth]{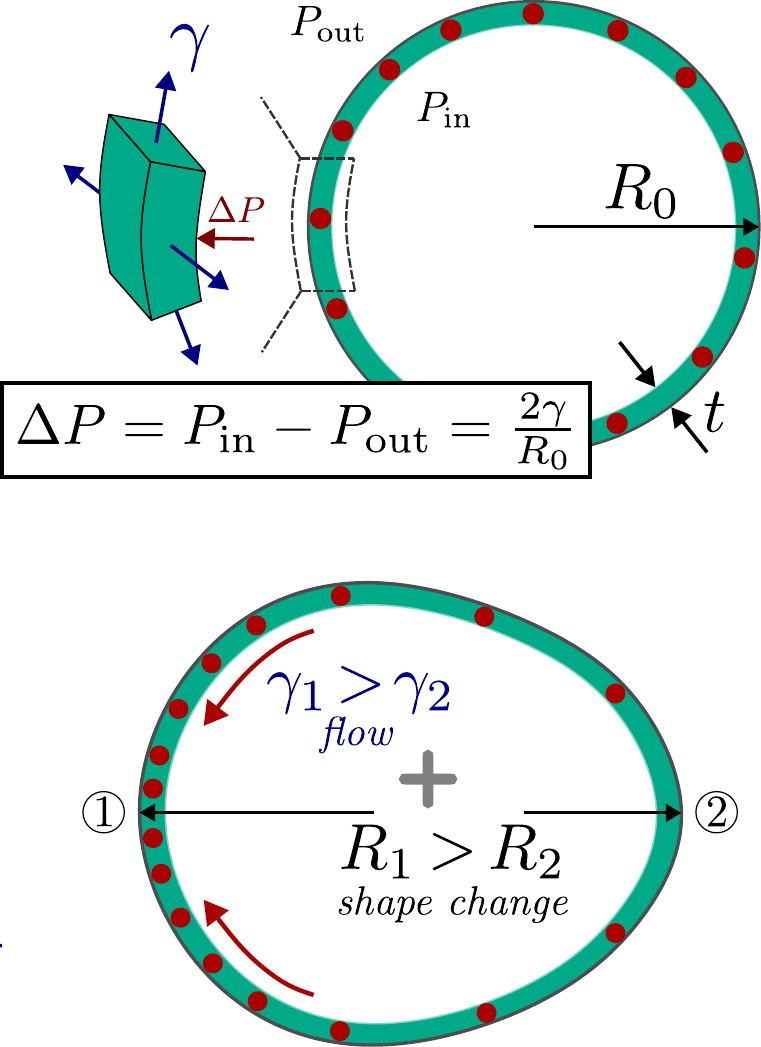}
    \caption{}\label{fig:animal-cells-b}
  \end{subfigure}
  \caption{\textbf{\textsf{(a)}} The actomyosin cortex is a thin layer of thickness $t \ll R_0$ at the inner face of the plasma membrane, where $R_0$ is the typical curvature radius of a cell. It is a network of actin protein filaments and binding proteins, among which myosin-2 molecular motors, who pull on the filaments via a chemical reaction. The network is under constant renewal: polymerisation occurs right under the plasma membrane, while depolymerisation occurs uniformly across the cortex. Adapted from~\cite{da2022viscous}. \textbf{\textsf{(b)}} According to Laplace's law, a uniform tension in the cortex gives rise to a spherical shape under constant hydrostatic pressure. By localising myosin activity, animal cells create gradients of contractility that lead, on long time scales, to tangential cortical flows and, on short time scales, to shape deformation with net cytoplasm movement (non-uniform cytoplasmic pressure). Inspired by~\cite{salbreux2012actin}, incorporating the appropriate curvature changes associated with increased local membrane tension.}
  \label{fig:animal-cells}
\end{figure}

This work centres upon the morphogenesis of single animal cells. Animal cells maintain their shape primarily through an effective surface tension, adopting shapes akin to soap bubbles~\cite{hayashi2004surface}. The main contributor to this effective surface tension in animal cells is the actomyosin cortex~\cite{clark2014stresses}: a thin biological interface at the inner face of the plasma membrane, represented in Figure~\ref{fig:animal-cells-a}. The cortex is composed of bundled and cross-linked actin protein filaments that form a dense and overlapping three-dimensional meshwork of a few hundreds of nanometres in thickness~\cite{clark2013monitoring}. Myosin-2 molecular motors populate this network and pull on actin filaments by converting chemical into mechanical forces to generate contractile stresses in the meshwork. These internal stresses lead to an effective surface tension over the cell surface and, due to cell curvature, create hydrostatic pressure in the cytoplasm following Young-Laplace's law. In addition, cortical actin is under constant renewal, through assembly and disassembly of the filaments. This material turnover remodels the whole network in about 30-60 s and releases any elastic stresses accumulated in that period~\cite{kelkar2020mechanics}. Being constantly out of thermodynamic equilibrium, the cortex is thus an \emph{active} system: It behaves as an elastic solid membrane at short time scales (<1 min), but flows like a viscous liquid at longer time scales~\cite{prost2015active}.

Cortical flows, induced by gradients of cortical tension, are the fundamental mechanism underlying shape change in animal cells~\cite{bray1988cortical,mayer2010anisotropies,turlier2014furrow,salbreux2017mechanics}. This is illustrated in~Figure~\ref{fig:animal-cells-b} as follows: A uniform tension in the cortex gives rise to a spherical shape under constant hydrostatic pressure proportional to the surface tension and curvature (here, the inverse of the radius). This is in agreement with normal force balance given by the Young-Laplace law. However, animal cells are able to spatiotemporally \emph{localise} myosin concentration and motor activity within the cortex, using complex cascades of chemical reactions, known as signalling pathways~\cite{Pires-daSilva2003,bement2024patterning}. This can create local imbalances of contractility that, at the time scale of minutes, give rise to Marangoni-like flows, pulling cortical components away from regions of relaxation toward regions of contraction. Moreover, on short time scales, normal force balance implies that gradients of surface tension must necessarily be corresponded with non-uniform cytoplasmic pressures and changes in curvature. While this explains how animal cells exert shape control with the actomyosin cortex, it also reveals that cortex-mediated cell deformation produces a net movement of the cytoplasm~\cite{mogilner2018intracellular}.

The cytoplasm is a gel-like substance that is crowded with various macromolecules, organelles and a dynamic network of protein filaments, known as the cytoskeleton~\cite{luby1999cytoarchitecture}. It has a \textmu m-scale viscosity of around 0.1-1 Pa$\cdot$s, about 100 to 1,000 times the one of water~\cite{hiramoto1969mechanical,najafi2023size,valberg1987magnetic}, albeit much lower than the one estimated for the cortex of $10^5$-$10^6$ Pa$\cdot$s~\cite{turlier2014furrow}. Despite the cytoplasm being a crowded medium at the microscopic scale, fast rearrangement of cytoplasmic components in the cell is essential for key cellular functions such as cell division, migration and polarisation. Many large animal cells (above 50 \textmu m in diameter) often accomplish this with cortical contractions that induce bulk intracellular flows~\cite{hiramoto1958quantitative,mogilner2018intracellular,shamipour2021cytoplasm,lu2023go}. Classical observations of this phenomenon include the establishment and maintenance of PAR polarity in the \emph{C. elegans} zygote~\cite{mittasch2018non,illukkumbura2020patterning}, the distribution of nuclei in the \emph{Drosophila} fly syncytial embryo~\cite{von1994actin,deneke2019self,hernandez2023two}, the 3D migration of a cell in a fluid \cite{farutin2019crawling,le2020actomyosin}, and the asymmetric spindle positioning model in mouse oocytes~\cite{yi2011dynamic,chaigne2013soft,chaigne2015narrow,wang2020symmetry,Liao2024}. Interestingly, the reverse coupling mechanism has also been reported, where cytoplasmic queues via diffusive transport and/or intracellular flows trigger flows in the actomyosin cortex~\cite{deneke2019self,mittasch2018non}. Despite these descriptions, the spatiotemporal interactions between cortical and intracellular flows—and their roles in cytoplasmic reorganization and cell shape—remain insufficiently characterized in most cases.

To address this, scholarly efforts combining biophysical experiments and numerical modelling are growing and effectively helping to better understand cortex-cytoplasm interactions~\cite{Liao2024,Barnhart2015,bhatnagar2023axis}. The \ac{fem} is frequently adopted to approximate the \acp{pde} that model these type of systems~\cite{torres2022interacting,da2022viscous}, given its flexibility to deal with intricate geometries, non-standard boundary conditions and nonlinearities~\cite{johnson1987numerical}. Yet, many surface-bulk couplings in living cells involve large distortions of the cell surface. This frames the problem within a classical dilemma in \ac{fe} theory and practice: selecting the description of motion~\cite{malvern1969introduction,Donea2004}. This choice constrains the relationship between the deforming body and the computational mesh underlying the \ac{fe} approximation. As a result, it determines the ability of the \ac{fe} method to deal with large deformations.

\begin{figure}[h!]
  \centering
  \begin{subfigure}{0.24\textwidth}
    \centering
    \includegraphics[height=4.5cm]{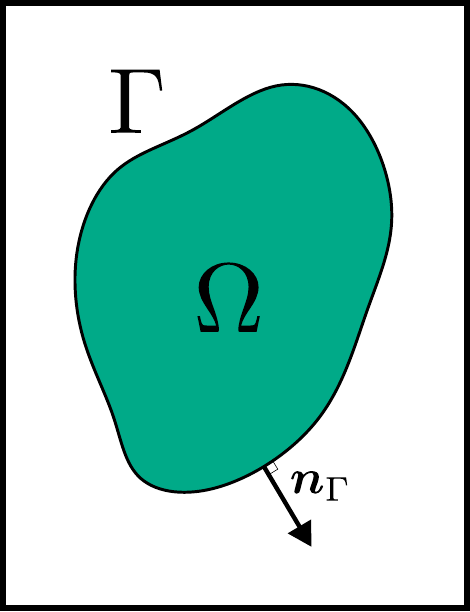}
    \caption{}\label{fig:domain-and-mesh-a}
  \end{subfigure}
  \begin{subfigure}{0.24\textwidth}
    \centering
    \includegraphics[height=4.5cm]{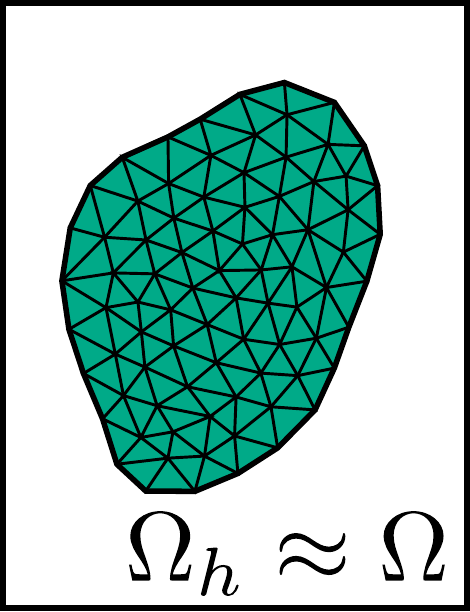}
    \caption{}\label{fig:domain-and-mesh-b}
  \end{subfigure}
  \begin{subfigure}{0.24\textwidth}
    \centering
    \includegraphics[height=4.5cm]{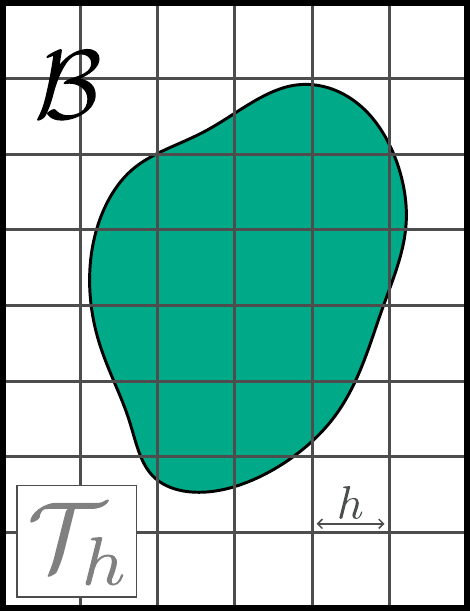}
    \caption{}\label{fig:domain-and-mesh-c}
  \end{subfigure}
  \begin{subfigure}{0.24\textwidth}
    \centering
    \includegraphics[height=4.5cm]{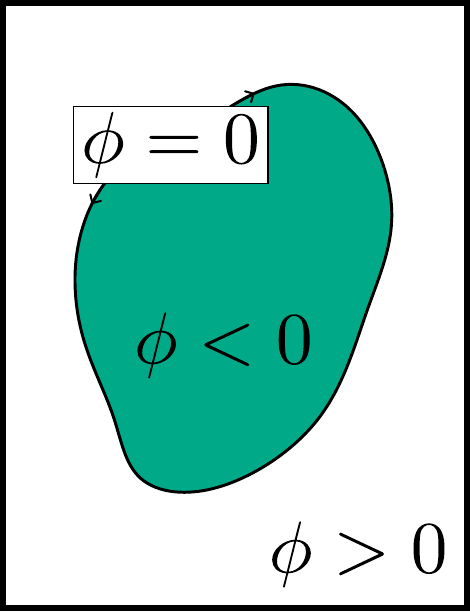}
    \caption{}\label{fig:domain-and-mesh-d}
  \end{subfigure}
  \caption{Geometry and mesh definitions. \textbf{\textsf{(a)}} The cell cortex is represented by a closed surface $\Gamma$ enclosing the cytoplasm $\Omega$. $\boldsymbol{n}_\Gamma$ is the outwards pointing unit normal on $\Gamma$. \textbf{\textsf{(b)}} A typical \emph{body-fitted} mesh (approximately) fitting to the boundary of the problem. \textbf{\textsf{(c)}} A \emph{fixed-grid} of an unfitted \ac{fe} analysis. The problem geometry $\Omega$ is embedded into an easy-to-generate mesh $\mathcal{T}_h$, usually covering a trivial region, such as a bounding box $\mathcal{B}$ of $\Omega$. \textbf{\textsf{(d)}} A \emph{level-set function} $\phi$ encoding the surface as its zero isosurface $\Gamma = \{ \phi = 0 \}$ and the bulk domain as its negative region $\Omega = \{ \phi < 0 \}$.} 
  \label{fig:domain-and-mesh}
\end{figure}

Most, if not all, approaches to describe large distortions with \acp{fe} sit between the two main perspectives of motion in continuum mechanics: Lagrangian and Eulerian \cite{Donea2004}. In the Lagrangian viewpoint, the computational grid follows the moving continuum, the grid nodes being permanently identified with the same material points \cite{bennett2006lagrangian}. A natural choice is to use a mesh fitting to the moving boundaries and interfaces in the continuum, i.e., a \emph{body-fitted} mesh (Figure~\ref{fig:domain-and-mesh-b}). This perspective is common in nonlinear solid mechanics~\cite{bonet1997nonlinear}. While it allows for an easy tracking of free surfaces and interfaces, frequent remeshing is needed to avoid numerical pollution from excessively distorted elements in the mesh \cite{khan2020surface,da2022viscous}. This issue does not appear in the Eulerian viewpoint. In this case, material particles and mesh nodes are decoupled: the computational grid stays fixed, while the continuum moves with respect to the grid (Figure~\ref{fig:domain-and-mesh-c}). \emph{Fixed-grid} techniques are predominant in fluid mechanics~\cite{anderson1995computational}. Since the mesh no longer fits to moving boundaries and interfaces, tracking their motion becomes more difficult. Besides, convective effects appear, due to the relative motion between the deforming material and the computational grid.

The solid-fluid duality in the mechanics of the actomyosin cortex is rather the rule than the exception in animal cell and tissue interfaces~\cite{turlier2014furrow,torres2019modelling,Voigt2019,da2022viscous}. At first glance, there is no clear choice to describe the motion of this type of surfaces and their interactions with surrounding fluids. Hence, it comes as no surprise to see the variety of \ac{fe} methods in the field reflecting this open-ended question. At one end, there are body-fitted methods~\cite{de2021numerical,wittwer2023computational,Ganesan2009,Edelmann2024,hernandez2024bulk}, relying on \ac{ale}~\cite{Dziuk2013,Mokbel2020,Rangarajan2015} descriptions to cope with the mesh deformation. At the other end, there are fixed-grid methods with diffuse or sharp representations of the moving boundary or interface; they are usually grounded, respectively, on phase-field~\cite{kloppe2024phase,teigen2011diffuse} or level-set~\cite{Frachon2023} methods. In between, there are hybrid approaches, coupling a body-fitted method for the interface with a fixed-grid one for the bulk phases~\cite{Barrett2014,barrett2017finite}. 

Observe that, of the above works, only body-fitted \ac{ale} and diffuse-interface approaches explicitly address applications in cell biology. In particular, as far as the authors know, the most robust and advanced technology in evolving sharp-interface methods remains to be transposed for modelling surface-bulk couplings in animal cells. Working to close this gap could be beneficial in the area for, at least, two reasons: (1) Sharp-interface methods have an apparent edge over body-fitted counterparts when it comes to model cell division, as they do not need to undergo complex numerical surgery for mesh scission~\cite{kovacs2024numerical}; (2) Phase-field models can be computationally prohibitive, especially in 3D, as they require small and adaptive mesh resolutions around the moving diffuse-interface~\cite{gomez2017computational,Bachini2024}.

Given this motivation, the main purpose of this work is to \emph{formulate a new sharp-interface \ac{fe} method for surface-bulk \acp{pde} with dynamic surfaces in fixed computational grids}. To approximate the coupled problem, we propose a novel partitioned scheme, in which the surface \acp{pde} are discretised with the trace \ac{fem}~\cite{Olshanskii2017} and the bulk \ac{pde} with the \ac{agfem}~\cite{badia2018aggregated}. Both methods fall into the class of so-called unfitted, embedded or immersed \ac{fe} methods~\cite{peskin1972flow,Mittal2005}. They are endowed with mathematical properties that ensure numerical stability and robustness, regardless of how the mesh overlaps the moving interface(s)~\cite{Hansbo2020,de2023stability}, which is an essential requirement in our context. Our contribution distinguishes itself from other unfitted methods in the literature~\cite{Frachon2023,Hansbo2016,Olshanskii2022} in two ways: (1) They solve the bulk problem with ghost penalty stabilisation~\cite{Burman2015}, while we choose \ac{agfem} to avoid locking issues that can appear with standard ghost penalty methods~\cite{badia2021linkAgFEM}. (2) We specifically target applications to cortex-cytoplasm fluid interactions in animal cells. For this reason, our model problem consists in a system of surface-bulk viscous flows, whose active driving forces are regulated by the surface transport of a diffusing species. Similar models have only been analysed with body-fitted or diffuse-interface methods~\cite{wittwer2023computational,aland2023phase}; in contrast to this work, they do not explicitly solve the surface PDE problem for the viscous flows.

The outline of this paper is as follows. First, we state the problem in Section~\ref{sec:model}. We introduce minimal notation in Section~\ref{subsec:notation} to write the dimensionless 3D governing equations of the problem in Section~\ref{subsec:equations}. The computational model is detailed in Section~\ref{sec:discrete}. We introduce first the embedded geometry setup in Section~\ref{subsec:embedded}. The discretisation in space in Section~\ref{subsec:space} adopts, for the bulk problem, the inf-sup stable pair formed by continuous quadratic velocities and discontinuous linear pressures; for the surface flow problem, continuous linear velocities; and, for the surface transport problem, continuous linear elements. Numerical analysis available for each individual discrete formulation underpin these choices~\cite{Badia2018Mixed,jankuhn2021trace}. The discretisation in time considers a simple backward Euler method and the moving surface is modelled with an evolving level-set method (Section~\ref{subsec:time}), that leverages high-order quadrature rules on implicit geometries~\cite{saye2022high} and extensions of the surface velocity to the whole grid by closest-point projections~\cite{saye2014high}. In the numerical experiments of Section~\ref{sec:experiments}, we analyse three challenging applications: (1) Self-organised shape emergence, as in~\cite{wittwer2023computational}, but at large hydrodynamic lengths, (2) relaxation dynamics of 3D bodies and (3) uni- and bi-lateral cell cleavage. Finally, concluding remarks and perspectives are listed in Section~\ref{sec:conclusions}.

%
%

\section{Model problem}
\label{sec:model}

We consider the minimal model of self-organization of the cell cortex introduced in~\cite{mietke2019minimal}. We restrict our analysis to a \emph{single cell}. The problem is given by a system of coupled surface-bulk viscous flow equations, where the surface phase represents the cortex and the bulk phase the cytoplasm. Flows are driven by gradients of cortical tension that are regulated by a molecular species diffusing on the cortex, representing the myosin molecular motors. This leads to a system of mixed-dimensional \acp{pde} in the three-dimensional space, where the surface and bulk \acp{pde} are posed in manifolds of dimensionality two and three, respectively.

As we seek to approximate the problem in a fixed three-dimensional grid, it is in our interest to formulate all governing equations in the \emph{global} Cartesian coordinates. In this way, we do not require \emph{local} coordinates or a parameterisation for the surface, which facilitates applying the level-set method or other fixed-grid techniques. To achieve this, we will use \ac{tdc} to express the surface \acp{pde} in terms of differential operators in Cartesian coordinates, like the bulk \acp{pde}. For the sake of brevity, we will skip most of the mathematical formalism underlying \ac{tdc}; the interested reader can find it in, e.g.,~\cite{delfour1996tangential,fries2020unified}.

\subsection{Preliminary definitions}
\label{subsec:notation}

Let $\Gamma(\mathrm{t})$ be a smooth, closed and time-evolving surface in $\mathbb{R}^3$. We denote by $\Omega(\mathrm{t})$ its enclosing volume ($\Gamma = \partial \Omega$). In our context, $\Gamma$ represents the fluid cortex and $\Omega$ the bulk cytoplasm of a single animal cell, see Figure~\ref{fig:domain-and-mesh-a}.



In what follows, we consider $\Gamma = \Gamma(\mathrm{t})$ and $\Omega = \Omega(\mathrm{t})$ for some fixed $\mathrm{t}$. We denote by $\boldsymbol{n}_{\Gamma}$ the outwards pointing unit normal vector on $\Gamma$ and by $\mathbf{P}(\boldsymbol{x})$ the normal projector onto the tangential space at $\boldsymbol{x} \in \Gamma$, defined by
\begin{equation}
  \mathbf{P}(\boldsymbol{x}) = \mathbf{Id} - \boldsymbol{n}_{\Gamma}(\boldsymbol{x}) \otimes \boldsymbol{n}_{\Gamma}(\boldsymbol{x}),
\end{equation}
where $\mathbf{Id}$ denotes the identity matrix. For illustration, the projection of a vector field $\boldsymbol{v} : \Gamma \to \mathbb{R}^3$ onto the tangent plane is given by $\mathbf{P} \boldsymbol{v}$ and for a second-order tensor function $\mathbf{A} \in \Gamma \to \mathbb{R}^{3 \times 3}$ the in-plane tensor is given by $\mathbf{P} \mathbf{A} \mathbf{P}$.

The \emph{tangential differential operators} intervening in our problem are defined as follows. The surface gradient $\nabla_\Gamma$ of a scalar function $u : \Gamma \to \mathbb{R}$ defined on the manifold is given by $\nabla_\Gamma u = ( \nabla u^e ) \mathbf{P}$, where $u^e$ denotes a smooth extension of $u$ in a tubular neighbourhood $\mathcal{U}$ of the manifold $\Gamma$. Given a vector field $\boldsymbol{v} : \Gamma \to \mathbb{R}^3$ on the manifold, the surface gradient is computed as $\nabla_\Gamma \boldsymbol{v} = \mathbf{P} \nabla \boldsymbol{v}^e \mathbf{P}$ and the surface divergence as $\mathrm{div}_\Gamma \boldsymbol{v} = \mathrm{tr}(\nabla_\Gamma \boldsymbol{v}) = \mathrm{tr}(\mathbf{P} \nabla \boldsymbol{v}^e \mathbf{P}) = \mathrm{tr}(\nabla \boldsymbol{v}^e \mathbf{P})$, where $\boldsymbol{v}^e$ is a smooth extension of $\boldsymbol{v}$ in $\mathcal{U}$. In particular, the surface Laplacian of a scalar field is given by $\Delta_\Gamma u = \mathrm{div}_\Gamma ( \nabla_\Gamma u )$. Finally, the surface divergence of a tensor field $\mathbf{A} : \Gamma \to \mathbb{R}^{3 \times 3}$ is given by $\mathrm{div}_\Gamma \mathbf{A} = \left[ \mathrm{div}_\Gamma(\boldsymbol{e}_1^T \mathbf{A}), \mathrm{div}_\Gamma(\boldsymbol{e}_2^T \mathbf{A}), \mathrm{div}_\Gamma(\boldsymbol{e}_3^T \mathbf{A}) \right]^T$, where $\boldsymbol{e}_i^T, \ i = 1,2,3$, refer to the canonical basis vectors.

The \emph{bulk strain-rate tensor} of a vector field $\boldsymbol{v} : \Omega \to \mathbb{R}^3$ is given by $\boldsymbol{\varepsilon}(\boldsymbol{v}) = \frac{1}{2} \left[ \nabla\boldsymbol{v} + (\nabla\boldsymbol{v})^T \right]$, whereas the \emph{surface strain-rate tensor} of a vector field $\boldsymbol{v} : \Gamma \to \mathbb{R}^3$ is computed as $\boldsymbol{\varepsilon}_\Gamma (\boldsymbol{v}) = \mathbf{P} \boldsymbol{\varepsilon} (\boldsymbol{v}^e) \mathbf{P}$. We conclude the preliminary definitions with the \emph{time derivative} $\partial_\mathrm{t}$ and the \emph{material derivative} ${\rm D}_\mathrm{t} (\bullet,\boldsymbol{v}) = \partial_\mathrm{t} (\bullet) + \boldsymbol{v} \cdot \nabla (\bullet)$.


\subsection{Problem statement}
\label{subsec:equations}

Our model problem is an extension of a classical mathematical model for two-phase flow with interfacial tension and viscosity~\cite{scriven1960dynamics}. It consists of an active Boussinesq-Scriven surface fluid that encloses a passive viscous bulk fluid, where a diffusing molecular species regulates the driving forces on the surface and creates a mechanochemical feedback.

We focus on cellular processes such as cell division, migration or polarisation, where time scales are long enough to assume the rheology of the cortex is viscous~\cite{mayer2010anisotropies,turlier2014furrow}. On the other hand, small characteristic length scales in living cells typically lead to low Reynolds and Womersley numbers ($\mathrm{Re},\ll 1,\,\,\mathrm{\alpha} \ll 1$)~\cite{mogilner2018intracellular,purcell1977life}. Therefore, neglecting gravitational and inertial forces, balance of momentum on the surface reads
\begin{equation}\label{eq:surface-momentum-balance}
  \mathrm{div}_\Gamma \mathbf{N}_\Gamma - \boldsymbol{f}^{\rm ext} = \boldsymbol{0}, \quad \rm{on} \ \Gamma,
\end{equation}
where $\mathbf{N}_\Gamma$ and $\boldsymbol{f}^{\rm ext}$ denote the surface stress tensor and the external forces per unit area. The surface (or membrane) stress tensor $\mathbf{N}_\Gamma$ obeys, generically, variants of the Boussinesq-Scriven constitutive law 
\begin{equation}\label{eq:surface-constitutive-law}
  \mathbf{N}_\Gamma = 2 \mu_\Gamma \boldsymbol{\varepsilon}_\Gamma(\boldsymbol{U}) + ( \lambda_\Gamma - \mu_\Gamma ) ( \mathrm{div}_\Gamma \, \boldsymbol{U} ) \mathbf{P} + \mathbf{N}_\Gamma^{\rm act}, \quad \rm{on} \ \Gamma,
\end{equation} 
where $\boldsymbol{U}$ denotes the surface velocity field. The first two terms yield the stress resultant owing to viscous effects, with $\mu_\Gamma > 0$ and $\lambda_\Gamma > \mu_\Gamma$ the surface shear and dilational viscosities. The last term $\mathbf{N}_\Gamma^{\rm act}$ is the contribution from active surface tension. We will show later that this term is the driving force of the system. For the cortex, which is a thin shell of actomyosin material, a dimensional reduction of the 3D constitutive equations to 2D membrane stress resultants yields the relation $\lambda_{\Gamma} = 3\mu_{\Gamma}$ under the assumption of cortical incompressibility \cite{da2022viscous}. Additionally, both viscous and active bending contributions have been shown numerically to be negligible \cite{da2022viscous}.
We further neglect mechanical contributions from the plasma membrane—specifically, the lipid bilayer’s tangential fluidity and transverse elasticity—including its bending rigidity and passive surface tension. These effects are typically one to several orders of magnitude smaller than the surface stresses generated by the cortex \cite{dimova2014recent,rangamani2022many} and are therefore assumed to have minimal impact on the system’s mechanical behavior.

Equations~\eqref{eq:surface-momentum-balance} and~\eqref{eq:surface-constitutive-law} imply the key assumption of the cortex being a nonmaterial interface. This means that balance of mass is trivial~\cite{bothe2010two}, albeit the surface divergence $\mathrm{div}_\Gamma \, \boldsymbol{U}$ does not generally vanish. If assuming a material interface, balance of mass can be accounted for in terms of cortical density~\cite{torres2019modelling} or thickness~\cite{turlier2014furrow}. In this case, material turnover also contributes to the active tension~\cite{da2022viscous}. 

The bulk phase is described as an incompressible Stokes fluid, thereby balance of momentum and mass on the bulk reads
\begin{equation}\label{eq:bulk-momentum-and-mass-balance}
  \begin{aligned}
    \mathrm{div} ( \boldsymbol{\sigma}(\boldsymbol{u},p) ) &= \boldsymbol{0}, \quad &\rm{in} \ \Omega, \\
    \mathrm{div} \, \boldsymbol{u} &= 0, \quad &\rm{in} \ \Omega,
  \end{aligned}
\end{equation}
where $\boldsymbol{u}$ and $p$ denote the bulk velocity and pressure fields. The bulk stress tensor $\boldsymbol{\sigma}$ is given by 
\begin{equation}\label{eq:bulk-constitutive-law}
  \boldsymbol{\sigma}(\boldsymbol{u},p) = 2\mu_\Omega \boldsymbol{\varepsilon}(\boldsymbol{u}) - p \mathbf{Id}, \quad \rm{in} \ \Omega.
\end{equation}
Supposing no-slip between the surface and bulk fluids on the surface simplifies the jump conditions to 
\begin{equation}\label{eq:bulk-jump-conditions}
  \begin{aligned}
    \boldsymbol{u} &= \boldsymbol{U}, \quad &\rm{on} \ \Gamma, \\
    \boldsymbol{\sigma}(\boldsymbol{u},p) \boldsymbol{n}_\Gamma &= \boldsymbol{f}^{\rm ext}, \quad &\rm{in} \ \Omega,
  \end{aligned}
\end{equation}
such that the surface is subject to the traction force from the passive fluid. At this point, it becomes clear that, due to the mechanical coupling between the surface and bulk fluids, surface flows and deformations generated by active tension set the passive bulk fluid into motion.

The equations for the coupled surface-bulk viscous flows are combined with the equations for the surface transport of the molecule regulating active stress:
\begin{equation}\label{eq:surface-transport}
  {\rm D}_\mathrm{t}(C,\boldsymbol{U}) + C ( \mathrm{div}_\Gamma \, \boldsymbol{U} ) - D_\Gamma \Delta_\Gamma C = k_{\rm on} \overline{c}_\Omega - k_{\rm off} C, \quad \rm{on} \ \Gamma.
\end{equation}
Here, $C$ denotes area concentration of the molecular species. Equation~\eqref{eq:surface-transport} accounts for advection (in the material derivative ${\rm D}_{\rm t}$), concentration changes due to local expansion or contraction of the surface (second term) and surface diffusion with constant diffusivity $D_\Gamma$. The right-hand side describes the exchange of molecules between the surface and bulk fluids, with attachment rate $k_{\rm on}$ from the bulk into the surface and detachment rate $k_{\rm off}$ from the surface to the bulk concentration $\overline{c}_\Omega$. For simplicity, $\overline{c}_\Omega$ is assumed homogeneous and constant (the cortex is supposed to be in contact with a chemical bath). 

We can now define the active surface tension $\mathbf{N}_\Gamma^{\rm act}$ of Equation~\eqref{eq:surface-constitutive-law}, which completes the mechanochemical coupling between the diffusing species and the surface-bulk flows:
\begin{equation}\label{eq:active-surface-tension}
  \mathbf{N}_\Gamma^{\rm act} ( C ) = \xi f(C,c_{\rm eq}) \mathbf{P}, \quad \text{in} \ \Gamma, \quad \text{where} \ f(C,c_{\rm eq}) = \frac{2 C^2}{C_{\rm eq}^2+C^2}.
\end{equation}
Equation~\eqref{eq:active-surface-tension} models active tension as a monotonously increasing and saturating Hill function $f$ of $C$, scaled by a coefficient $\xi$, measuring myosin motor activity. The constant $C_{\rm eq} = (k_{\rm on}/k_{\rm off}) \overline{c}_\Omega$ represents the surface concentration at equilibrium of surface/bulk protein exchange. 

As in~\cite{stone1990effects,wittwer2023computational}, it is interesting to decompose the force term of the (active) tension in Equation~\eqref{eq:surface-momentum-balance} in the local coordinate system of the surface as 
\begin{equation}\label{eq:active-forces}
  \mathrm{div}_\Gamma \mathbf{N}_\Gamma^{\rm act} = \xi f'(C,c_{\rm eq}) \nabla_\Gamma C - \xi f(C,c_{\rm eq}) H \boldsymbol{n}_\Gamma,
\end{equation}
where $H$ is the (doubled) mean curvature of the surface. The effects of inhomogeneous active tension are apparent in Equation~\eqref{eq:active-forces}: The first term isolates the Marangoni effect, acting as a tangential force from regions of low to high surface concentration. The second term captures the normal forces driving the dynamic changes of surface shape, which balance the bulk pressure jump in accordance with Laplace's law deriving from Equation \eqref{eq:bulk-jump-conditions}.

To conclude, we adimensionalise the governing equations with the initial cell radius $R$, the diffusion time $\tau_D = R^2/D_\Gamma$ and the concentration at equilibrium $C_{\rm eq}$ as characteristic scales. Furthermore, we assume here that $\lambda_\Gamma = \mu_\Gamma$, that is, surface shear and dilational viscosities coincide. This leads to the nondimensional system:

\vspace{0.15cm}
\noindent\textbf{Surface flow and molecular transport}
\begin{align}
  \mathrm{div}_\Gamma \left[ 2 \boldsymbol{\varepsilon}_\Gamma(\boldsymbol{U}) + \mbox{\textit{Pe}} f(C,1) \mathbf{P} \right] - \frac{2 R}{L_\eta}\boldsymbol{\varepsilon}(\boldsymbol{u}) \boldsymbol{n}_\Gamma + p \boldsymbol{n}_\Gamma &= \boldsymbol{0}, &\text{for} \ t > 0, &\enskip \text{on} \ \Gamma(t), \label{eq:full-non-dimensional-1}\\
  \partial_\mathrm{t}C + \boldsymbol{U}\cdot\nabla_{\Gamma}C +  C ( \mathrm{div}_\Gamma \, \boldsymbol{U} ) - \Delta_\Gamma C + \tau_D k_{\rm off} (C-1) &= 0, &\text{for} \ t > 0, &\enskip \text{on} \ \Gamma(t), \label{eq:full-non-dimensional-2}\\
  \intertext{\textbf{Bulk viscous flow}}
\frac{2 R}{L_\eta} \mathrm{div} ( \boldsymbol{\varepsilon} ( \boldsymbol{u} ) ) - \nabla p &= \boldsymbol{0}, &\text{for} \ t > 0, &\enskip \text{in} \ \Omega(t), \label{eq:full-non-dimensional-3}\\
  \mathrm{div} \, \boldsymbol{u} &= 0, &\text{for} \ t > 0, &\enskip \text{in} \ \Omega(t), \label{eq:full-non-dimensional-4}\\
  \boldsymbol{u} &= \boldsymbol{U}, &\text{for} \ t > 0, &\enskip \text{on} \ \Gamma(t), \label{eq:full-non-dimensional-5}\\
  \intertext{\textbf{Surface evolution}}
  \partial_\mathrm{t} \Gamma - ( \boldsymbol{U} \cdot \boldsymbol{n}_\Gamma ) \boldsymbol{n}_\Gamma &= 0,  &\text{for} \ t > 0, &\enskip \text{on} \ \Gamma(t), \label{eq:full-non-dimensional-8} \\
  \intertext{\textbf{Initial conditions}}
  C(\boldsymbol{x},0) &= C_0(\boldsymbol{x}), &\text{for} \ \boldsymbol{x} \ \text{in} & \ \Gamma(0), \label{eq:full-non-dimensional-7}\\
  \Gamma(0) &= \Gamma_0.\label{eq:full-non-dimensional-9}
\end{align}
Note that the system of Equations~\eqref{eq:full-non-dimensional-1}-\eqref{eq:full-non-dimensional-9} is closed by a law for the evolution of the surface, which is dictated by the normal component of the surface velocities. Moreover, in order to uniquely determine the pressure, we additionally require that $\int_\Omega p = 0$. 

The complete system is characterised by three nondimensional parameters: (1) A Péclet-like number $\mbox{\textit{Pe}} = \xi R^2 / \lambda_\Gamma D_\Gamma$, comparing active to diffusive transport; (2) a ratio $L_\eta / R$, comparing the cell size $R$ with a hydrodynamic length scale $L_\eta =  \lambda_\Gamma / \mu_\Omega$ that relates surface to bulk viscosity; and (3) $\tau_D k_{\rm off}$ comparing the time for surface concentration homogenization via diffusion and attachment/detachment kinematics.

\section{Discrete formulation}
\label{sec:discrete}

In this section, we describe an \emph{unfitted} FE approximation of Problem~\eqref{eq:full-non-dimensional-1}-\eqref{eq:full-non-dimensional-9}. Here, unfitted implies, in a broad sense, that the discretisation is defined in a fixed computational grid and uses a sharp representation of the evolving geometry. We proceed as follows. We start with a typical setup for \emph{fixed-grid} analysis, based on a level-set \emph{sharp-interface} representation. We approximate the continuous model with a partitioned scheme outlined in~Figure~\ref{fig:coupling-scheme}. It is composed of four building blocks for each subproblem: (a) bulk flows --Equations~\eqref{eq:full-non-dimensional-3}-\eqref{eq:full-non-dimensional-5}--, (b) surface flows --Equation~\eqref{eq:full-non-dimensional-1}--, (c) surface molecular transport --Equation~\eqref{eq:full-non-dimensional-2}-- and (d) surface evolution --Equation~\eqref{eq:full-non-dimensional-8}--. We describe the full solving strategy in a bottom-to-top fashion: First, we formulate stable and optimal \ac{fe} discretisations in space for the three PDE (sub)systems. Next, we describe the update of the surface shape. We conclude with the coupling of the individual blocks. For readability, we initially assume the shape of $\Omega$ (thus $\Gamma$) does not vary in time, until dealing with shape dynamics in Section~\ref{subsec:time}.

\subsection{Embedded geometry setup}
\label{subsec:embedded}

Mesh and geometry configuration are illustrated in Figure~\ref{fig:domain-and-mesh-c}. Let $\mathcal{T}_h$ be a quasi-uniform \emph{background grid} in $\mathbb{R}^3$, composed of either $n$-simplices or $n$-cubes and with a characteristic mesh size $h = \max_{T \in \mathcal{T}_h}\{\mathrm{diam}(T)\}$. We assume that $\mathcal{T}_h$ meshes a bounding-box $\mathcal{B}$ of $\Omega$, covering but not necessarily fitting to the boundary $\Gamma$. To circumvent minor technical details, we require that $\mathcal{T}_h$ is fine enough to resolve the curvature of $\Gamma$. A precise statement of this assumption needs further formalism that we skip here, for the sake of conciseness; it can be usually found in specialised works such as~\cite{de2017condition,gurkan2019stabilized}.

The sharp-interface representation is carried out with the level-set method~\cite{osher1988fronts}. Accordingly, the problem geometry is implicitly defined as the zero isosurface of a smooth \emph{level-set function} $\phi(\boldsymbol{x}) : \mathbb{R}^3 \to \mathbb{R}$, that verifies $\Gamma = \{ \boldsymbol{x} : \phi(\boldsymbol{x}) = 0\}$ and $\Omega = \{ \boldsymbol{x} : \phi(\boldsymbol{x}) < 0\}$, see Figure~\ref{fig:domain-and-mesh-d}. The level-set method is the simplest way to track surfaces without an explicit parametrisation. It is also a reasonable choice in our transient context, because surface dynamics are driven by the mechanics of the cortex --Equations~\eqref{eq:full-non-dimensional-8}-\eqref{eq:full-non-dimensional-9}. Therefore, in practice, we will only need to define the level set function $\phi^0$ of the initial cell configuration $\Gamma^0$, a simple sphere or spheroid, in most situations.

\begin{figure}[h!]
  \centering
  \begin{subfigure}{0.24\textwidth}
    \centering
    \includegraphics[height=4.5cm]{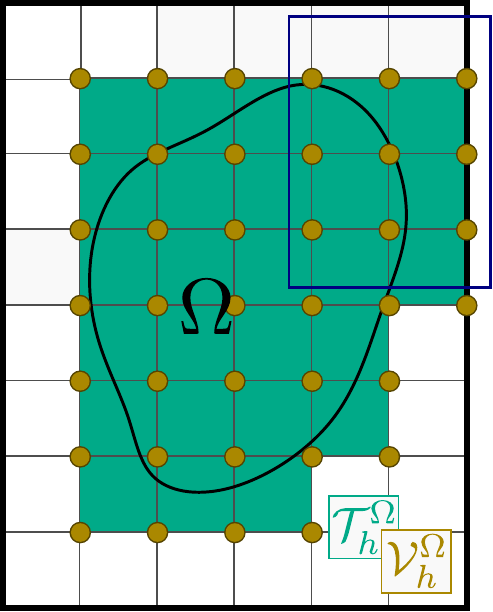}
    \caption{}\label{fig:unfitted-approximation-a}
  \end{subfigure}
  \begin{subfigure}{0.24\textwidth}
    \centering
    \includegraphics[height=4.5cm]{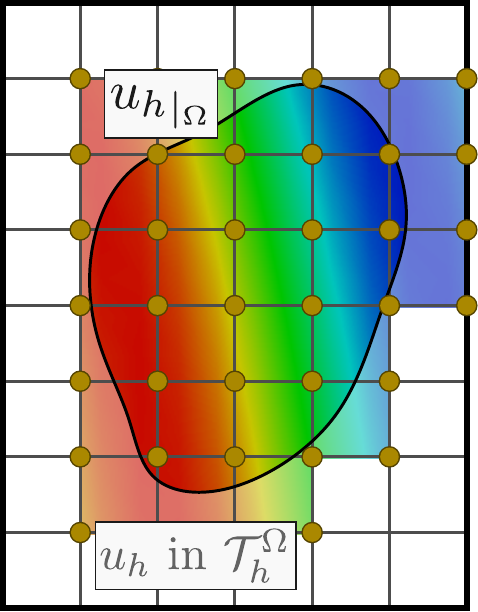}
    \caption{}\label{fig:unfitted-approximation-b}
  \end{subfigure}
  \begin{subfigure}{0.24\textwidth}
    \centering
    \includegraphics[height=4.5cm]{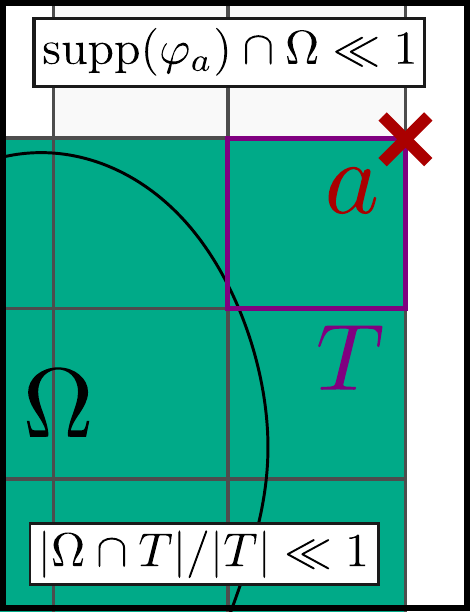}
    \caption{}\label{fig:unfitted-approximation-c}
  \end{subfigure}
  \begin{subfigure}{0.24\textwidth}
    \centering
    \includegraphics[height=4.5cm]{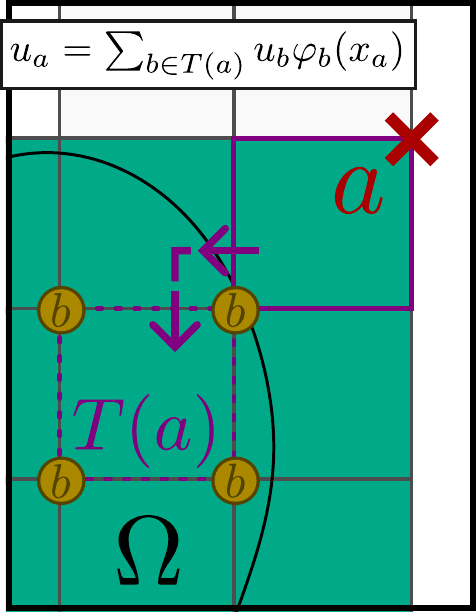}
    \caption{}\label{fig:unfitted-approximation-d}
  \end{subfigure}
  \caption{Illustration of an unfitted \ac{fe} approximation for a bulk \ac{pde} in $\Omega$. \textbf{\textsf{(a)}} The \ac{fe} discretisation is built on the $\Omega$-\emph{active mesh} $\mathcal{T}_h^\Omega$ formed by the grid cells intersecting $\Omega$. The simplest scenario is to approximate the problem with a linear Lagrangian \ac{fe} space $\mathcal{V}_h^\Omega$, where the shape functions and degrees of freedom (in yellow circles) are associated to the vertices of $\mathcal{T}_h^\Omega$. \textbf{\textsf{(b)}} The \ac{fe} solution $u_h$ of the problem is found in $\mathcal{T}_h^\Omega$ and its restriction $u_h{_|{_\Omega}}$ is the approximation to the continuous problem. \textbf{\textsf{(c)}} The \emph{small cut cell problem}. Illustration of a badly cut cell $T$ with a very small cut portion in $\Omega$. If $a$ is a \ac{dof} of $\mathcal{V}_h^\Omega$, we see that its support inside $\Omega$ is so small that it can lead to ill-conditioning (see Equation~\eqref{eq:condition-number}). \textbf{\textsf{(d)}} A way to fix the small cut cell problem, proposed in, e.g., the Aggregated \ac{fem}~\cite{badia2018aggregated}, is to extrapolate the \ac{dof} value $u_a$ in terms of the \ac{dof} values of an interior cell $T(a)$. Purple arrows indicate the path mapping $a$ to $T(a)$ according to a cell aggregation scheme \added{described in Figure~\ref{fig:cell-aggregation}}.}
  \label{fig:unfitted-approximation}
\end{figure}

Our main idea to build the \ac{fe} approximation amounts to \emph{extend} the discrete problem from the region of interest to a subset of the cells in the fixed-grid. We illustrate the procedure for the bulk problem in Figures~\ref{fig:unfitted-approximation-a} and~\ref{fig:unfitted-approximation-b}. The first step is to find all grid cells that intersect $\Omega$, this gives the subset $\mathcal{T}_h^\Omega = \{ T \in \mathcal{T}_h : T \cap \Omega \neq \emptyset \}$, referred to as the $\Omega$-\emph{active mesh}. Next, we build the \ac{fe} discretisation in $\mathcal{T}_h^\Omega$. For instance, if we consider standard linear \ac{fe} spaces, the shape functions and \acp{dof} will be associated to the vertices of $\mathcal{T}_h^\Omega$. The \ac{fe} solution of the problem will thus be defined in $\mathcal{T}_h^\Omega$ and its restriction to $\Omega$ yields the sought-after approximation of the continuous problem. We proceed analogously for the surface problems, where the subset $\mathcal{T}_h^\Gamma = \{ T \in \mathcal{T}_h : T \cap \Gamma \neq \emptyset \} \subseteq \mathcal{T}_h^\Omega$ is referred to as the $\Gamma$-\emph{active} or \emph{cut mesh}. 

We readily see that discrete extensions provide a route to massively simplify the mesh generation step of \ac{fe} problems in general, by enabling their resolution on a fixed grid. However, this comes at a high cost. Unfitted methods suffer from three main drawbacks: (1) numerical integration on cut cells requires dedicated procedures; (2) essential (e.g.~Dirichlet) boundary conditions must be weakly imposed; and (3) naive discrete extensions often lead to (almost) singular linear systems, the so-called \emph{small cut-cell problem}. All these issues and possible remedies have been extensively reviewed in, e.g.,~\cite{de2023stability,schillinger2015finite}.

Of all three challenges, the small cut-cell problem is widely viewed as the most critical one, involving the biggest effort for its mitigation or elimination~\cite{Burman2015,de2021numerical}. Let us give an intuitive, yet general idea of the issue. When working with moving boundaries and interfaces, it is often impractical, if not almost impossible, to tune with manual intervention how the grid overlaps the geometry. In particular, we must cope with any type of intersection between the grid cells and the geometry, including arbitrarily small ones, see example in Figure~\ref{fig:unfitted-approximation-c}. In this situation, we may have basis functions with very reduced support inside the geometry. If using a classical \ac{fe} method, these basis functions lead to rows and columns in the linear system with values close to zero, a clear case of ill-conditioning.

In fact, rigorous mathematical analysis for the bulk problem in a fixed-grid~\cite{de2017condition} proves that the condition number $\kappa$ of a standard \ac{fe} system $\mathbf{A}$ for the Laplace operator satisfies
\begin{equation}
  \label{eq:condition-number}
  \kappa(\mathbf{A}) \sim \left[ \min_{T \in \mathcal{T}_h^\Omega} \frac{| \Omega \cap T |}{| T |} \right]^{-2q+1-2/d},
\end{equation}
where the base is the (relative) volume of the smallest cut fraction and, in the exponent, $q$ is the order of the \ac{fe} approximation and $d$ is the space dimension. Compared to the $h^{-2}$ scaling for the body-fitted case, Equation~\eqref{eq:condition-number} exposes the severity of the small cut-cell problem, especially in high-order methods. We expect similar estimates for surface problems.

\subsection{Unfitted FE approximations}
\label{subsec:space}

Given the above ill-conditioning issues and their special relevance in our context, we discretise Problem~\eqref{eq:full-non-dimensional-1}-\eqref{eq:full-non-dimensional-9} with unfitted \ac{fe} methods that are not affected by the small cut-cell problem. This means, in particular, that they are geometrically robust as to how the computational grid overlaps the geometry. To reduce the notational burden, \emph{we initially omit the time step superscripts of the viscous flow \ac{fe} problems}. We refer to Appendix~\ref{app:axisymmetric} for the 2D rotationally symmetric version of the 3D discrete problem described in this section.

\subsubsection{Bulk viscous flows}

Let $(\boldsymbol{u}_h,p_h)$ and $(\boldsymbol{v}_h,q_h)$ denote the pairs of trial and test functions of a suitable velocity-pressure \ac{fe} space. The bilinear forms arising from balance of momentum and mass (Equations~\eqref{eq:full-non-dimensional-3}-\eqref{eq:full-non-dimensional-4}) are given by
\[
  a_h(\boldsymbol{u}_h,\boldsymbol{v}_h) = \frac{2 R}{L_\eta} \int_\Omega \boldsymbol{\varepsilon} (\boldsymbol{u}_h) : \boldsymbol{\varepsilon} (\boldsymbol{v}_h) \ \mathrm{d}\Omega
  \qquad \text{and} \qquad
  b_h(\boldsymbol{u}_h,p_h) = - \int_\Omega p_h ( \nabla \cdot \boldsymbol{u}_h ) \ \mathrm{d}\Omega.
\]
As there are no mesh nodes on the boundary $\Gamma$, it is not straightforward to enforce the Dirichlet condition of Equation~\eqref{eq:full-non-dimensional-5} in the usual \emph{strong} sense, that is, by prescribing the value of the condition at the boundary nodes. For this reason, we impose it in a \emph{weak} sense, that is, by augmenting the variational formulaton with additional terms~\cite{fernandez2004imposing}. Usage of the widely employed Nitsche method~\cite{benzaken2024constructing,nitsche1971uber} gives rise to the forms
\[
  \begin{aligned}
    i_h(\boldsymbol{u}_h,p_h,\boldsymbol{v}_h,q_h) &= \int_\Gamma \frac{\alpha}{h} \ \boldsymbol{u}_h \cdot \boldsymbol{v}_h - \left[ \frac{2 R}{L_\eta} \boldsymbol{\varepsilon} ( \boldsymbol{u}_h ) \boldsymbol{n}_\Gamma - p_h \boldsymbol{n}_\Gamma \right] \cdot \boldsymbol{v}_h - \left[ \frac{2 R}{L_\eta} \boldsymbol{\varepsilon} ( \boldsymbol{v}_h ) \boldsymbol{n}_\Gamma - q_h \boldsymbol{n}_\Gamma \right] \cdot \boldsymbol{u}_h
    \ \mathrm{d}\Gamma \\
    \text{and} \quad j_h(\boldsymbol{v}_h,q_h;\boldsymbol{U}_h) &= \int_\Gamma \frac{\alpha}{h} \ \boldsymbol{U}_h \cdot \boldsymbol{v}_h - \left[ \frac{2 R}{L_\eta} \boldsymbol{\varepsilon} ( \boldsymbol{v}_h ) \boldsymbol{n}_\Gamma - q_h \boldsymbol{n}_\Gamma \right] \cdot \boldsymbol{U}_h
    \ \mathrm{d}\Gamma,
  \end{aligned}
\]
where $\alpha > 0$ is a large-enough coefficient to ensure coercivity of the weak form and $\boldsymbol{U}_h$ represents the discrete velocity at the surface, thereby marking the coupling with the \ac{fe} problem for the surface flow.

In order to state the discrete problem, it remains to choose the \ac{fe} space for the velocity-pressure pair $(\boldsymbol{u}_h,p_h)$. Our goal is to leverage the Lagrangian \ac{fe} space $\mathcal{V}_h^\Omega$ in $\mathcal{T}_h^\Omega$ formed by the pair $\mathcal{Q}_2 \times \mathcal{P}_1$ of (element-wise) continuous quadratic velocities and discontinuous linear pressures (restricted to zero mean global pressures). If we were in the body-fitted case, $\mathcal{V}_h^\Omega$ would be a classical choice, endowed with a discrete inf-sup stability condition~\cite{ern2004theory}. Unfortunately, $\mathcal{V}_h^\Omega$ does not preserve this key stability property in the unfitted case and, in presence of small cut elements, suffers from the ill-conditioning issues explained in Section~\ref{subsec:embedded}.

A natural approach to avert the numerical instabilities is to eliminate from the linear system all \acp{dof} linked to basis functions with too small support inside $\Omega$, subsequently referred to as \emph{ill-posed} \acp{dof}. This can be achieved, among others, with the \emph{Aggregated Finite Element Method} (\ac{agfem})~\cite{badia2018aggregated}. The main idea of \ac{agfem} is to \emph{extrapolate} the values of ill-posed \acp{dof} in terms of the values of \acp{dof} sitting on a cell with large-enough intersection with $\Omega$. For simplicity, we directly assume this cell is $\Omega$-\emph{interior}, that is, strictly inside $\Omega$. The extrapolation amounts to enforce linear \ac{dof} constraints of the form
\begin{equation}\label{eq:agfem-constraint}
  u_a = \sum_{b \in \ T(a)} u_b \varphi_b(x_a),
\end{equation}
in which $u_a$ represents the value of a generic ill-posed \ac{dof} $a$, and $b$ iterates over the \ac{dof} values $u_b$ and shape functions $\varphi_b$ of an interior cell $T(a) \in \mathcal{T}_h^\Omega$. An example of this type of constraint is illustrated in Figure~\ref{fig:unfitted-approximation-d}.

\begin{figure}[h!]
  \centering
  \begin{subfigure}{0.24\textwidth}
    \centering
    \includegraphics[height=4.5cm]{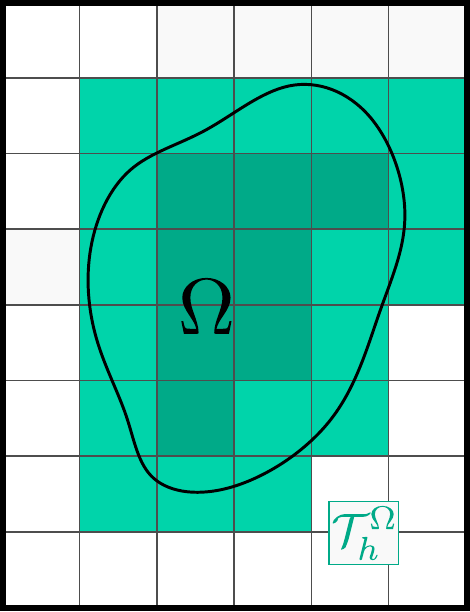}
    \caption{}\label{fig:cell-aggregation-a}
  \end{subfigure}
  \begin{subfigure}{0.24\textwidth}
    \centering
    \includegraphics[height=4.5cm]{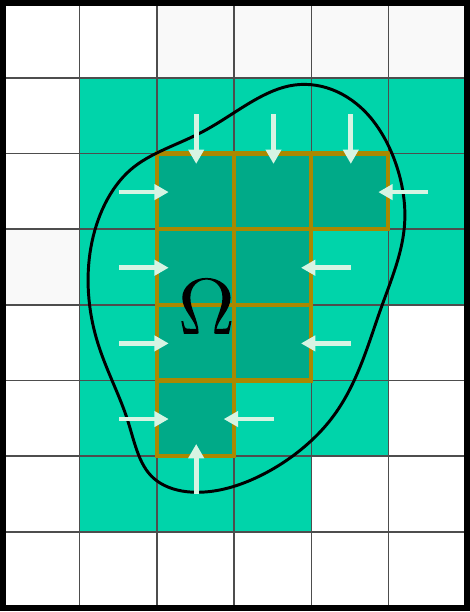}
    \caption{}\label{fig:cell-aggregation-b}
  \end{subfigure}
  \begin{subfigure}{0.24\textwidth}
    \centering
    \includegraphics[height=4.5cm]{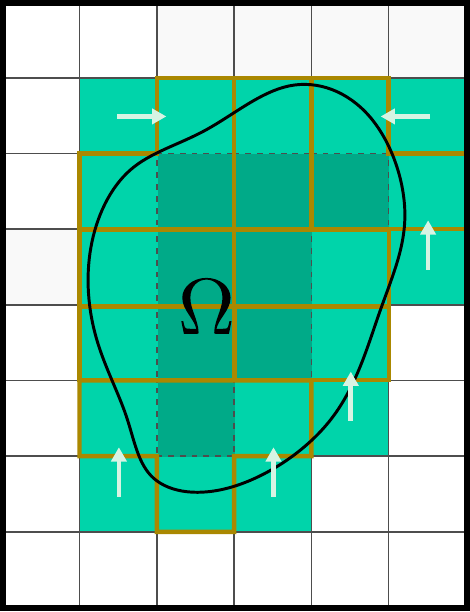}
    \caption{}\label{fig:cell-aggregation-c}
  \end{subfigure}
  \begin{subfigure}{0.24\textwidth}
    \centering
    \includegraphics[height=4.5cm]{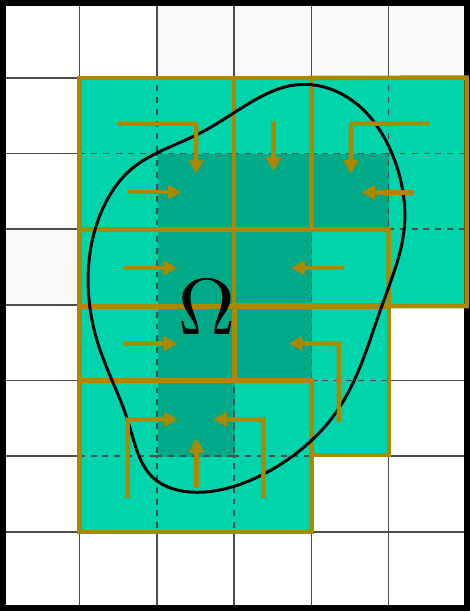}
    \caption{}\label{fig:cell-aggregation-d}
  \end{subfigure}
  \caption{\added{Illustration of the cell aggregation scheme in $\mathcal{T}_h^\Omega$. \textbf{\textsf{(a)}} The goal of the scheme is to map every cut cell (\emph{light green}) to a (unique) interior cell (\emph{dark green}). This is achieved by iteratively growing cell aggregates (\emph{yellow borders}) from interior to cut cells. \textbf{\textsf{(b)}} Given initial aggregates, corresponding to the interior cells, we find all facet-connected cut cell neighbours to the aggregates (\emph{white arrows}). \textbf{\textsf{(c)}} Next, we merge all found cut cells into their aggregates and repeat the procedure until aggregating all cut cells into a single interior cell. \textbf{\textsf{(d)}} The cell paths (\emph{yellow arrows}) formed in the aggregates yield the sought-after map. We refer to~\cite{badia2018aggregated,verdugo2019distributed} for the algorithmic description of the scheme.}}
  \label{fig:cell-aggregation}
\end{figure}

We readily see that to define and evaluate the above constraints in $\mathcal{V}_h^\Omega$, we need to detect and gather all its ill-posed \acp{dof} and assign to each one of them a suitable interior cell. For this, we resort to an automatic and easy-to-parallelise \emph{cell aggregation} scheme~\cite{verdugo2019distributed}, \added{in Figure~\ref{fig:cell-aggregation}}, which derives this map from cell paths linking badly cut cells to interior cells. Note that, for good numerical accuracy and stability, cell paths should be as short as possible and connected through interior or cut facets~\cite[Lemma 2.2]{badia2018aggregated}.

Thanks to cell aggregation, we can apply Equation~\eqref{eq:agfem-constraint} to all ill-posed \acp{dof}, pruning all basis functions associated with badly cut cells. This results in a restriction of the original \ac{fe} space $\mathcal{V}_h^\Omega$ to the so-called \emph{aggregated \ac{fe} space} $\mathcal{V}_{h,\rm ag}^\Omega$. In this way, aggregated \ac{fe} spaces fix the ill-conditioning issues of vanilla unfitted \ac{fem}; they are stable, regardless of the configuration of the cut cells, and retain the optimal approximation properties and condition number bounds of body-fitted \ac{fe} methods. We refer to, e.g.,~\cite{badia2018aggregated,burman2022cutfem} for fundamental theoretical analysis and results about Ag\ac{fem}. 

Our particular realisation of $\mathcal{V}_{h,\rm ag}^\Omega$ for the Stokes problem follows the robust and stable mixed formulation introduced in~\cite{Badia2018Mixed}, opting for a serendipity extension and no pressure stabilisation. Hence, the discretisation of the bulk problem given by Equations~\eqref{eq:full-non-dimensional-3}-\eqref{eq:full-non-dimensional-5} finally reads: \emph{Find $(\boldsymbol{u}_h,p_h) \in \mathcal{V}_{h,\rm ag}^\Omega$ such that}
\begin{equation}
  \label{eq:discrete-bulk-problem}
  a_h(\boldsymbol{u}_h,\boldsymbol{v}_h) + b_h(\boldsymbol{v}_h,p_h) + b_h(\boldsymbol{u}_h,q_h) + i_h(\boldsymbol{u}_h,p_h,\boldsymbol{v}_h,q_h) = j_h(\boldsymbol{v}_h,q_h;\boldsymbol{U}_h), \enskip \forall\, (\boldsymbol{v}_h,q_h) \in \mathcal{V}_{h,\rm ag}^\Omega.
\end{equation}

\subsubsection{Surface viscous flows}

Let now $\boldsymbol{U}_h$ and $\boldsymbol{V}_h$ denote the trial and test functions of a vector-valued \ac{fe} space, that will be later defined. From balance of momentum (Equation~\eqref{eq:full-non-dimensional-1}), the bilinear form is
\[
  A_h^U(\boldsymbol{U}_h,\boldsymbol{V}_h) = \int_\Gamma 2 \, \boldsymbol{\varepsilon}_\Gamma (\boldsymbol{U}_h) : \boldsymbol{\varepsilon}_\Gamma ( \boldsymbol{V}_h ) + \rho \, \boldsymbol{U}_h \cdot \boldsymbol{V}_h \ \mathrm{d}\Gamma
\]
with a very small friction term premultiplied by $\rho \! \ll \! 1$ required, when $\Gamma$ is a closed surface, to eliminate Killing vector fields~\cite{sakai1996riemannian}. The right-hand side terms
\[
  \begin{aligned}
    F_h^{\rm act}(\boldsymbol{V}_h;C_h) &= \int_\Gamma \mathrm{div}_\Gamma ( \mbox{\textit{Pe}} f(C_h,1) \mathbf{P} ) \cdot \boldsymbol{V}_h  \ \mathrm{d}\Gamma \\
    F_h^{\rm cyt}(\boldsymbol{V}_h;\boldsymbol{u}_h,p_h) &= - \int_\Gamma \left[ \frac{2 R}{L_\eta} \boldsymbol{\varepsilon} ( \boldsymbol{u}_h ) \boldsymbol{n}_\Gamma - p_h \boldsymbol{n}_\Gamma \right] \cdot \boldsymbol{V}_h \ \mathrm{d}\Gamma
  \end{aligned}
\]
correspond to the active $F_h^{\rm act}$ and cytoplasm reaction $F_h^{\rm cyt}$ forces. We observe that the former couples the problem to the molecular surface transport via $C_h$ and the latter to the bulk flows via $(\boldsymbol{u}_h,p_h)$.

For surface problems, robustness to cut location is attained by perturbing the problem with a stabilisation term, in the spirit of trace \ac{fe} methods~\cite{Olshanskii2017}. In particular, we adopt the \emph{normal derivative volume stabilisation}, since it leads to optimal discretisation errors and condition number bounds for the vector Laplacian problem on \emph{fixed} surfaces~\cite{jankuhn2021trace,burman2018cut}:
\[
  S_h^U(\boldsymbol{U}_h,\boldsymbol{V}_h) = \int_{\mathcal{N}_h^\Gamma} \frac{\beta}{h} \ \boldsymbol{\varepsilon}_\Gamma (\boldsymbol{U}_h) \boldsymbol{n}_\Gamma \cdot \boldsymbol{\varepsilon}_\Gamma (\boldsymbol{V}_h) \boldsymbol{n}_\Gamma \ \mathrm{d}\Omega,
\]
where $\mathcal{N}_h^\Gamma = \bigcup_{T \in \mathcal{T}_h^\Gamma} \overline{T}$ is the domain covered by $\mathcal{T}_h^\Gamma$ and $\beta > 0$ is a large-enough coefficient to ensure well-posedness. We observe that $S_h^U(\boldsymbol{U}_h,\boldsymbol{V}_h)$ implies integrating the basis functions in the whole cells cut by $\Gamma$, which necessarily eliminates the instabilities from small intersections of $\Gamma$ with $T \in \mathcal{T}_h^\Gamma$. On the other hand, the effect of this term is to extend the solution from $\Gamma$ to $\mathcal{N}_h^\Gamma$, by \emph{weakly} imposing that the value on the surface is constant through the normal $\boldsymbol{n}_\Gamma$. In doing so, it crucially fixes another key problem of unfitted \ac{fem} for surface \acp{pde}: the fact that bulk basis functions defined in $\mathcal{T}_h^\Gamma$ restricted on $\Gamma$ are, in general, linearly dependent. We refer to, e.g.,~\cite[Figure 6]{schollhammer2021higher} for a clear illustration of the linear dependency in bilinear quadrilateral cells.

The discretisation of the surface flow problem is completed with the choice of a suitable \ac{fe} space. In this case, we take a standard vector-valued linear Lagrangian \ac{fe} space $\mathcal{V}_h^\Gamma$. In order to segregate the surface and bulk problems for the iterative scheme, we must further restrict the rigid body modes of the surface and account for the compatibility with bulk incompressibility~\eqref{eq:full-non-dimensional-4}. Hence, the problem is approximated in the reduced \ac{fe} space
\[
  \mathcal{V}_{h,*}^\Gamma = \left\{ \, \boldsymbol{v}_h \in \mathcal{V}_h^\Gamma : \int_\Gamma \boldsymbol{v}_h = 0, \int_\Gamma \nabla \times \boldsymbol{v}_h = 0, \int_\Gamma \boldsymbol{v}_h \cdot \boldsymbol{n}_\Gamma = 0 \, \right\}
\]
and the discrete surface flow problem given by Equation~\eqref{eq:full-non-dimensional-1} reads: \emph{Find $\boldsymbol{U}_h \in \mathcal{V}_{h,*}^\Gamma$ such that}
\begin{equation}
  \label{eq:discrete-surface-flow-problem}
  A_h^U(\boldsymbol{U}_h,\boldsymbol{V}_h) + S_h^U(\boldsymbol{U}_h,\boldsymbol{V}_h) = F_h^{\rm act}(\boldsymbol{V}_h;C_h) + F_h^{\rm cyt}(\boldsymbol{V}_h;\boldsymbol{u}_h,\boldsymbol{v}_h), \quad \forall\, \boldsymbol{V}_h \in \mathcal{V}_{h,*}^\Gamma.
\end{equation}

\subsubsection{Surface molecular transport}

For the \ac{fe} formulation of the surface transport problem given by Equations~\eqref{eq:full-non-dimensional-2} and~\eqref{eq:full-non-dimensional-7}, we introduce minimal notation for the discretisation in time. Given the time domain $[0,T]$, we assume 
a uniform partition with time step $\Delta \mathrm{t} = T / N$ in $N$ time intervals $I_n = [\mathrm{t}^{n-1},\mathrm{t}^n)$, $\mathrm{t}^n = n \Delta \mathrm{t}$, $n = 1, \ldots, N$. In what follows, we use the supercript $n$ to denote evaluation at $t^n$, e.g., $\boldsymbol{U}_h^n = \boldsymbol{U}_h(\mathrm{t}^n)$ or $\Gamma^n := \Gamma(\mathrm{t}^n)$.

We derive the discrete problem for an implicit Euler scheme in time and a linear scalar-valued (transient) Lagrangian \ac{fe} space $\mathcal{W}_h^\Gamma$. Letting $C_h^n$ and $D_h^n$ represent trial and test functions of $\mathcal{W}_h^\Gamma$, we introduce the forms
\[
  M_h(C_h^n,D_h^n) = \int_{\Gamma^n} \! C_h^n D_h^n \ \mathrm{d}\Gamma,
  \quad A_h^C(C_h^n,D_h^n) = \int_{\Gamma^n} \! \nabla C_h^n \cdot \nabla D_h^n \ \mathrm{d}\Gamma,
  \quad \text{and} \quad L_h(D_h^n) = \int_{\Gamma^n} \! D_h^n \ \mathrm{d}\Gamma,
\]
representing unit mass, diffusion and source terms of the problem. Like the previous case, we stabilise the problem with the form
\[
  S_h^C(C_h^n,D_h^n) = \int_{\mathcal{N}_h^{\Gamma^n}} \! \frac{\gamma}{h} \ ( \nabla C_h^n \cdot \boldsymbol{n}_\Gamma ) ( \nabla D_h^n \cdot \boldsymbol{n}_\Gamma ) \ \mathrm{d}\Omega,
\]
where $\gamma > 0$ is a large-enough penalty coefficient. The particular choice of $S_h^C(C_h^n,D_h^n)$ is on the grounds of numerical stability and optimal error estimates proven in, e.g.,~\cite{lehrenfeld2018stabilized}. Finally, the coupling term with $\boldsymbol{U}_h^n$ reads
\[
  B_h(C_h^n,D_h^n;\boldsymbol{U}_h^n) = \int_{\Gamma^n} \left( \boldsymbol{U}_h \cdot \nabla_\Gamma C_h^n \right) D_h^n + \left( \mathrm{div}_\Gamma \boldsymbol{U}_h^n \right) C_h^n D_h^n \ \mathrm{d}\Gamma,
\]
where the fact that $C_h^n$ is weakly extended along the normal allows us to consider the surface gradient in the convective term of the material derivative. With these ingredients, the discrete surface transport problem reads: \emph{For $C_h^0 \in \mathcal{W}_h^\Gamma$ approximating the initial condition~\eqref{eq:full-non-dimensional-7}, find $C_h^n \in \mathcal{W}_h^\Gamma$, $n = 1, \ldots, N$ such that}
\begin{equation}
  \label{eq:discrete-surface-transport-problem}
  \begin{aligned}
    \left( \frac{1}{\Delta \mathrm{t}} + \tau_D k_{\rm off} \right) M_h(C_h^n,D_h^n) + B_h(C_h^n,D_h^n;\boldsymbol{U}_h^n) + A_h^C(C_h^n,D_h^n) + S_h^C(C_h^n,D_h^n) &= \\ \frac{1}{\Delta \mathrm{t}} M_h(C_h^{n-1},D_h^n) + \tau_D k_{\rm off} L_h(D_h^n), \qquad \forall\, D_h^n \in \mathcal{W}_{h}^\Gamma.&
  \end{aligned}
\end{equation}
We observe that the above formulation conserves mass only approximately, due to the discretisation error in time and evaluating $C_h^{n-1}$ on $\Gamma^n$ (see end of Section~\ref{subsec:coupling}). However, in the numerical experiments of Section~\ref{sec:experiments}, we notice that the error in mass conservation is reasonably controlled by the small time increments required to capture the problem nonlinearities. In other situations, we could require a conservative scheme as in, e.g.~\cite{olshanskii2014error,myrback2024high}. On the other hand, we use a SUPG variant~\cite{olshanskii2014stabilized} to stabilise high-activity cases ($\mbox{\textit{Pe}} \gg 1$) that lead to advection-dominated transport.

\subsubsection{Numerical integration}\label{subsubsec:integration}

Except for the stabilisation terms, all integrals involved in the above weak formulations are defined in either $\Omega$ or $\Gamma$. Clearly, their evaluation on cut cells cannot be achieved with standard quadrature rules, because we need to restrict the integration to the portions $\Omega \cap T$ or $\Gamma \cap T$, $T \! \in \! \mathcal{T}_h^\Gamma$. There are many strategies to adapt the local integration on cut cells, reviewed in, e.g.,~\cite{de2023stability,saye2022high}; they attempt, in most cases, to strike a good balance between accuracy and computational cost. In this work, we adopt the high-order quadrature rules introduced in~\cite{saye2022high} for curved surfaces and volumes implicitly defined by the level sets of multivariate polynomials. Underpinning our choice is the fact that shape dynamics are governed by the discrete \ac{fe} unknown $\boldsymbol{U}_h$, which calls for the convenient representation of the evolving surface with piecewise polynomial level-sets (see next section).

\subsection{Surface evolution}
\label{subsec:time}

We outline now the evolving level-set method for shape evolution, as governed by Equations~\eqref{eq:full-non-dimensional-8} and~\eqref{eq:full-non-dimensional-9}. First, we use the level-set description of the interface $\Gamma(\mathrm{t})$, introduced in Section~\ref{subsec:embedded}, to recast the system into the classic initial value formulation~\cite{osher1988fronts}
\begin{equation}
  \label{eq:level-set-evolution-continuous}
  \begin{aligned}
    \partial_t \phi + U_{\perp}^{\rm e} \| \nabla \phi \| = 0, \qquad \qquad &\text{in} \ \mathcal{B} \times [0,T], \\
    \phi(\boldsymbol{x},0) = \phi_0(\boldsymbol{x}), \qquad \qquad &\text{in} \ \mathcal{B} \times \{0\}.
  \end{aligned}
\end{equation}
The above system describes the (pure) advection of the level-set field $\phi$ in the bounding-box $\mathcal{B}$ with speed function $U_{\perp}^{\rm e}$, which denotes an appropriate extension of the surface normal velocities $\boldsymbol{U} \cdot \boldsymbol{n}_\Gamma$ from $\Gamma$ to $\mathcal{B}$.

In order to accommodate our numerical integration strategy, see Section~\ref{subsubsec:integration}, we consider a \ac{fe} discretisation of the level-set function $\phi$ in $\mathcal{B}$. According to this, we assume $\mathcal{V}_h^\mathcal{B}$ is a scalar-valued Lagrangian \ac{fe} space of order two, the same as the bulk velocity approximation. Considering an explicit Euler scheme in time leads to the update rule: \emph{For $\phi^0 \in \mathcal{V}_h^\mathcal{B}$ approximating the initial condition $\phi_0$, find $\phi^n \in \mathcal{V}_h^\mathcal{B}$, $n = 1, \dots, N$ such that}
\begin{equation}
  \label{eq:discrete-level-set-evolution}
  \phi^n = \phi^{n-1} - \Delta \mathrm{t} \, U_{\perp}^{\mathrm{e},n-1} \| \nabla \phi^{n-1} \| = 0, \qquad \phi^n \in \mathcal{V}_h^\mathcal{B}, \qquad n \geq 1.
\end{equation}
We observe that, at each time step, the rule above only involves updating the nodal values of the discrete level-set function. In order to evaluate $U_{\perp}^{\mathrm{e},n-1}$, we compute the value of $\boldsymbol{U} \cdot \boldsymbol{n}_\Gamma$ at the closest point projection to $\Gamma$ of each node of $\mathcal{V}_h^\mathcal{B}$. To this end, we resort to the high-order algorithms devised in~\cite{saye2014high}. In our numerical tests, we do not require to stabilise the scheme with, e.g., upwind methods, because the time steps are rather small. We further leverage closest point projections to represent $\phi^n$ as nodal projections of signed distance functions; in this way, we can assume $\| \nabla \phi^{n-1} \| \approx 1$. 


\subsection{Solution strategy}
\label{subsec:coupling}

\begin{figure}[h!]
  \centering
  \includegraphics[width=\textwidth]{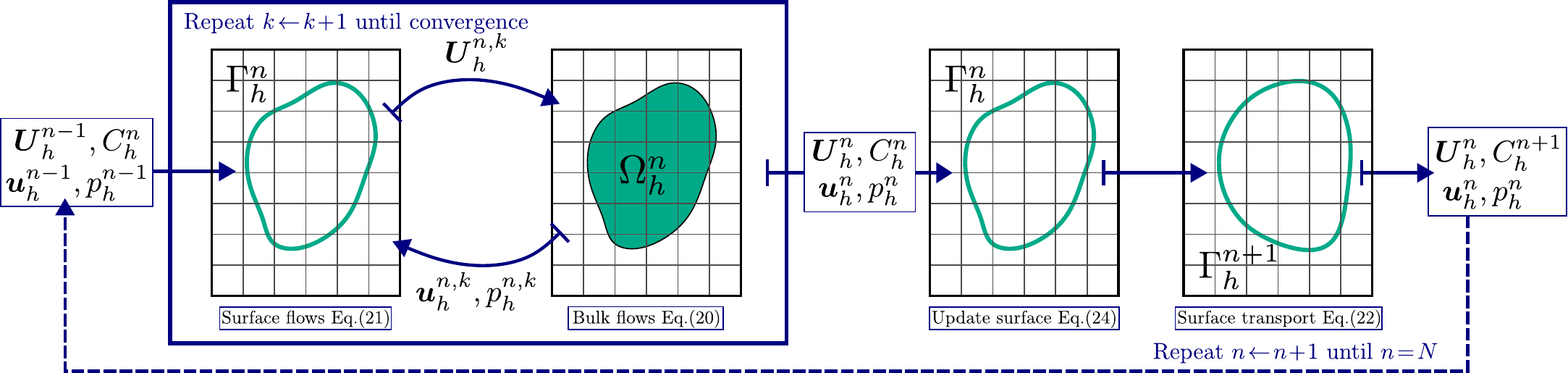}
  \caption{Schematic representation of the coupled iterative scheme described in Algorithm~\ref{alg:main-loop}.}
  \label{fig:coupling-scheme}
\end{figure}

The coupling scheme of the full transient problem is represented in~Figure~\ref{fig:coupling-scheme}. We consider an explicit coupling in time (by substitution) between the surface transport and the coupled surface-bulk flow problem. The latter is also solved with an iterative surface-vs-bulk partitioned scheme. We then update the surface shape with the resulting surface velocities and close the simulation loop to compute the solution at the following time step. This can be expressed altogether in Algorithm~\ref{alg:main-loop}, which encodes the main simulation loop.

\begin{algorithm}
  \caption{Coupled iterative scheme for the \ac{fe} approximation of Problem~\eqref{eq:full-non-dimensional-1}-\eqref{eq:full-non-dimensional-9}.}\label{alg:main-loop}
  \begin{algorithmic}[1]
    \State Given $\boldsymbol{U}_h^{-1}$, $\boldsymbol{u}_h^{-1}$, $p_h^{-1}$, $C_h^0$ and $\phi_h^0$,
    \For{$n = 0, \ldots, N$}
    \State $k \gets 0$
    \State $\boldsymbol{U}_h^{n,0}, \boldsymbol{u}_h^{n,0}, p_h^{n,0} \gets \boldsymbol{U}_h^{n-1}, \boldsymbol{u}_h^{n-1}, p_h^{n-1}$
    \Repeat
    \State $k \gets k+1$
    \State{$\boldsymbol{U}_h^{n,k} \gets$ Solve \eqref{eq:discrete-surface-flow-problem} with $\boldsymbol{u}_h^{n,k-1}$, $p_h^{n,k-1}$ and $C_h^n$ in $\Gamma_h^n = \{ \phi_h^n \equiv 0 \}$}\label{line:carry-bulk}
    \State{$\boldsymbol{u}_h^{n,k}, p_h^{n,k} \gets$ Solve \eqref{eq:discrete-bulk-problem} with $\boldsymbol{U}_h^{n,k}$ in $\Omega_h^n = \{ \phi_h^n < 0 \}$}
    \Until{$\boldsymbol{U}_h^{n,k} \approx \boldsymbol{U}_h^{n,k-1}$, $\boldsymbol{u}_h^{n,k} \approx \boldsymbol{u}_h^{n,k-1}$} and $p_h^{n,k} \approx p_h^{n,k-1}$
    \State $\boldsymbol{U}_h^n, \boldsymbol{u}_h^n, p_h^n \gets \boldsymbol{U}_h^{n,k}, \boldsymbol{u}_h^{n,k}, p_h^{n,k}$
    \If{$n < N$}
    \State{$\phi_h^{n+1} \gets$ Update \eqref{eq:discrete-level-set-evolution} with $\phi_h^n$ and $\boldsymbol{U}_h^n$ in $\mathcal{B}$}
    \State{$C_h^{n+1} \gets$ Solve \eqref{eq:discrete-surface-transport-problem} with $\boldsymbol{U}_h^n$ and $C_h^n$ in $\Gamma_h^{n+1} = \{ \phi_h^{n+1} \equiv 0 \}$}\label{line:carry-surface}
    \EndIf
    \EndFor
  \end{algorithmic}
\end{algorithm}

Algorithm~\ref{alg:main-loop} exposes the need to carry information between consecutive discrete surfaces. In particular, the bulk unknowns $(\boldsymbol{u}_h^{n-1},p_h^{n-1})$, found in $\Omega_h^{n-1}$, have to be evaluated in $\Gamma_h^n = \partial \Omega^n$ (line~\ref{line:carry-bulk}); while the surface unknowns $\boldsymbol{U}_h^n$, $C_h^n$, found in $\Gamma_h^n$, have to be evaluated in $\Gamma_h^{n+1}$ (line~\ref{line:carry-surface}). In order to ensure this, we follow a similar approach to the one detailed in, e.g.,~\cite{lehrenfeld2018stabilized} for surface and~\cite{lehrenfeld2019eulerian,burman2022eulerian} for bulk problems. The main idea is to extend the discrete problems further away from the cells cut by the current surface/boundary, until they cover a region wide enough to contain the next surface (which is not known). This is achieved by formulating the unfitted surface \ac{fe} problems in $\mathcal{T}_{h,n}^\Gamma = \left\{ T \in \mathcal{T}_h : T \cap \{ | \phi_h^n | < \delta^n \} \neq \emptyset \right\}$ and the unfitted bulk \ac{fe} problem in $\mathcal{T}_{h,n}^\Omega = \left\{ T \in \mathcal{T}_h : T \cap \{ \phi_h^n < \delta^n \} \neq \emptyset \right\}$, where $\delta^n \propto \Delta {\rm t} \, \| \boldsymbol{U}_h^{n-1} \|_\infty$ is a width parameter. We refer to the above references for a complete description and discussion of this strategy.

\added{Regarding the stability of the coupling scheme, numerical experiments in Section~\ref{sec:experiments} adopt rather small time steps $\Delta {\rm t} \approx 10^{-4}-10^{-5}$ to correctly capture fast shape changes and, in particular, sharp nonlinear transitions of self-organised shape emergence, as in Figure~\ref{fig:pattern}. Hence, we reasonably assume stability of the scheme for such small time increments. For larger time steps, a rigorous stability analysis must be carried out.}

\section{Numerical experiments}
\label{sec:experiments}


Numerical implementation of the unfitted \ac{fe} formulation in Section~\ref{sec:discrete} was carried out using the \texttt{Gridap.jl}~\cite{Badia2020,Verdugo2022} \ac{fe} software ecosystem, written in the Julia programming language. The code is available at the \href{https://github.com/ericneiva/SurfaceBulkViscousFlows}{\texttt{SurfaceBulkViscousFlows}} GitHub repository and supplemented with demonstrators of the numerical examples that follow. The unfitted \ac{fe} tools are provided by \texttt{GridapEmbedded.jl}~\cite{gridapembedded_0_9_5}. It interfaces to the C++ \texttt{algoim}~\cite{algoim} library, via a Julia wrapper~\cite{algoimwrapper}, to compute quadratures for domains implicitly defined by multivariate polynomials~\cite{saye2022high} and closest point projections on implicit surfaces~\cite{saye2014high}. We solve our linear systems with the parallel sparse direct solver MUMPS~\cite{MUMPS:1,MUMPS:2} available at the PETSc library v3.15.2~\cite{petsc-web-page,petsc-user-ref,petsc-efficient}. Numerical results are postprocessed with ParaView v5.10.0~\cite{AHRENS2005717}.

In all cases, (1) we discretise the bounding box $\mathcal{B}$ of the problem with Cartesian grids of uniform mesh size; (2) the initial level-set $\phi_0 \in \mathcal{V}_h^{\mathcal{B}}$ is defined by the closest-point projections of the \ac{fe} nodes of $\mathcal{V}_h^{\mathcal{B}}$ on the surface of analysis (typically a sphere); and (3) the stabilisation coefficients are fixed at $\alpha = 20.0$, $\beta = \gamma = 10.0$ (on the conservative side) and $\rho = 10^{-3}$. We include examples with both the 2D rotationally symmetric and 3D versions of the nondimensional system given by Equations~\eqref{eq:full-non-dimensional-1}-\eqref{eq:full-non-dimensional-9}.


\subsection{Verification examples}
\label{subsec:verification}

We carry out two convergence tests to validate our code implementation. In the first experiment, we consider a steady-state 2D axisymmetric problem in a \emph{fixed geometry with known analytical solution}. To this end, we transform the fluid sphere problem with two bulk phases of~\cite[Section 4-21]{happel2012low} into a surface-bulk viscous flow problem. This leads to a coupled surface-bulk Stokes problem where (for this experiment only) the surface is material and, as a result, obeys the surface incompressibility condition. Using the method of manufactured solutions, we set the data of the problem such that the general solution for the surface and bulk phases, given by the stream functions in polar coordinates, reads
\[
  \begin{aligned}
    \psi_\Gamma &= \sin^2 \theta \left( - \frac{1}{2} B r + C r^2 + \frac{D}{r} \right), \quad &r = 1 \\
    \psi_\Omega &= \sin^2 \theta \left( \frac{1}{10} E r^4 + G r^2 \right), \quad &r < 1
  \end{aligned}
\]
with
\[
  B = \frac{3}{2} \frac{1 + \frac{2}{3} \sigma}{1 + \sigma}, \quad C = \frac{1}{2}, \quad D = \frac{1}{4} \frac{1}{1 + \sigma}, \quad E = \frac{5}{2} \frac{\sigma}{1 + \sigma}, \quad G = - \frac{1}{4} \frac{\sigma}{1 + \sigma}, \quad \text{and} \ \sigma = \frac{\mu_\Gamma}{\mu_\Omega}.
\]

\begin{figure}[!ht]
  \centering
  \begin{subfigure}{0.32\textwidth}
    \centering
    \includegraphics[width=\textwidth]{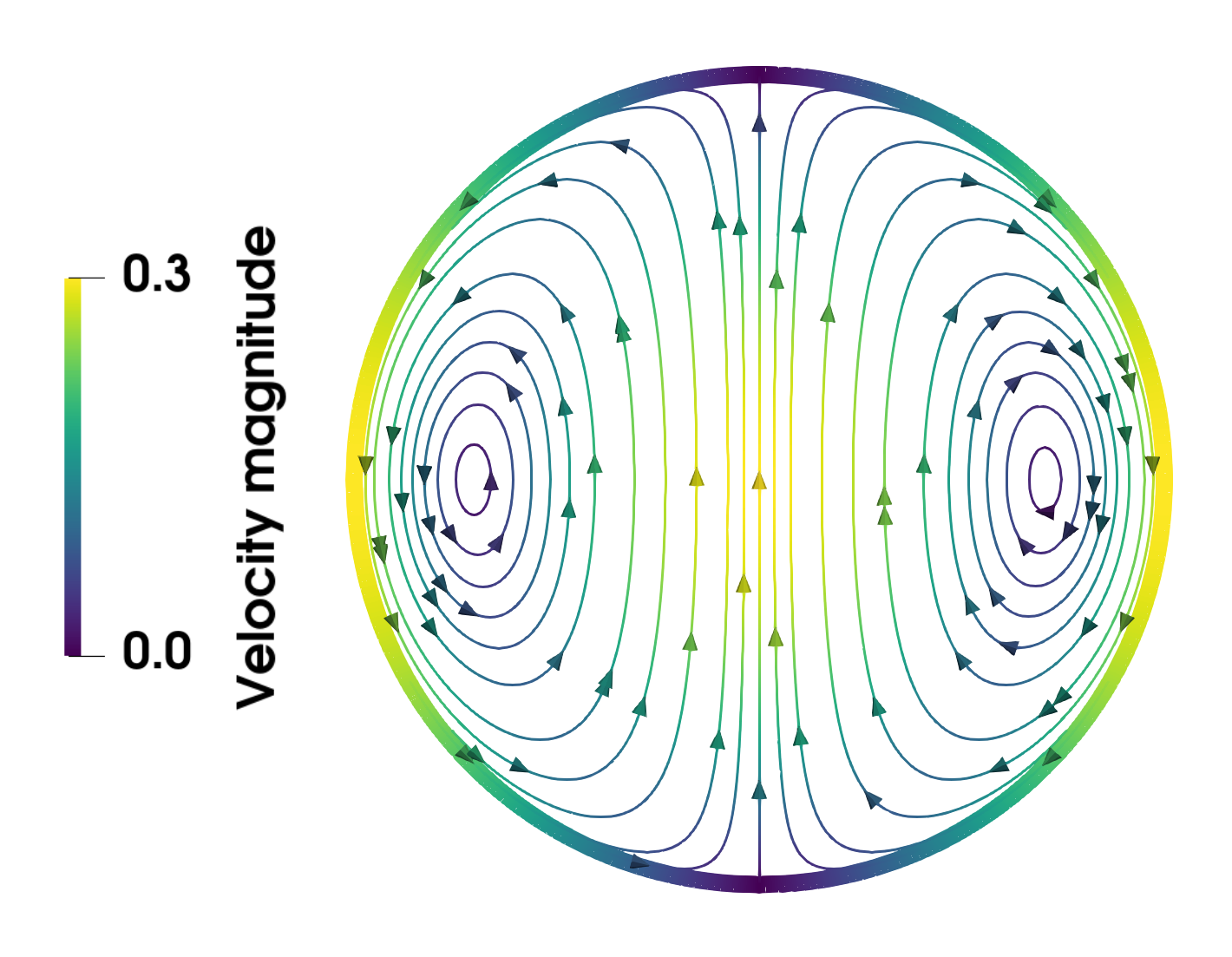}
    \caption{}\label{fig:verification-a}
  \end{subfigure}
  \begin{subfigure}{0.32\textwidth}
    \centering
    \includegraphics[width=\textwidth]{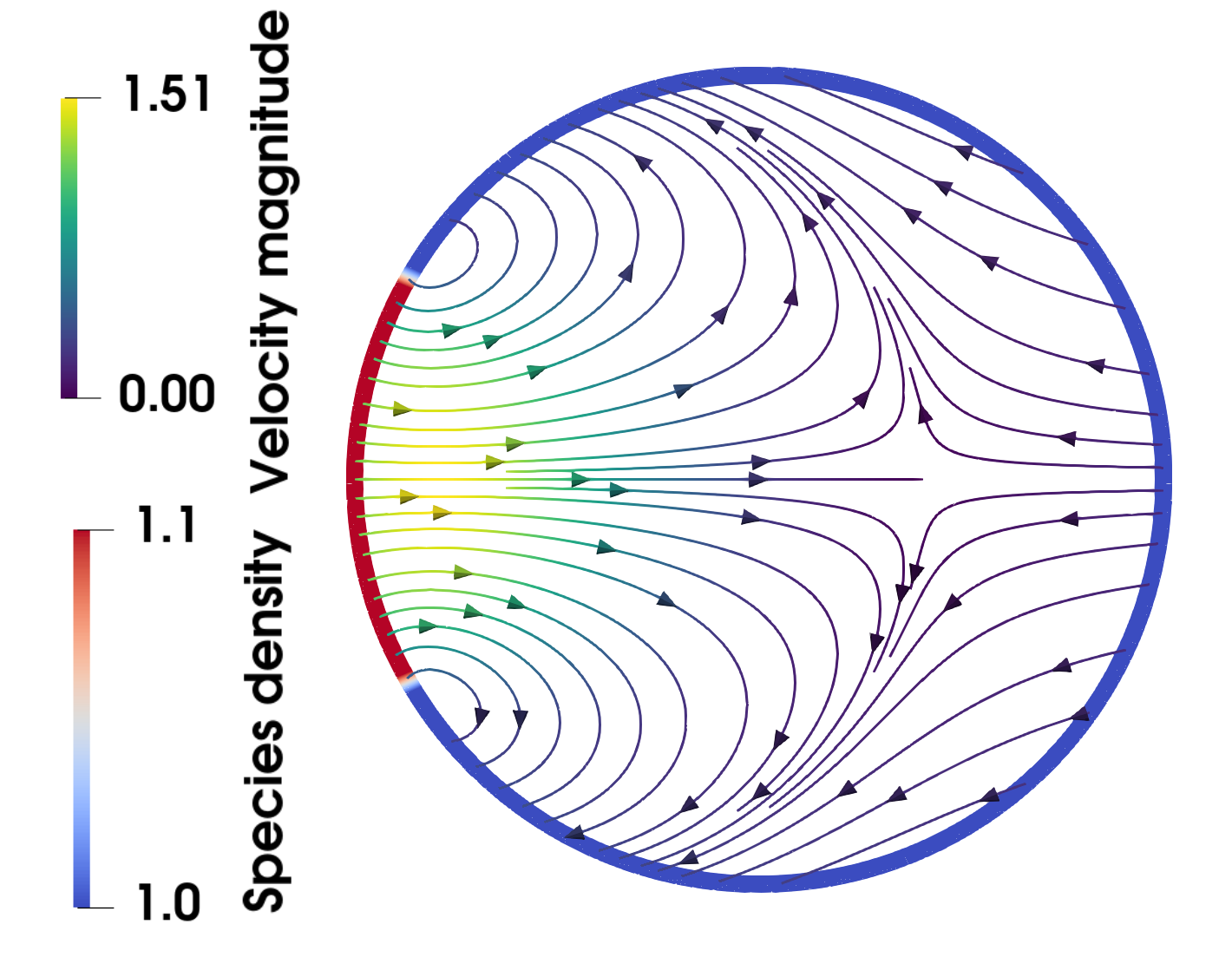}
    \caption{}\label{fig:verification-b}
  \end{subfigure}
  \begin{subfigure}{0.32\textwidth}
    \centering
    \includegraphics[width=\textwidth]{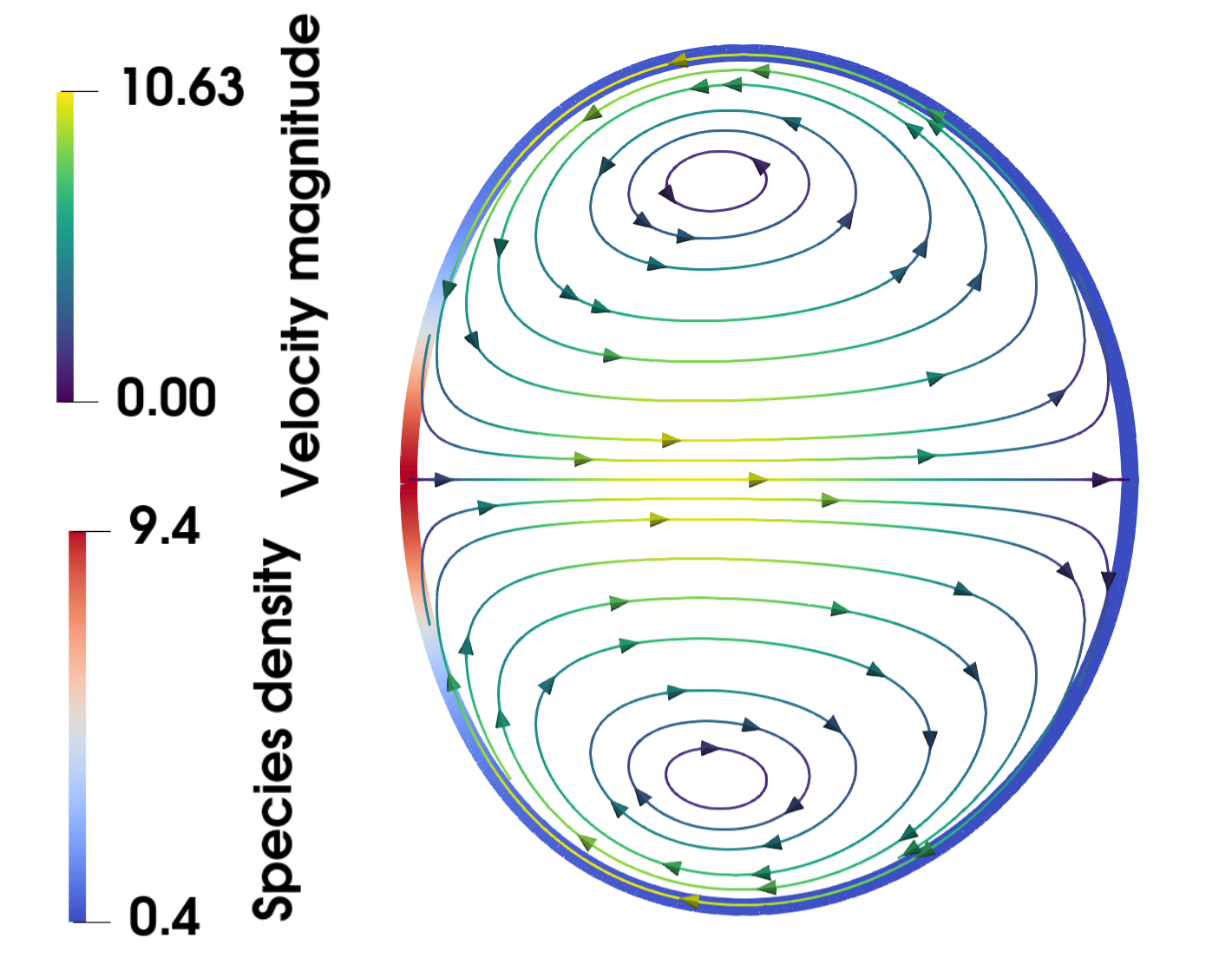}
    \caption{}\label{fig:verification-c}
  \end{subfigure} \\ \vspace{0.25cm}
  \begin{subfigure}{0.52\textwidth}
    \centering
    \includegraphics[width=0.9\textwidth]{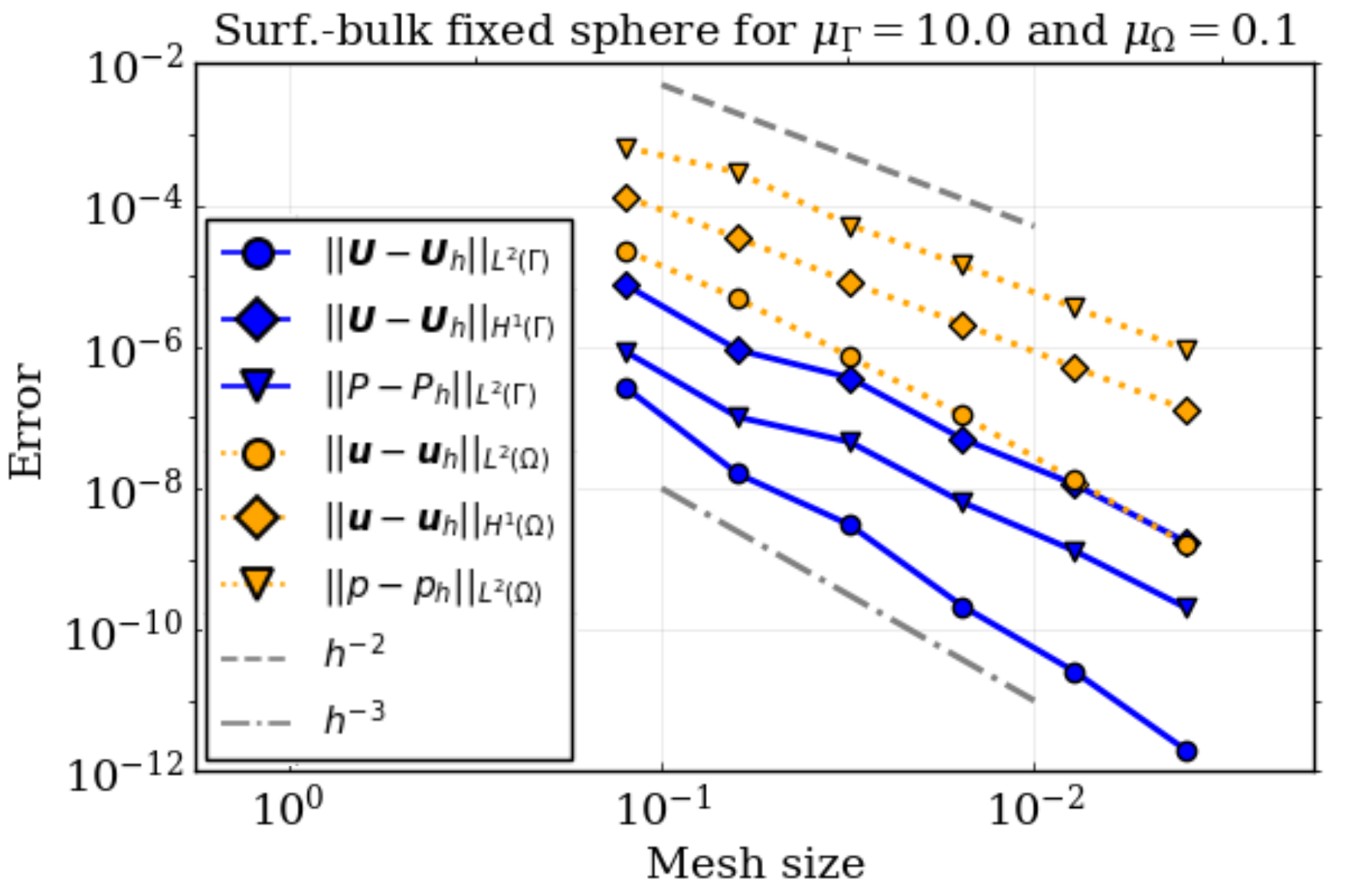}
    \caption{}\label{fig:verification-d}
  \end{subfigure}
  \begin{subfigure}{0.44\textwidth}
    \centering
    \includegraphics[width=0.9\textwidth]{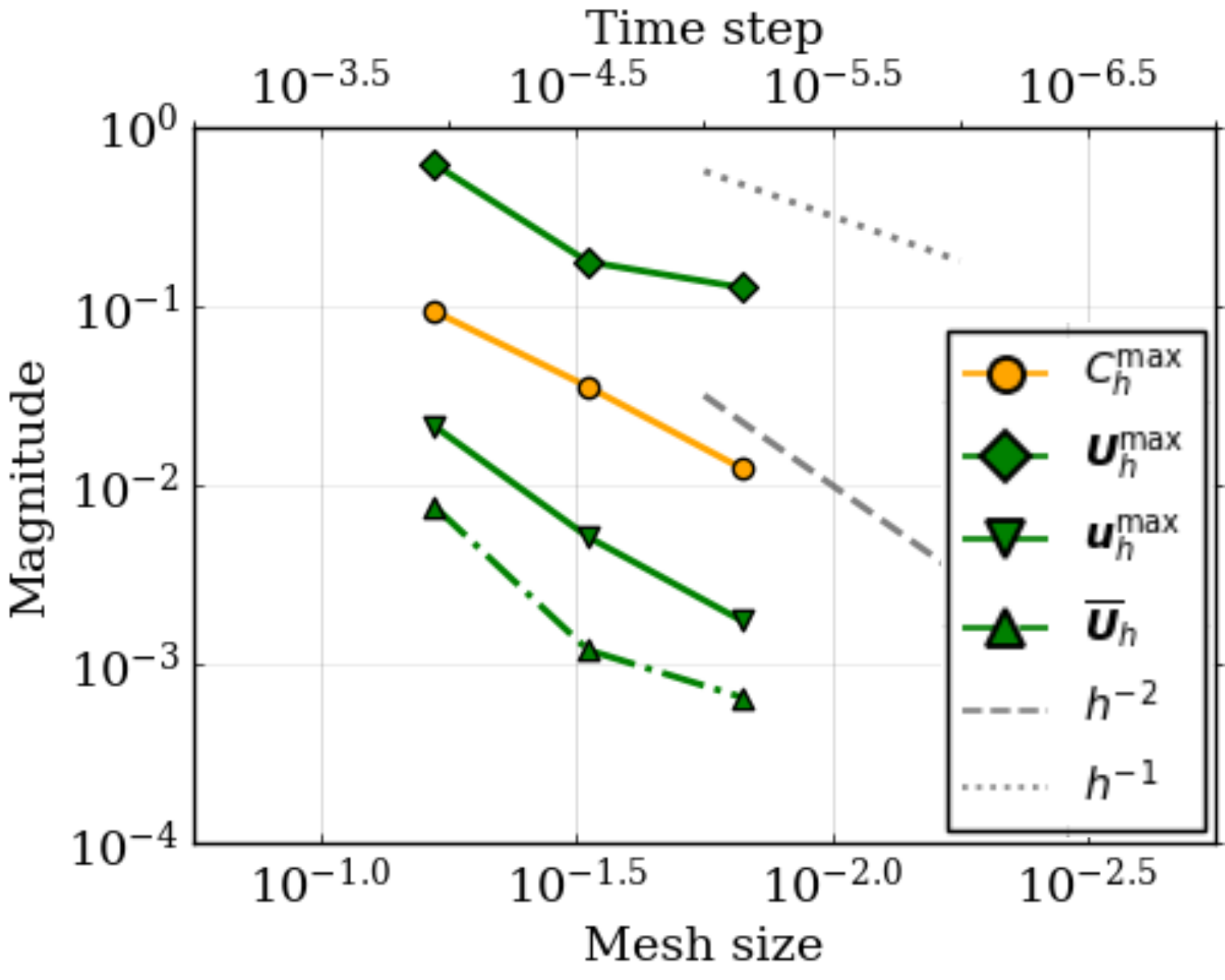}
    \caption{}\label{fig:verification-e}
  \end{subfigure}
  \caption{\textbf{Verification examples.} \emph{2D axisymmetric surface-bulk fluid sphere problem adapted from~\cite[Section 4-21]{happel2012low}:} \textbf{\textsf{(a)}} Velocity magnitude and streamlines of the steady-state solution. \textbf{\textsf{(d)}} The rate of decay of the approximation error under uniform refinements obeys theoretical estimates~\cite{Badia2018Mixed,jankuhn2021error}. \emph{2D axisymmetric dynamic surface:} \textbf{\textsf{(b)}} Initial species concentration and velocity fields. Species concentration at the left sextant is 10\% higher than elsewhere, setting the system out of equilibrium. \textbf{\textsf{(c)}} Equilibrium shape and solution in the comoving frame at $T = 3.0$. \textbf{\textsf{(e)}} Decay of the experimental errors with uniform refinements. We observe rates of convergence close to quadratic order for all quantities. \emph{Note that the results are reflected across the axis of symmetry for illustration purposes.}}
  \label{fig:verification}
\end{figure}

\added{The numerical approximation adopts \ac{fe} pairs with proven inf-sup stability:} The bulk phase is discretised with Ag\ac{fem}, using the inf-sup stable pair $\mathcal{Q}_2 \times \mathcal{P}_1$~\cite{Badia2018Mixed}, as described in Section~\ref{subsec:space}. In contrast, the surface phase is discretised with a consistent trace \ac{fe} formulation for the surface Stokes problem~\cite{jankuhn2021error} using the inf-sup stable pair $\mathcal{Q}_2 \times \mathcal{Q}_1$. Futher details of the implementation can be found in the \href{https://github.com/ericneiva/SurfaceBulkViscousFlows/blob/main/examples/Verification/VerificationInFixedSphere.jl}{code demonstrator} of this example. As shown in Figure~\ref{fig:verification-d} optimal convergence rates of the iterative scheme are observed under uniform mesh refinements.

The second experiment deals with a dynamic surface modelled by the 2D axisymmetric version of Equations~\eqref{eq:full-non-dimensional-1}-\eqref{eq:full-non-dimensional-9} (see Appendix~\ref{app:axisymmetric} for the discrete formulation). We consider a bounding-box $\mathcal{B} = [-1.2,1.2] \times [0.0,1.2]$ embedding a sphere of radius $R=1$ centred at the origin. The values of the three nondimensional parameters are $\mbox{\textit{Pe}} = 30.0$, $L_\eta = 10^4$ and $\tau_D k_{\rm off} = 10.0$.

The initial density field of the molecular species (see Figure~\ref{fig:verification-b}) is set to $1.0$ everywhere, except on a sextant, where the value is increased to $1.1$. This initial bias implies the active tension on the surface is non-homogeneous, which cannot happen in a sphere. As a result, the system is forced into motion to find the equilibrium shape of Figure~\ref{fig:verification-c}. In this process, the Marangoni effect increases the concentration on the sextant, while decreasing it on the other side. Meanwhile, the surface flows generate a reverse fountain flow in the bulk. At steady-state, the system has deformed into a bean-like shape, where the region of higher surface tension is flatter than the opposite side, the body travels at constant speed in the direction of the flows at the axis of symmetry, and the flow field in the comoving frame is tangent to the surface. This swimming mechanism mediated by Marangoni forces has been identified in the spontaneous self-propulsion of active colloids and droplets~\cite{zottl2016emergent}.

Setting $h_0 = 0.08$ and $\Delta \mathrm{t}_0 = 0.004$ as the initial mesh size and time step, we study the convergence under simultaneous uniform refinements in space and time, using an explicit Euler time integration scheme until $T = 0.5$. In front of incomplete theoretical results and limited (semi-)analytical examples, we limit our analysis to the \emph{experimental} order of convergence of key quantities at steady-state: maximum concentration $C_h^{\max}$, maximum surface $\boldsymbol{U}_h^{\max}$ and bulk $\boldsymbol{u}_h^{\max}$ velocities, and average travelling speed $\overline{\boldsymbol{U}}_h$. The \emph{experimental} order of convergence for each monitored quantity $\square$ is computed as
\begin{equation}
  \label{eq:experimental-order-of-convergence}
  {\rm eoc}_i (\square) := \frac{ \ln | \square_{i-1} - \square_{i} | - \ln | \square_{i} - \square_{i+1} | }{ \ln h_i - \ln h_{i+1} },
\end{equation}
where $\ln h_i - \ln h_{i+1} = 2$ due to imposing uniform mesh refinements. Figure~\ref{fig:verification-e} reports almost quadratic experimental rates of convergence for the four quantities.

\subsection{Self-organised shape emergence}
\label{subsec:self}

We now study the mechanochemical instabilities at the heart of the system formed by Equations~\eqref{eq:full-non-dimensional-1}-\eqref{eq:full-non-dimensional-9}, guiding spontaneous polarisation and pattern formation. Linear stability analysis~\cite{mietke2019minimal} of the rotationally symmetric model has shown the instabilities are regulated by the Péclet number $\mbox{\textit{Pe}}$, relating active and diffusive surface transport, and the hydrodynamic length $L_\eta$, coupling the surface and bulk fluids.

Given an infinitesimal perturbation of a homogeneous concentration on the surface, a large enough Péclet number $\mbox{\textit{Pe}}$ will give rise to a positive feedback loop, whereby regions of higher concentration will emerge and grow in magnitude by virtue of the Marangoni effect. The system reaches a steady state when active and diffusive transport balance each other out, leading to a stationary shape and concentration field on the surface and a steady flow, following the gradient of concentration.

On the other hand, the hydrodynamic screening length $L_\eta$ categorises the stationary pattern. Symmetric or ring patterns appear when $L_\eta / R \ll 1$, while asymmetric or polar ones dominate when $L_\eta / R \gg 1$. In contrast to~\cite{mietke2019minimal,wittwer2023computational}, we restrict ourselves to the latter case to reflect the common situation in animal cells; where typical 3D viscosities and geometrical scales of the cortex (thickness) and the cytoplasm (radius) yield rather large hydrodynamic lengths, with $L_\eta / R \approx 10^{3-5}$, see Table~\ref{tab:hydrodynamic_length}. This type of polar instabilities, at large hydrodynamic lengths, have been suggested as a mechanism for sustained unidirectional cell motility in three-dimensional environments~\cite{hawkins2011spontaneous}.

\begin{table}[h!]
  \centering
  \renewcommand{\arraystretch}{1.1}
  \scriptsize
  \begin{tabular}{|c|c|c|c|}
    \hline
    \textbf{Notation} & \textbf{Quantity} & \textbf{Experimental values} & \textbf{References} \\ \hline
    $\lambda_\Gamma^{\rm 3D}$ & 3D actomyosin shear viscosity & $10^5-10^6$ Pa$\cdot$s & \cite{da2022viscous,turlier2014furrow} \\ \hline
    $\mu_\Omega$ & 3D cytoplasm viscosity (not crowded) & $0.1-1$ Pa$\cdot$s & \cite{hiramoto1969mechanical,najafi2023size,valberg1987magnetic} \\ \hline
    $R$ & cell radius (large cells) & $50-100$ \textmu m & \cite{hiramoto1958quantitative,shamipour2021cytoplasm,lu2023go} \\ \hline
    $e$ & cortical thickness & $0.2-2$ \textmu m & \cite{clark2013monitoring} \\ \hline
  \end{tabular}
  \vspace{0.2cm}
  \caption{\textbf{Self-organised shape emergence.} Experimental values reported in the literature for large animal cells leading to the estimate $L_\eta / R = \lambda_\Gamma / \mu_\Omega R = \lambda_\Gamma^{\rm 3D} e / \mu_\Omega R \approx 10^{3-5}$.}
  \label{tab:hydrodynamic_length}
\end{table}

As in the previous example, we adopt the 2D axisymmetric version of Equations~\eqref{eq:full-non-dimensional-1}-\eqref{eq:full-non-dimensional-9} with a bounding-box domain given by $\mathcal{B} = [-1.2,1.2] \times [0.0,1.2]$ embedding a unit sphere centred at the origin. The initial concentration field on the surface, assuming $c_{\rm eq} = 1$, is given by $C_h^0 = 1 + \epsilon_h \in \mathcal{W}_h^\Gamma$, where $\epsilon_h \in \mathcal{W}_h^\Gamma$ represents a random-valued \ac{fe} function such that $\max \{ | \epsilon_h | \} < 10^{-5}$ and $\int_\Gamma \epsilon_h \ \mathrm{d}\Gamma = 0$.

Using real spherical harmonics $Y_{lm}(\theta,\varphi)$, linear stability analysis of a homogeneous stationary state leads to a critical Péclet number~\cite{mietke2019minimal}
\begin{equation}
  \label{eq:critical-pe}
  \mbox{\textit{Pe}}_{\rm cr}^l = \frac{1}{c_{\rm eq} \partial_C f(C,c_{\rm eq})} \left( 1 + \frac{ \tau_D k_{\rm off} }{ l(l+1) } \right) \left[ l(l+1) + \left( (l-1)(l+2) + (1+2l) \frac{R}{L_\eta} \right) \right],
\end{equation}
where $l \geq 1$ denotes the index of the radial modes. Regardless of the azimuthal mode number $m$, modes with $l \geq 1$ become unstable at $\mbox{\textit{Pe}} = \mbox{\textit{Pe}}_{\rm cr}^l$. We initially assume $\tau_D k_{\rm off} = 10.0$. Since $c_{\rm eq} \partial_C f(C,c_{\rm eq}) = 1$, the critical Péclet number, according to Equation~\eqref{eq:critical-pe}, for the first polar mode is $\mbox{\textit{Pe}}_{\rm cr}^1 = 12 + 3/L_\eta \approx 12$. 

We use this theoretical value of $\mbox{\textit{Pe}}_{\rm cr}^1$ to tune the mesh size and time step for our analysis, in such a way that we correctly capture the stable to nonstable transition. To this end, we compute the first discrete Pearson correlation coefficient $r_1$ between the surface concentration field $C_h$ and the real spherical harmonic $Y_{10}(\theta)$. For $l \geq 1$, the discrete Pearson correlation coefficient $r_l$ is given by
\begin{equation}
  \label{eq:discrete-Pearson-coefficient}
  r_l = \frac{ \sum_{k=0}^{N_{\rm points}} ( C_k - C_{\rm mean} ) Y_{l0}(\theta_k) }{ \sqrt{\sum_{k=0}^{N_{\rm points}} ( C_k - C_{\rm mean} )^2 \ \sum_{k=0}^{N_{\rm points}} ( Y_{l0}(\theta_k) )^2} },
\end{equation}
where $k$ indexes the $N_{\rm points}$ quadrature points for the integration of the surface integrals, $C_k$ is the concentration on the $k$-th point, $C_{\rm mean}$ is the mean concentration on the quadrature points and $\theta_k$ is the polar angle of the $k$-th point.

\begin{figure}[!ht]
  \centering
  \begin{subfigure}{0.6\textwidth}
    \centering
    \includegraphics[width=\textwidth]{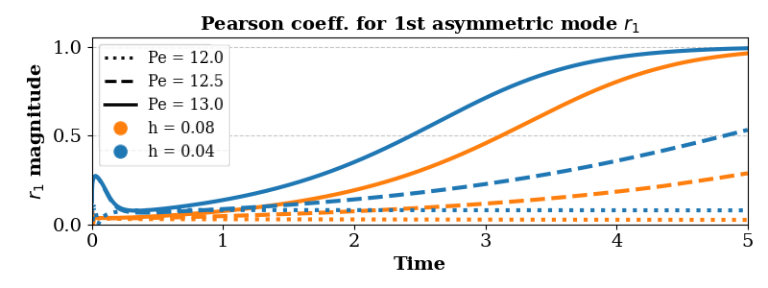}
    \caption{}\label{fig:tune-mesh-a}
  \end{subfigure} \\
  \begin{subfigure}{0.35\textwidth}
    \centering
    \includegraphics[width=\textwidth]{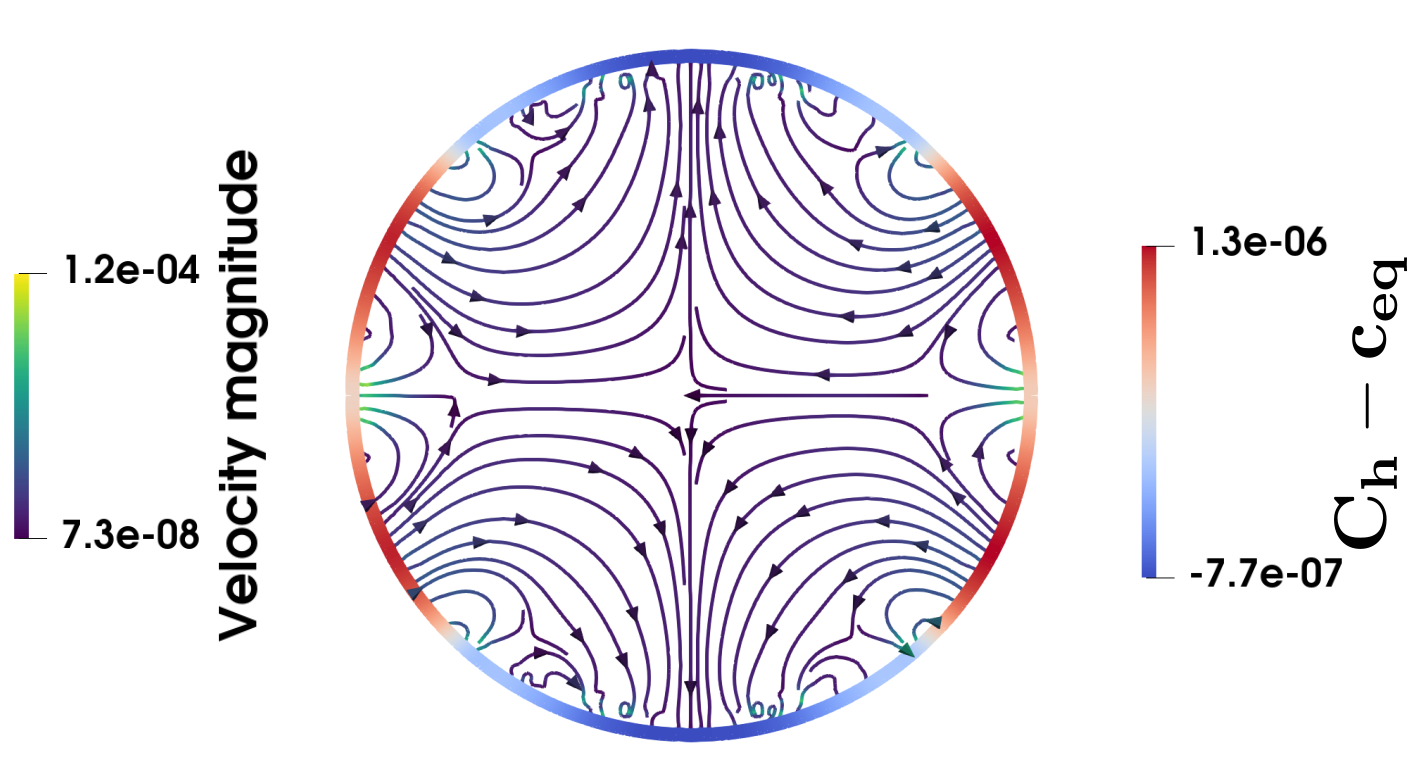}
    \caption{$\mbox{\textit{Pe}} = 12$}\label{fig:tune-mesh-b}
  \end{subfigure} \quad \quad \quad
  \begin{subfigure}{0.35\textwidth}
    \centering
    \includegraphics[width=\textwidth]{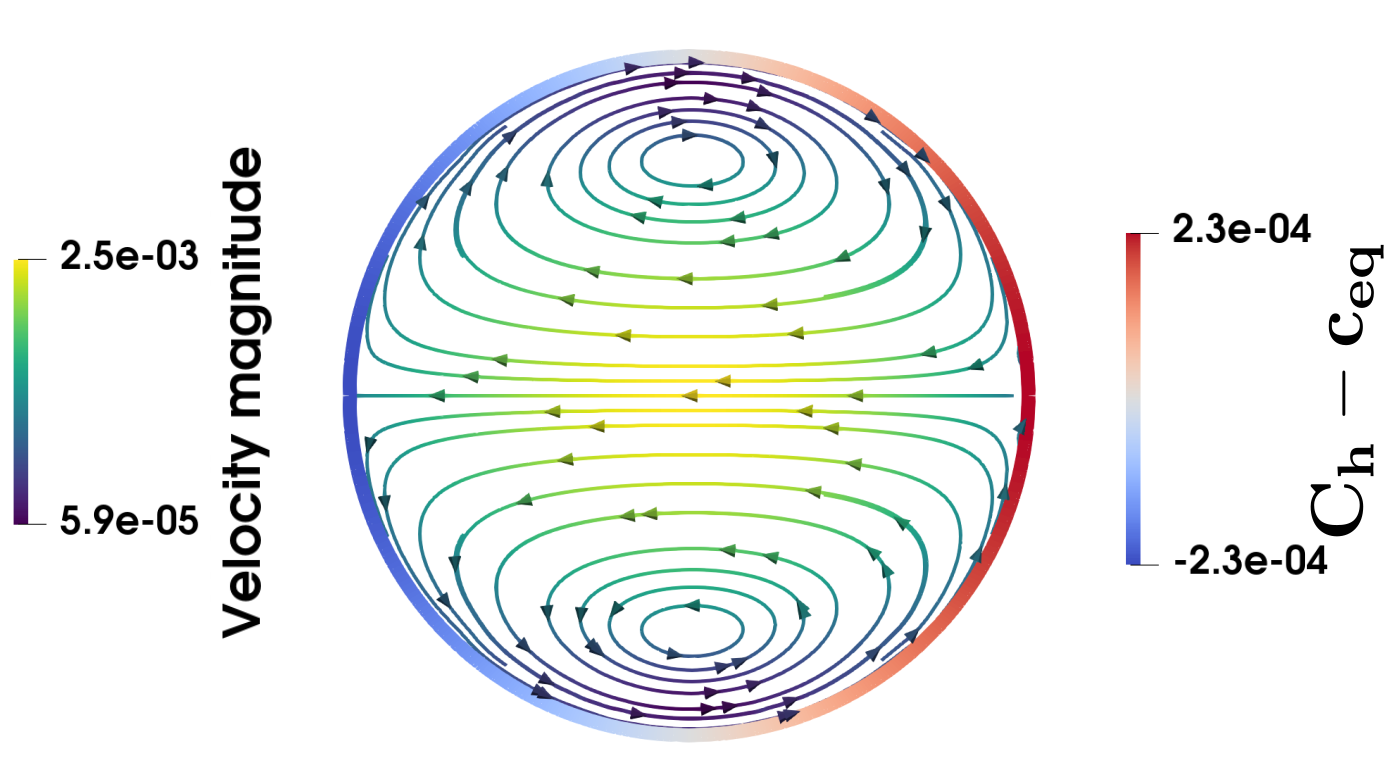}
    \caption{$\mbox{\textit{Pe}} = 13$}\label{fig:tune-mesh-c}
  \end{subfigure}
  \caption{\textbf{Self-organised shape emergence.} \emph{Tuning the mesh size for analysis of pattern formation.} Before the analysis reported in Figure~\ref{fig:pattern}, we found the minimum mesh size needed to correctly capture the stable to unstable transition in the linear regime, as predicted by Equation~\eqref{eq:critical-pe}. \textbf{\textsf{(a)}} Setting $h = 0.04$ provides a good balance between computational cost and correctly capturing the theoretically predicted appearance of the first instability mode. \textbf{\textsf{(b)}} At $\mbox{\textit{Pe}} = 11$ and $h = 0.04$, no instabilities appear and concentration heterogeneities slowly diffuse out. \textbf{\textsf{(c)}} At $\mbox{\textit{Pe}} = 13$ and $h = 0.04$, we observe a paradigmatic first order polar instability forming.}
  \label{fig:tune-mesh}
\end{figure}

Figure~\ref{fig:tune-mesh} shows the evolution of $r_1$ for a few mesh sizes. A value of $r_1$ close to 1 indicates that the first mode dominates in the harmonic decomposition, implying a polar instability. We observe that taking $h = 0.04$ captures better the stable to unstable transition than $h = 0.08$, while keeping a moderate computational cost. We will thus consider this mesh size for the remaining 2D cases of this subsection. We considered $\Delta t = 5.0 \cdot 10^{-5}$, as in the experiments that follow.

\begin{figure}[!ht]
  \centering
  \begin{subfigure}{0.4\textwidth}
    \centering
    \includegraphics[width=0.9\textwidth]{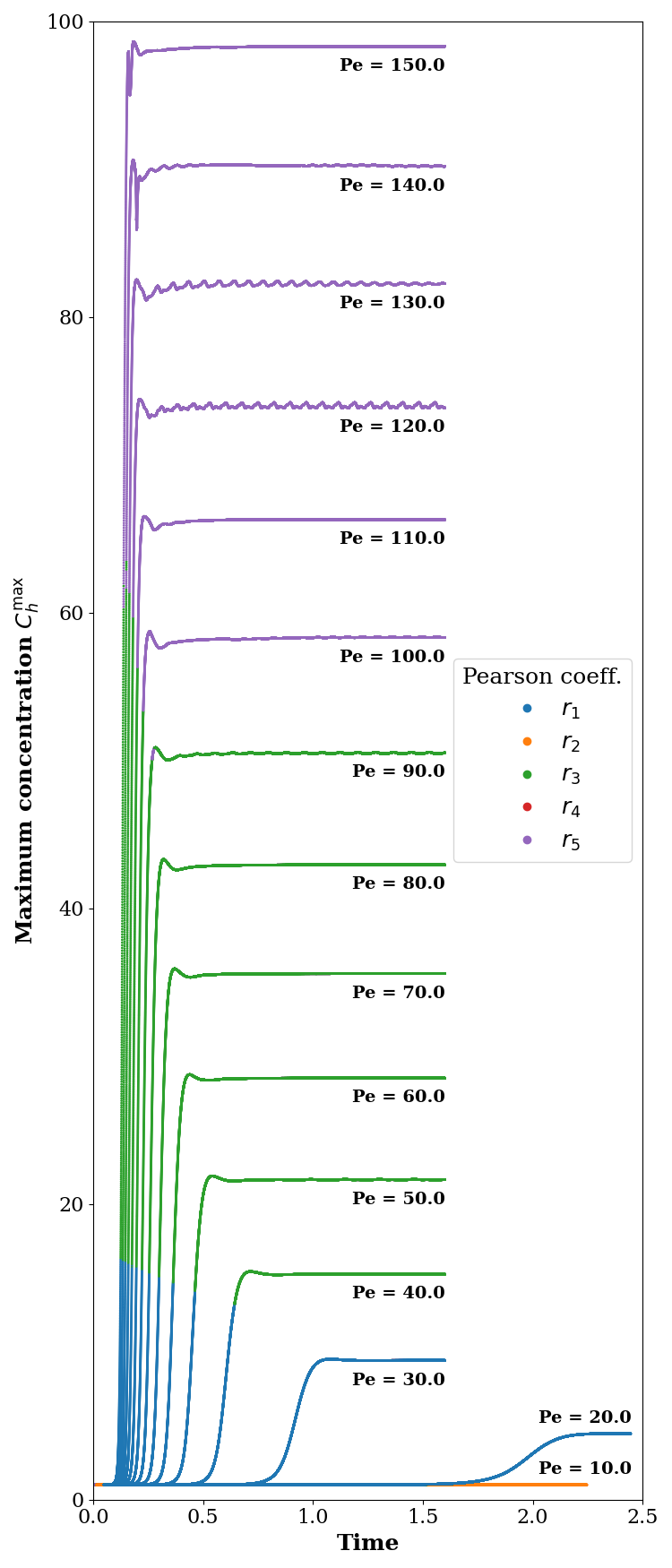}
    \caption{}\label{fig:pattern-a}
  \end{subfigure}
  \begin{subfigure}{0.4\textwidth}
    \centering
    \includegraphics[width=0.9\textwidth]{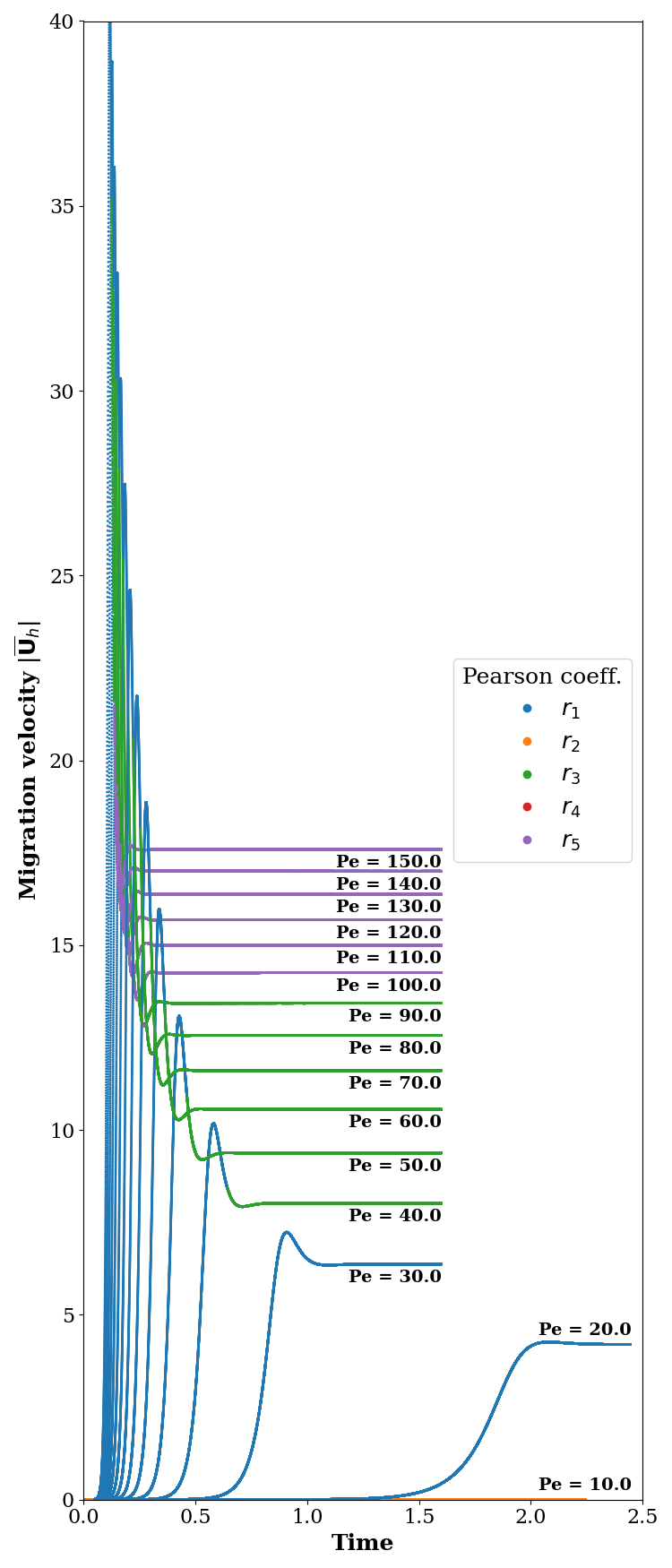}
    \caption{}\label{fig:pattern-b}
  \end{subfigure} \\[0.5cm]
  \begin{subfigure}{\textwidth}
    \centering
    \includegraphics[width=0.9\textwidth]{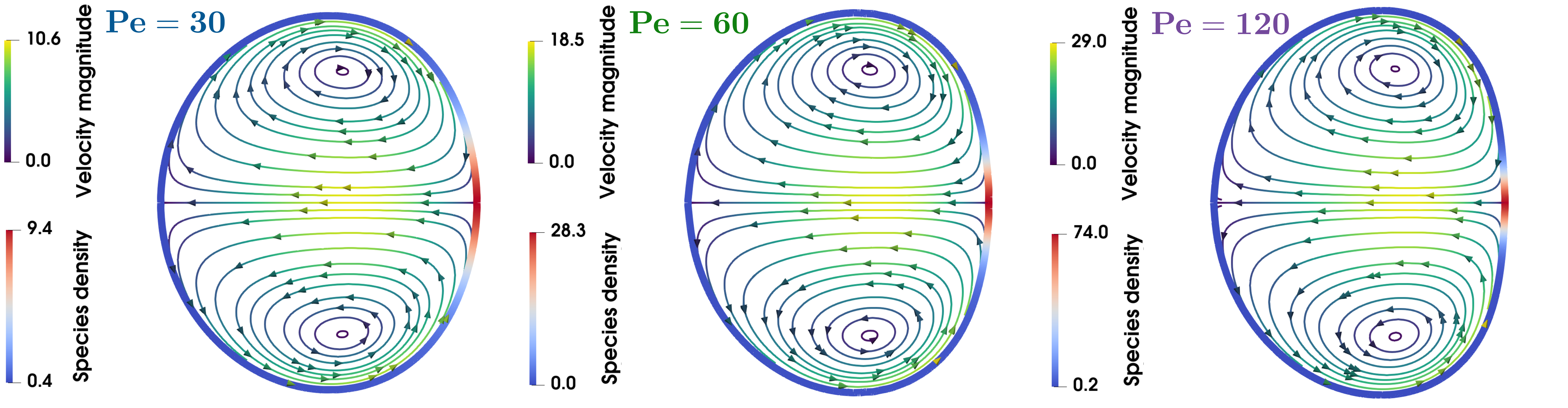}
    \caption{}\label{fig:pattern-c}
  \end{subfigure}
  \caption{\textbf{Self-organised shape emergence.} \emph{Pattern formation at large hydrodynamic lengths.} A perturbation of an homogeneous concentration field leads to spontaneous pattern formation at large enough $\mbox{\textit{Pe}}$ numbers. \textbf{\textsf{(a)}} Peak concentration and \textbf{\textsf{(b)}} migration velocities for $\mbox{\textit{Pe}} \in [10,150]$ and $L_\eta \in \{10^3,10^4,10^5\}$. The evolution of the leading mode is coloured along the plotted lines. At large hydrodynamic lengths, we do not observe differences in shapes, flow fields and magnitudes. For this reason, we only plot the results for $L_\eta = 10^4$. \textbf{\textsf{(c)}} Steady-state results obtained for leading modes 1-2-3 at, respectively, $\mbox{\textit{Pe}} = 30,60,120$.}
  \label{fig:pattern}
\end{figure}

Figure~\ref{fig:pattern} reports the evolution in time of maximum species concentration $C_h^{\max}$, travelling speed $\overline{\boldsymbol{U}}_h$ and dominant radial modes $r^{\max}$ for increasing values of $\mbox{\textit{Pe}}$ and hydrodynamic lengths $L_\eta \in \{10^3,10^4,10^5\}$. The dominant mode is identified as the maximum among the first seven discrete Pearson correlation coefficients, i.e., $r^{\max} = \max_{l = 0, \ldots, 6} r_l$. Simulations are run until a quasi-steady state is reached, in which all unknown values remain approximately constant and the flow field in the comoving frame is tangent to the surface. We prescribe $\Delta t = 5.0 \cdot 10^{-5}$, which is required to follow the steep nonlinear transition at large $\mbox{\textit{Pe}}$. As expected, all leading modes are asymmetric; increasing $\mbox{\textit{Pe}}$ reduces the time to enter the unstable regime and promotes the dominance of high order modes. The results also align well with Equation~\eqref{eq:critical-pe}, even far from the linear regime, showing almost no sensitivity to the value of the hydrodynamic length $L_\eta$ when $L_\eta / R \gg 1$. This suggests that the underlying migration mechanism is above all a cortical instability, with little influence of the cytoplasm.

\begin{figure}[!ht]
  \centering
  \begin{subfigure}{0.4\textwidth}
    \centering
    \includegraphics[width=\textwidth]{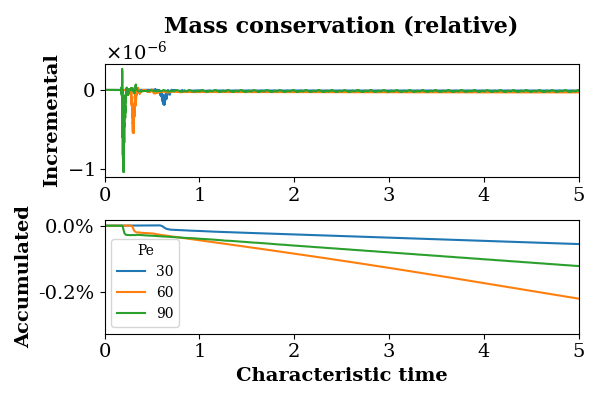}
    \caption{}\label{fig:mass-conservation-a}
  \end{subfigure}
  \begin{subfigure}{0.4\textwidth}
    \centering
    \includegraphics[width=\textwidth]{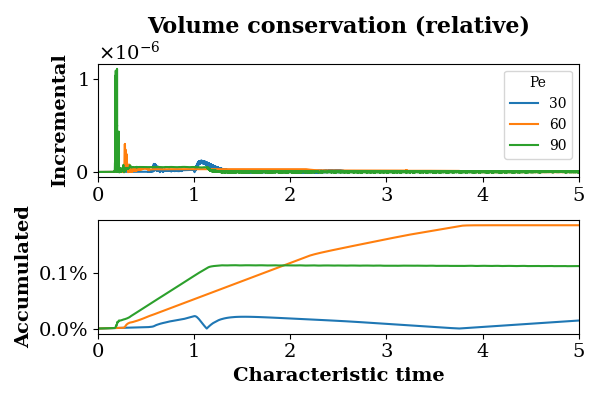}
    \caption{}\label{fig:mass-conservation-b}
  \end{subfigure}
  \caption{\textbf{Self-organised shape emergence.} \emph{Mass and volume conservation errors.} Setting $\tau_D k_{\rm off} = 0.0$, i.e., no mass exchange between the surface and the bulk, we study the evolution of the mass and volume conservation errors for $h = 0.04$, $\Delta t = 5.0 \cdot 10^{-5}$, $\mbox{\textit{Pe}} \in \{30,60,90\}$ and $L_\eta\! =\! 10^4$. \textbf{\textsf{(a)}} Mass and \textbf{\textsf{(b)}} volume conservation errors are non-negligible, but low enough to not significantly impact the results.} 
  \label{fig:mass-conservation}
\end{figure}

Supplementary experiments with $\tau_D k_{\rm off} = 0.0$, i.e., no mass exchange between the surface and the bulk, evaluate the mass conservation property of molecular species $C$ on the surface. Figure~\ref{fig:mass-conservation-a} details the evolution in time of the incremental and accummulated relative deviation from the initial surface mass for $\mbox{\textit{Pe}} \in \{30,60,90\}$ and $L_\eta\! =\! 10^4$. Despite not using a conservative scheme, the relative error in mass conservation is low enough to ensure it has no influence in the numerical results. We reach a similar conclusion when analysing the error in volume conservation in Figure~\ref{fig:mass-conservation-b}. 

\subsection{Relaxation dynamics}
\label{subsec:relaxation}

We continue analysing the mechanochemical instabilities of Equations~\eqref{eq:full-non-dimensional-1}-\eqref{eq:full-non-dimensional-9} focusing on a different driving force: curvature inhomogeneities~\cite{mietke2019self}. The linear stability analysis of Section~\ref{subsec:self} identifies the sphere as a stable shape for $\mbox{\textit{Pe}}\! <\! \mbox{\textit{Pe}}_{\rm cr}^l$, $l \geq 1$, in which diffusion, dominating over advection, homogenenises the concentration of the molecular species. On the other hand, recalling Equation~\eqref{eq:active-forces}, the decomposition of the active forces exposes the capability of mean curvature gradients alone to generate the Marangoni effect. Obviously, these cannot be balanced out by diffusive transport. Hence, in the diffusion-dominated regime, the system is brought to relax its shape towards a sphere.

\begin{figure}[h!]
  \centering
  \begin{subfigure}{0.32\textwidth}
    \centering
    \includegraphics[width=\textwidth]{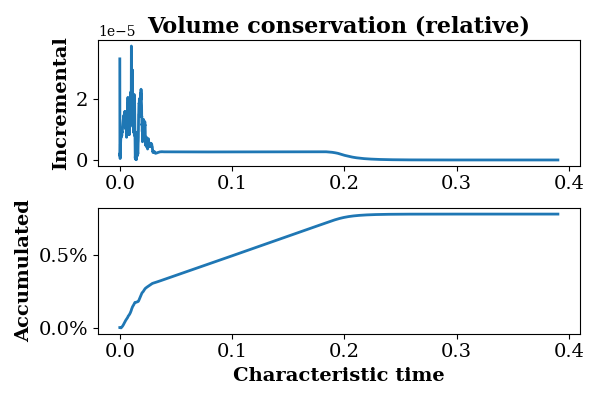}
    \caption{Popcorn relaxation}\label{fig:popcorn-relaxation-vols}
  \end{subfigure}
  \begin{subfigure}{0.32\textwidth}
    \centering
    \includegraphics[width=\textwidth]{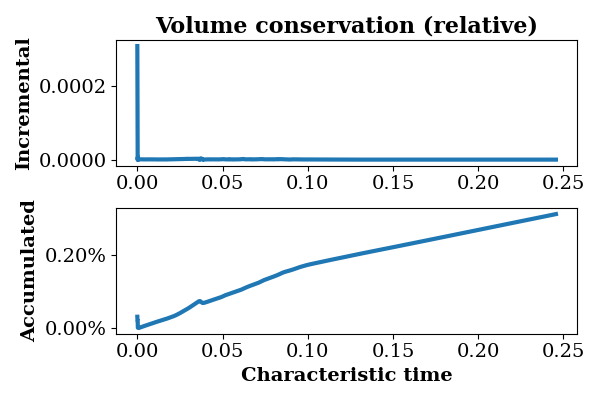}
    \caption{Pear relaxation}\label{fig:ellipse-relaxation-vols}
  \end{subfigure}
  \begin{subfigure}{0.32\textwidth}
    \centering
    \includegraphics[width=\textwidth]{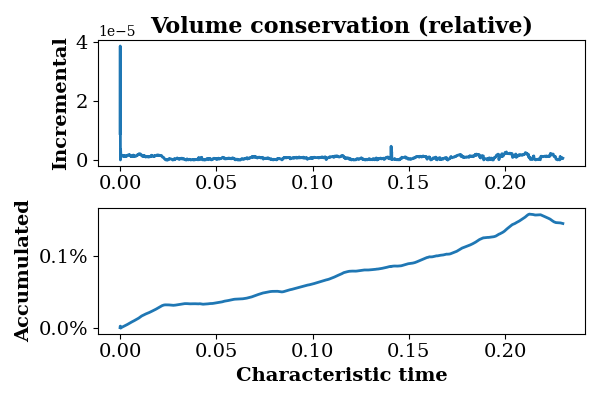}
    \caption{Torus relaxation}\label{fig:ellipse-relaxation-vols}
  \end{subfigure}
  \caption{\textbf{Relaxation dynamics.} We observe low errors in volume conservation for the three relaxed geometries.}
  \label{fig:relaxation-vols}
\end{figure}

\begin{figure}[h!]
  \centering
  \begin{subfigure}[t]{0.08\textwidth}
    \centering
    \includegraphics[width=1.0\textwidth]{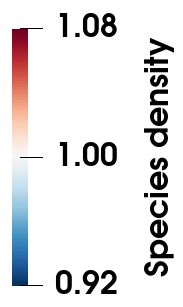}
  \end{subfigure}
  \begin{subfigure}[t]{0.2\textwidth}
    \centering
    \includegraphics[width=1.0\textwidth]{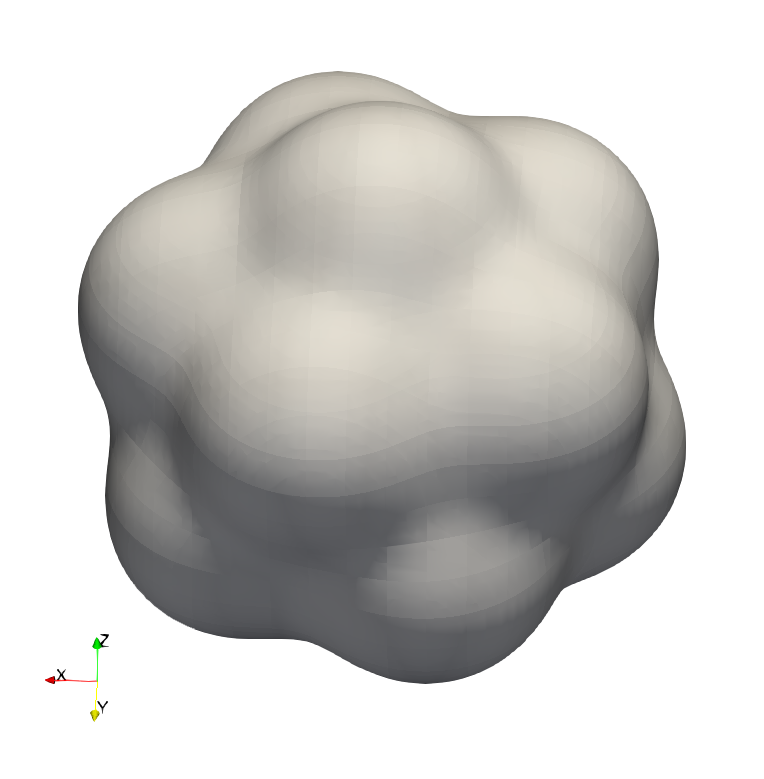}
    \caption{$\mathrm{t} = 0.0$}\label{fig:popcorn-relaxation-a}
  \end{subfigure}
  \begin{subfigure}[t]{0.2\textwidth}
    \centering
    \includegraphics[width=1.0\textwidth]{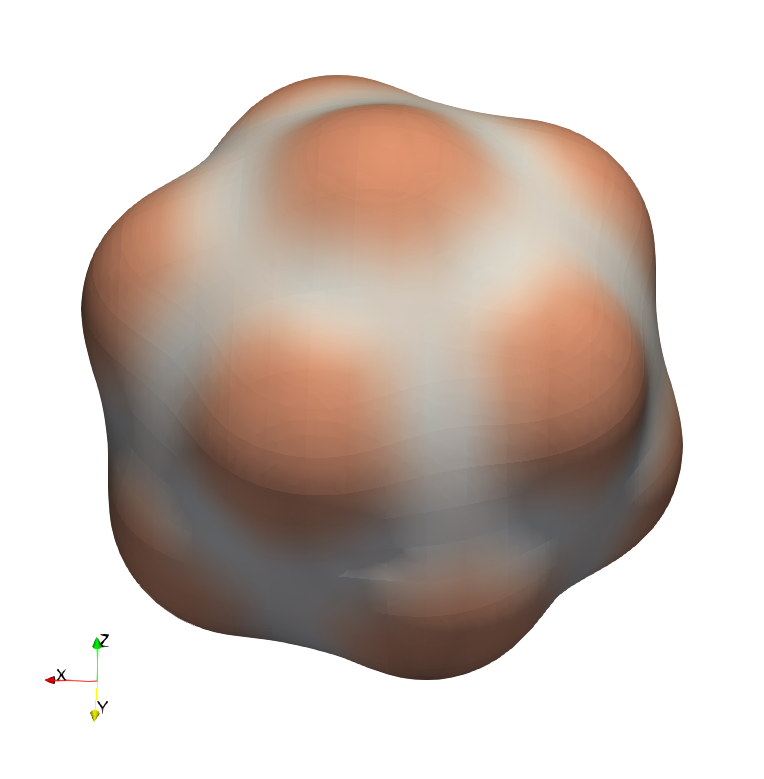}
    \caption{$\mathrm{t} = 0.005$}\label{fig:popcorn-relaxation-b}
  \end{subfigure}
  \begin{subfigure}[t]{0.2\textwidth}
    \centering
    \includegraphics[width=1.0\textwidth]{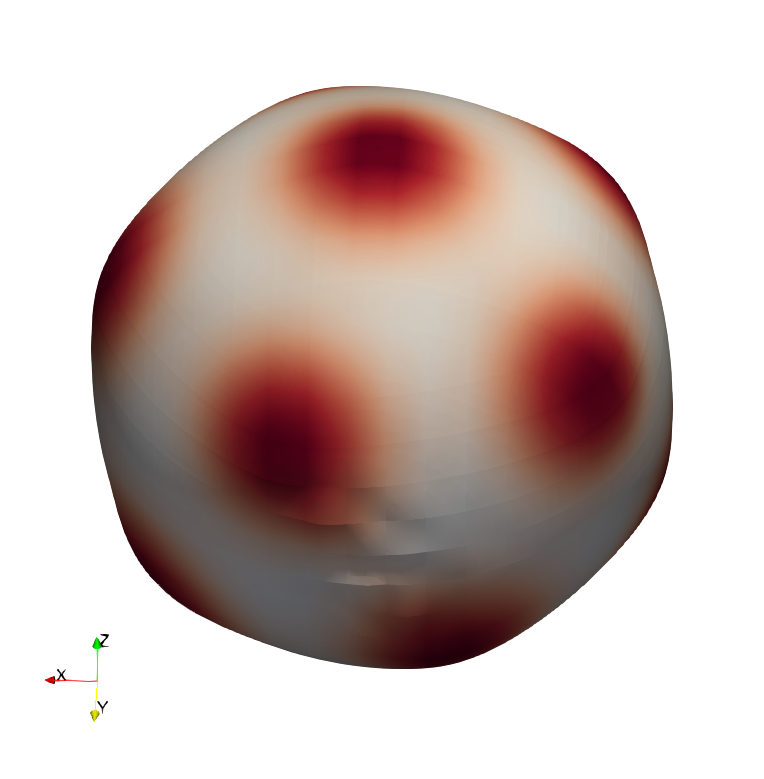}
    \caption{$\mathrm{t} = 0.02$}\label{fig:popcorn-relaxation-c}
  \end{subfigure}
  \begin{subfigure}[t]{0.2\textwidth}
    \centering
    \includegraphics[width=1.0\textwidth]{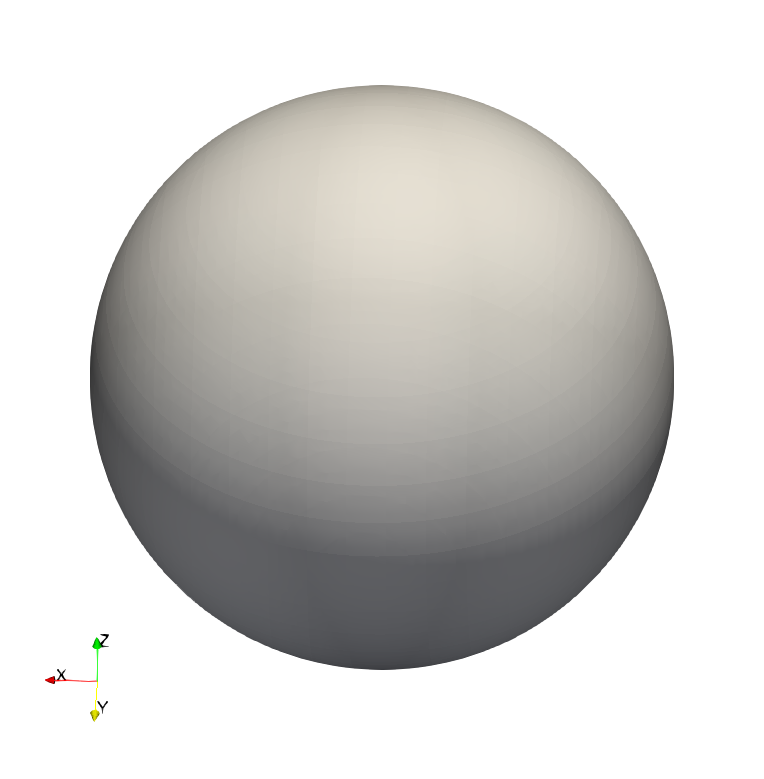}
    \caption{$\mathrm{t} = 0.2$}\label{fig:popcorn-relaxation-d}
  \end{subfigure} \\
  \begin{subfigure}[t]{0.08\textwidth}
    \centering
    \includegraphics[width=1.0\textwidth]{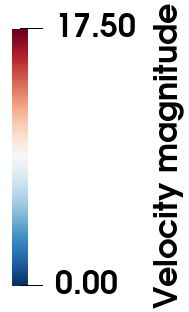}
  \end{subfigure}
  \begin{subfigure}[t]{0.2\textwidth}
    \centering
    \includegraphics[width=1.0\textwidth]{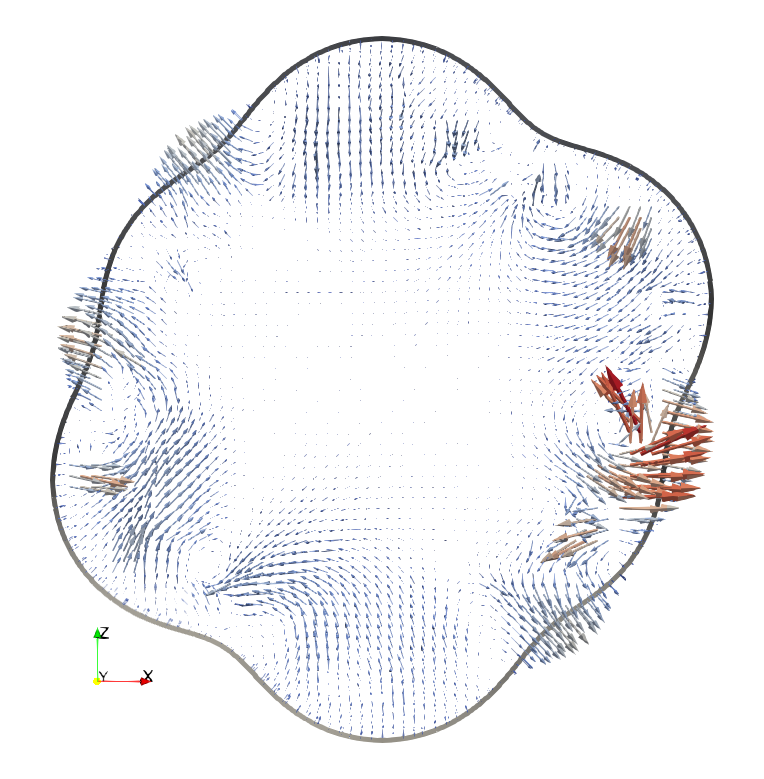}
    \caption{$\mathrm{t} = 0.0$}\label{fig:popcorn-relaxation-e}
  \end{subfigure}
  \begin{subfigure}[t]{0.2\textwidth}
    \centering
    \includegraphics[width=1.0\textwidth]{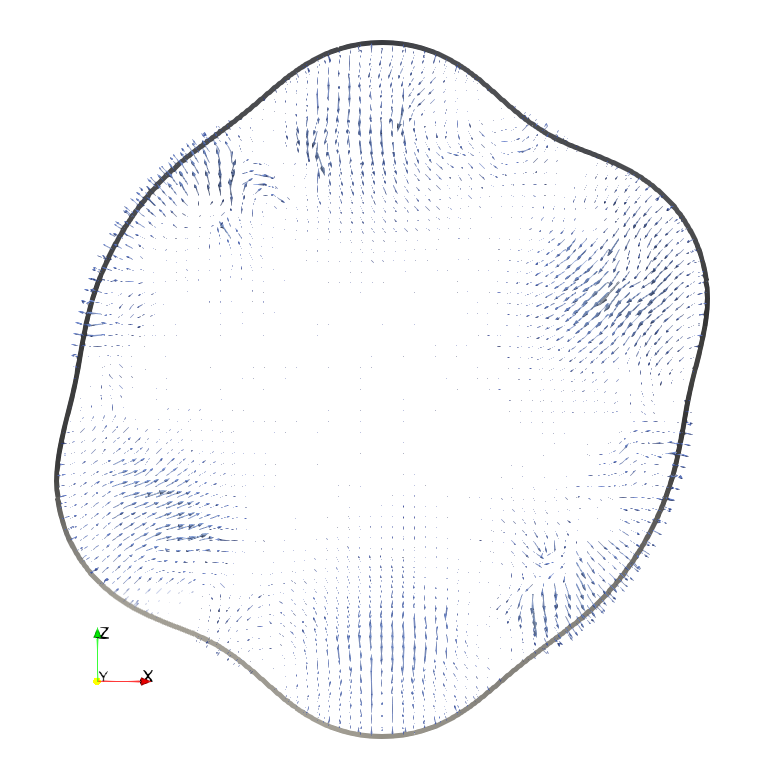}
    \caption{$\mathrm{t} = 0.005$}\label{fig:popcorn-relaxation-f}
  \end{subfigure}
  \begin{subfigure}[t]{0.2\textwidth}
    \centering
    \includegraphics[width=1.0\textwidth]{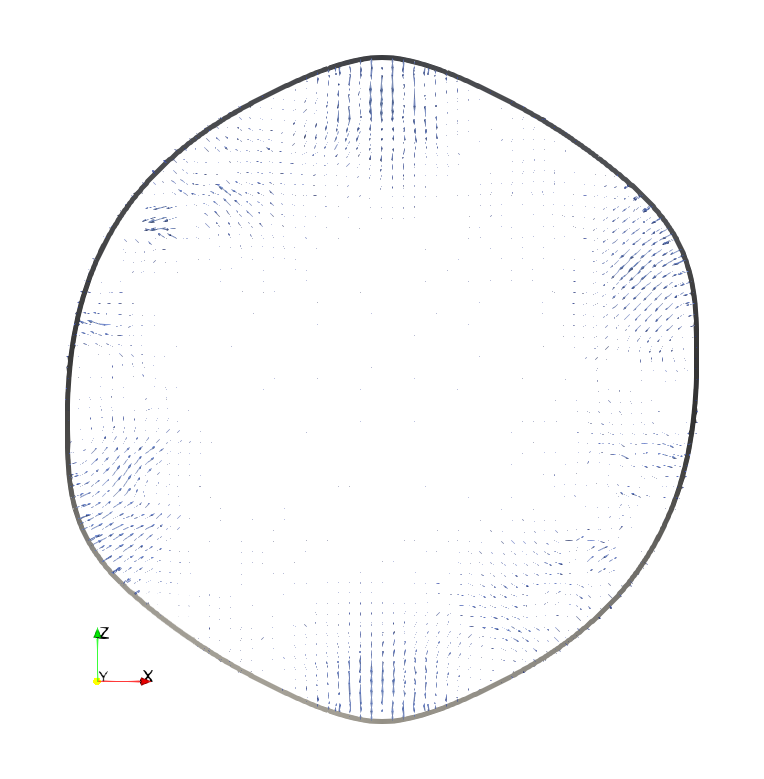}
    \caption{$\mathrm{t} = 0.02$}\label{fig:popcorn-relaxation-g}
  \end{subfigure}
  \begin{subfigure}[t]{0.2\textwidth}
    \centering
    \includegraphics[width=1.0\textwidth]{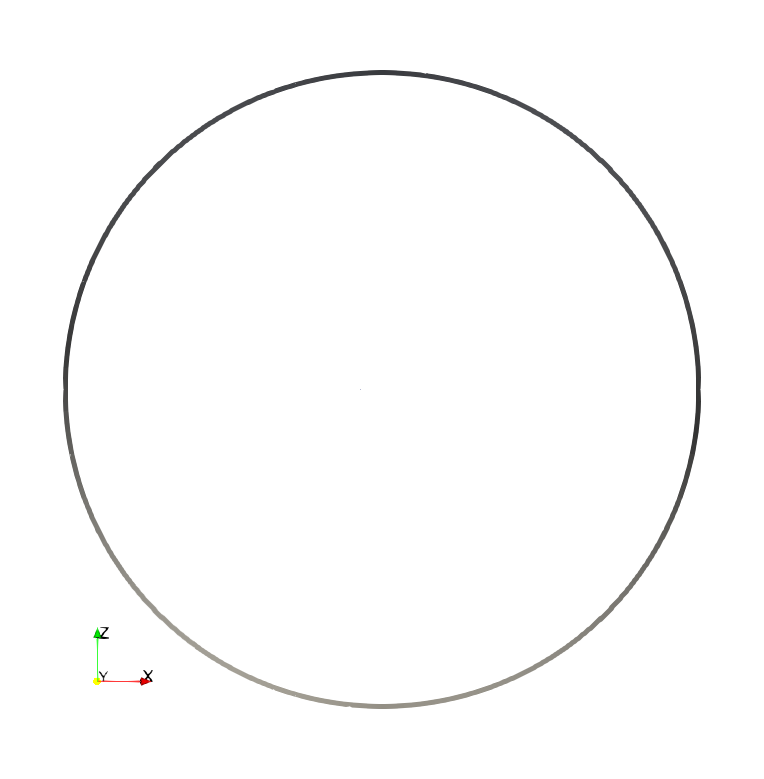}
    \caption{$\mathrm{t} = 0.2$}\label{fig:popcorn-relaxation-h}
  \end{subfigure} \\
  \begin{subfigure}[t]{0.08\textwidth}
    \centering
    \includegraphics[width=1.0\textwidth]{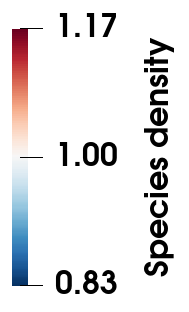}
  \end{subfigure}
  \begin{subfigure}[t]{0.2\textwidth}
    \centering
    \includegraphics[width=1.0\textwidth]{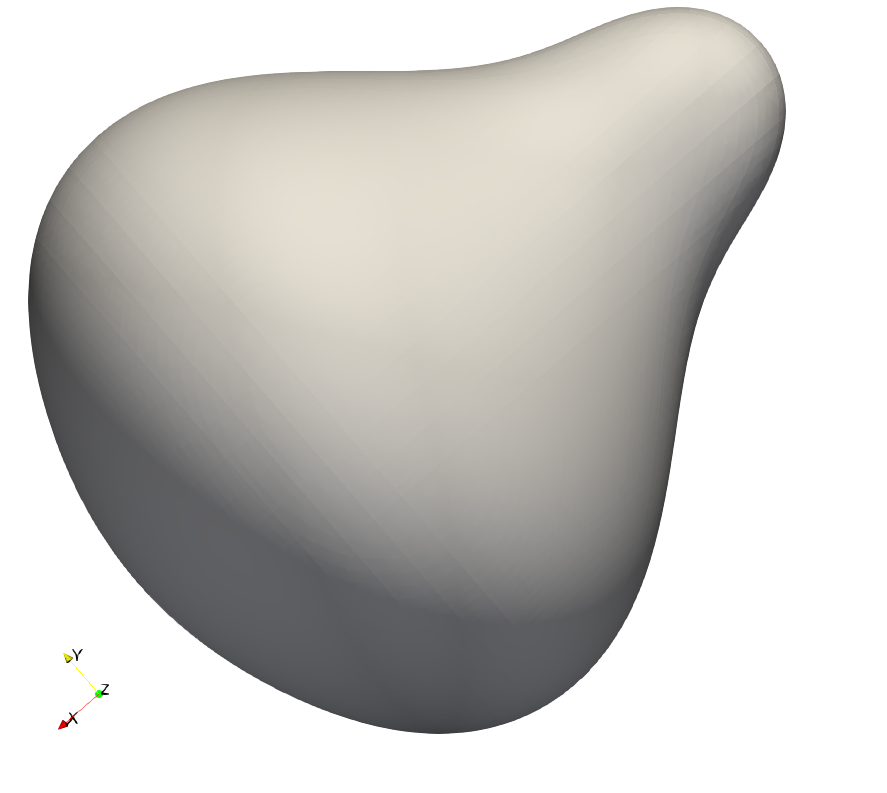}
    \caption{$\mathrm{t} = 0.0$}\label{fig:pear-relaxation-a}
  \end{subfigure}
  \begin{subfigure}[t]{0.2\textwidth}
    \centering
    \includegraphics[width=1.0\textwidth]{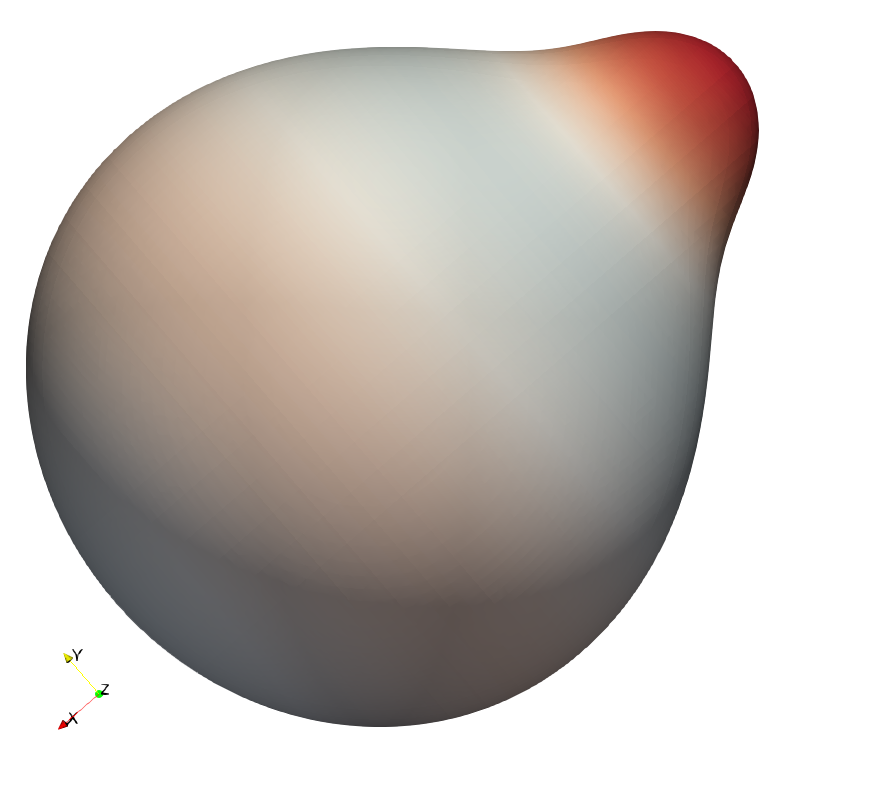}
    \caption{$\mathrm{t} = 0.03$}\label{fig:pear-relaxation-b}
  \end{subfigure}
  \begin{subfigure}[t]{0.2\textwidth}
    \centering
    \includegraphics[width=1.0\textwidth]{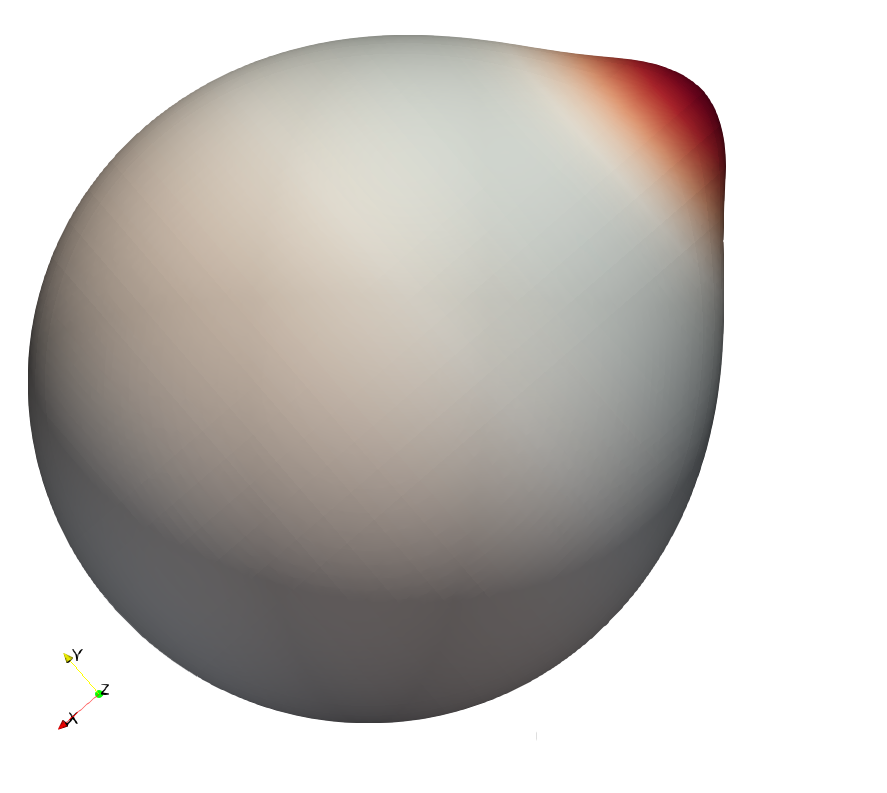}
    \caption{$\mathrm{t} = 0.06$}\label{fig:pear-relaxation-c}
  \end{subfigure}
  \begin{subfigure}[t]{0.2\textwidth}
    \centering
    \includegraphics[width=1.0\textwidth]{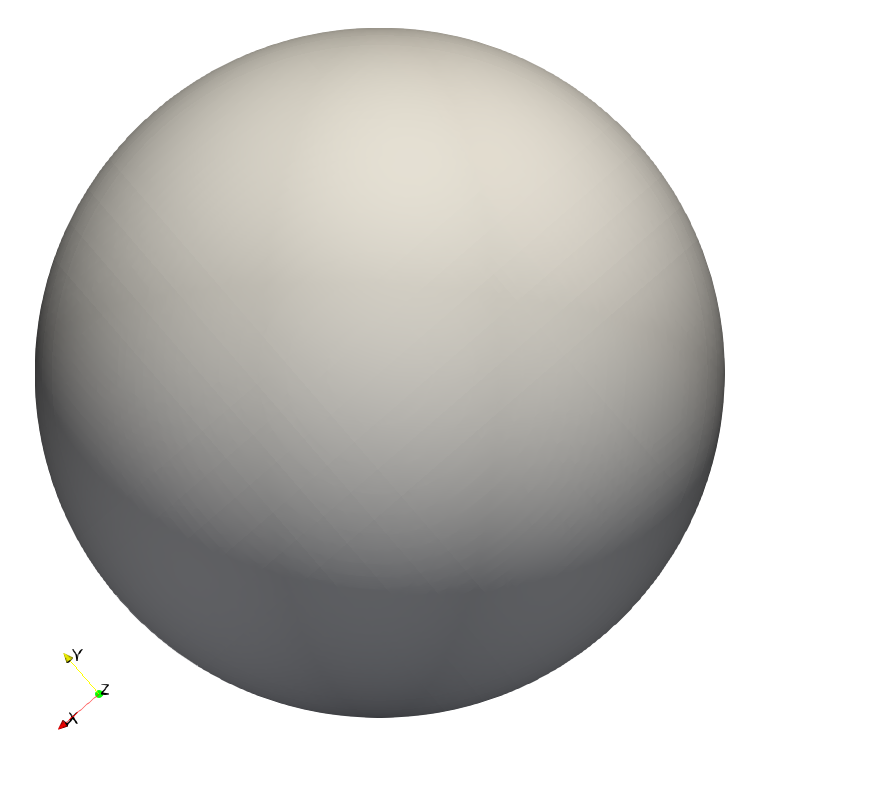}
    \caption{$\mathrm{t} = 0.24$}\label{fig:pear-relaxation-d}
  \end{subfigure} \\
  \begin{subfigure}[t]{0.08\textwidth}
    \centering
    \includegraphics[width=1.0\textwidth]{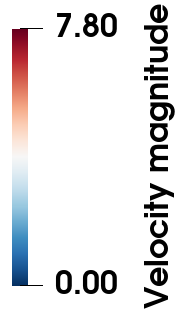}
  \end{subfigure}
  \begin{subfigure}[t]{0.2\textwidth}
    \centering
    \includegraphics[width=1.0\textwidth]{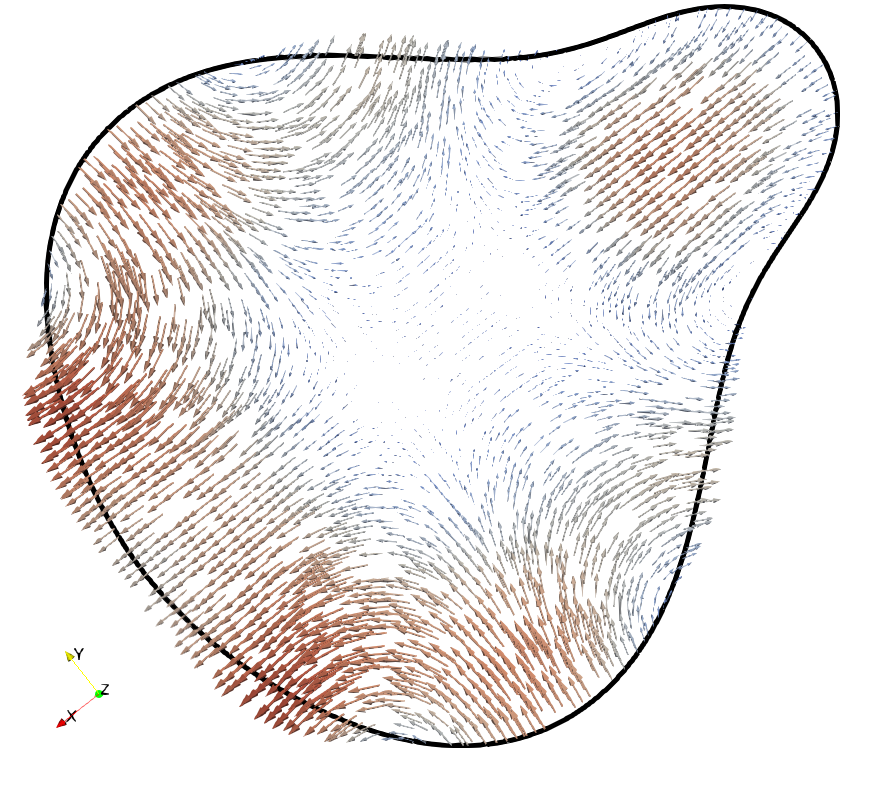}
    \caption{$\mathrm{t} = 0.0$}\label{fig:pear-relaxation-e}
  \end{subfigure}
  \begin{subfigure}[t]{0.2\textwidth}
    \centering
    \includegraphics[width=1.0\textwidth]{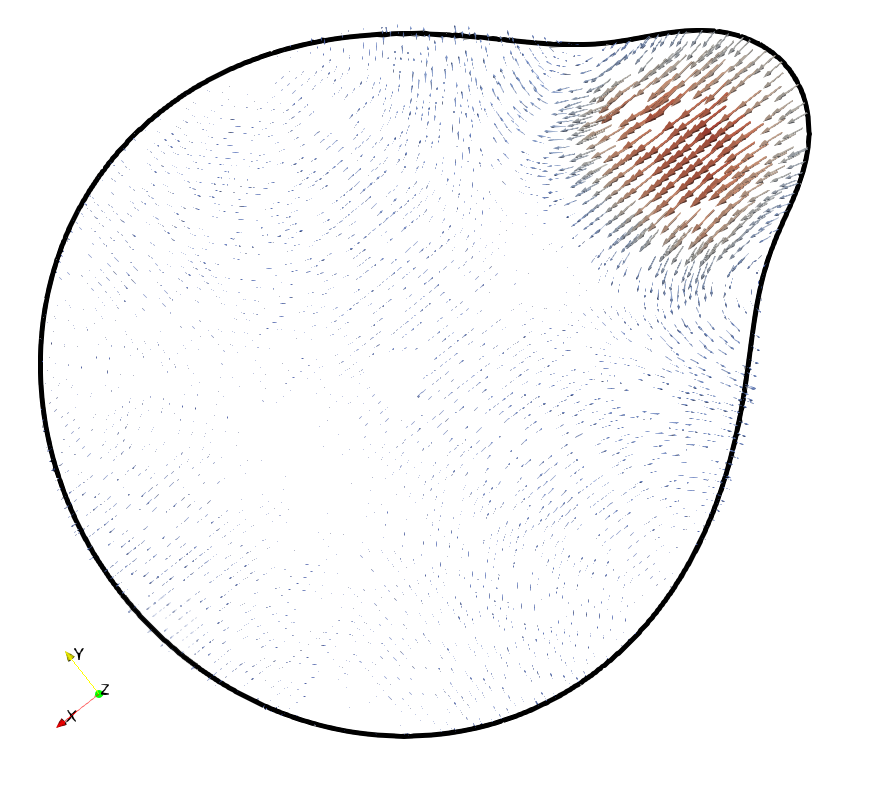}
    \caption{$\mathrm{t} = 0.03$}\label{fig:pear-relaxation-f}
  \end{subfigure}
  \begin{subfigure}[t]{0.2\textwidth}
    \centering
    \includegraphics[width=1.0\textwidth]{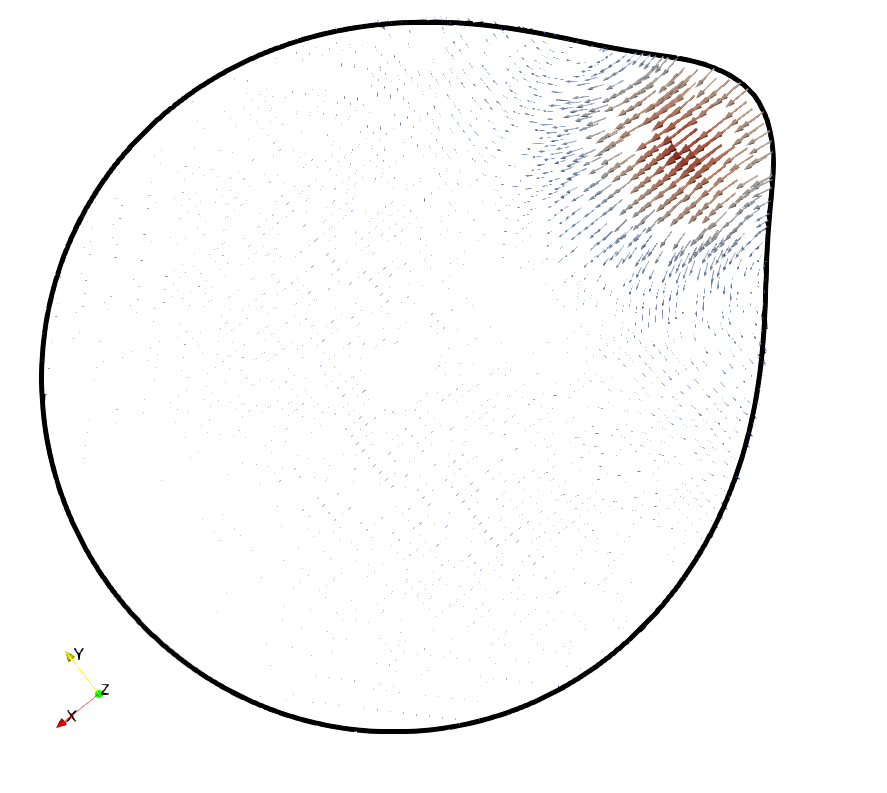}
    \caption{$\mathrm{t} = 0.06$}\label{fig:pear-relaxation-g}
  \end{subfigure}
  \begin{subfigure}[t]{0.2\textwidth}
    \centering
    \includegraphics[width=1.0\textwidth]{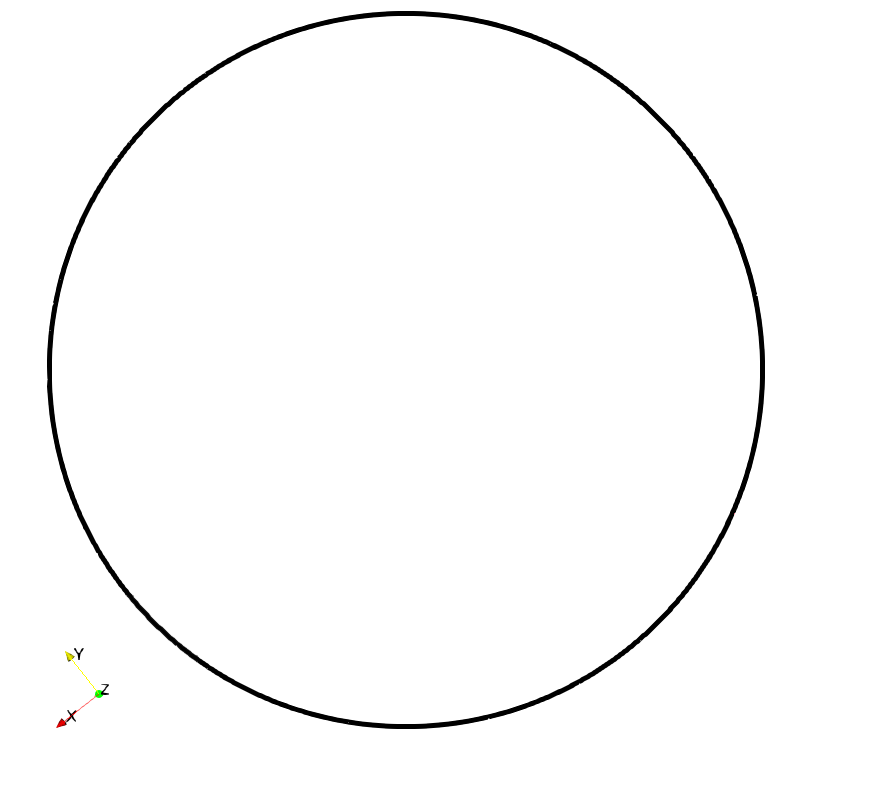}
    \caption{$\mathrm{t} = 0.24$}\label{fig:pear-relaxation-h}
  \end{subfigure} \\
  \begin{subfigure}[t]{0.08\textwidth}
    \centering
    \includegraphics[width=1.0\textwidth]{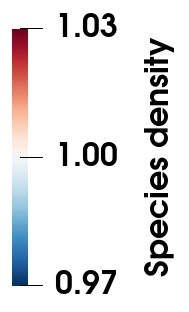}
  \end{subfigure} 
  \begin{subfigure}[t]{0.2\textwidth}
    \centering
    \includegraphics[width=1.0\textwidth]{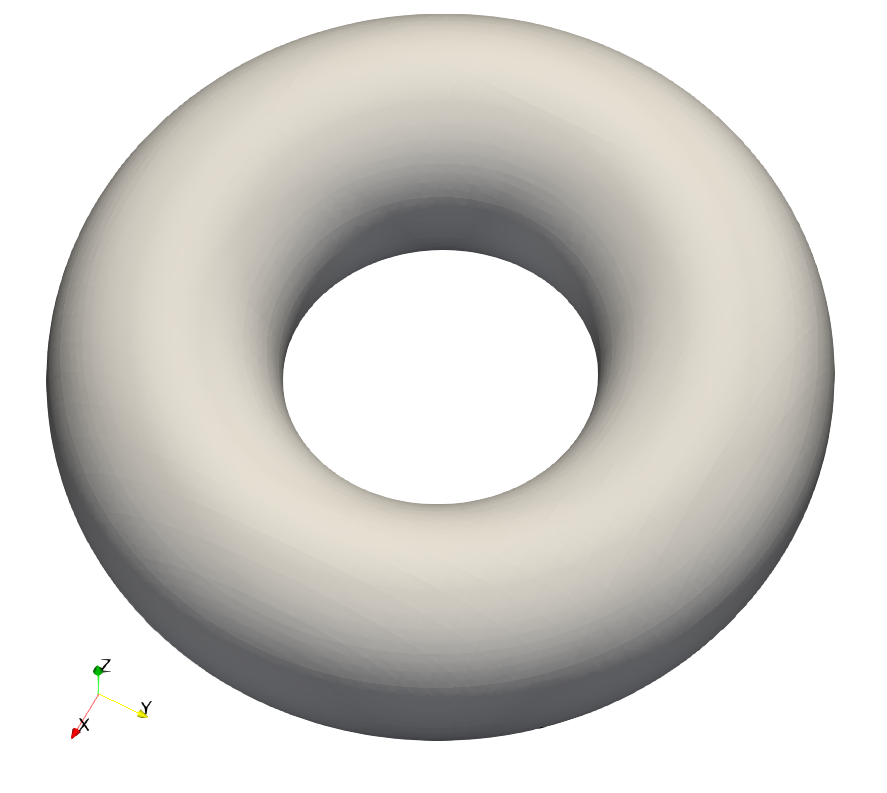}
    \caption{$\mathrm{t} = 0.0$}\label{fig:torus-relaxation-a}
  \end{subfigure}
  \begin{subfigure}[t]{0.2\textwidth}
    \centering
    \includegraphics[width=1.0\textwidth]{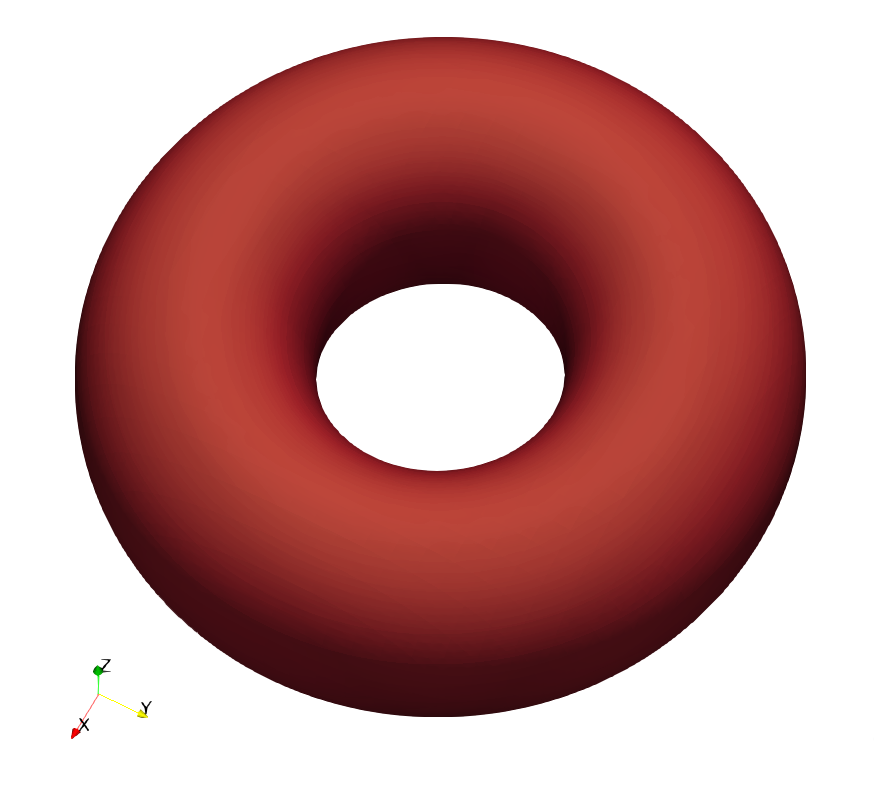}
    \caption{$\mathrm{t} = 0.06$}\label{fig:torus-relaxation-b}
  \end{subfigure}
  \begin{subfigure}[t]{0.2\textwidth}
    \centering
    \includegraphics[width=1.0\textwidth]{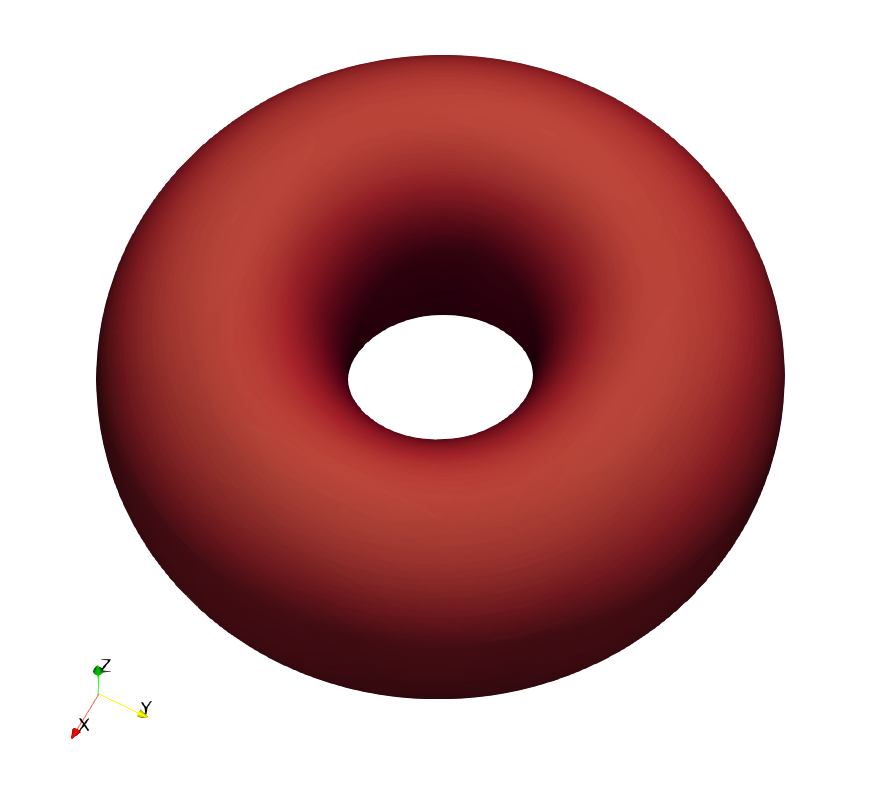}
    \caption{$\mathrm{t} = 0.12$}\label{fig:torus-relaxation-c}
  \end{subfigure}
  \begin{subfigure}[t]{0.2\textwidth}
    \centering
    \includegraphics[width=1.0\textwidth]{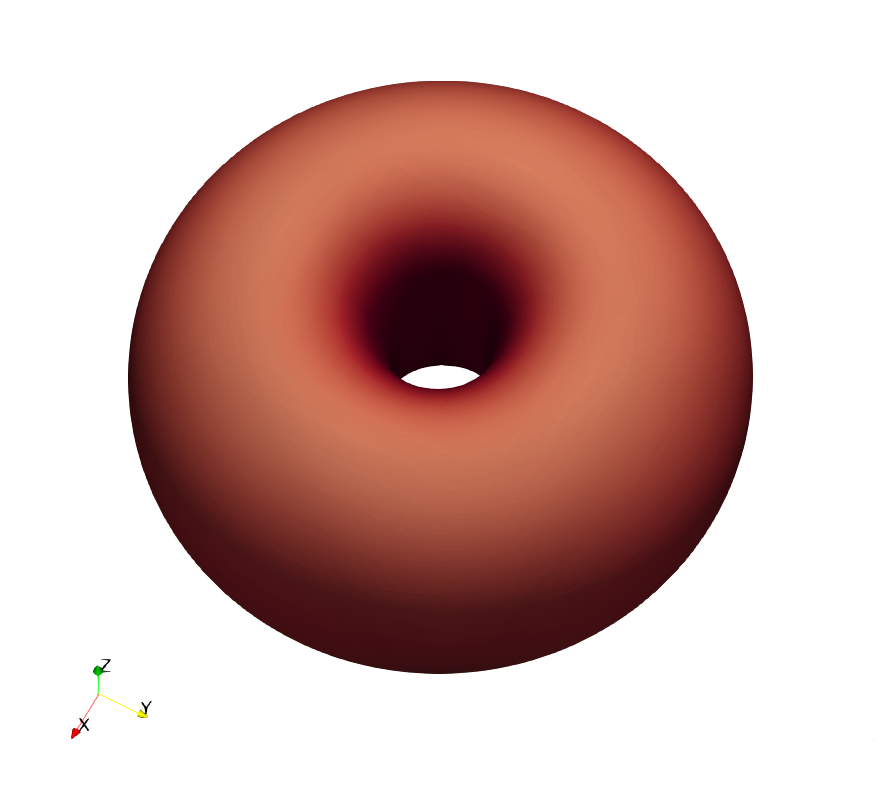}
    \caption{$\mathrm{t} = 0.23$}\label{fig:torus-relaxation-d}
  \end{subfigure} \\
  \begin{subfigure}[t]{0.08\textwidth}
    \centering
    \includegraphics[width=1.0\textwidth]{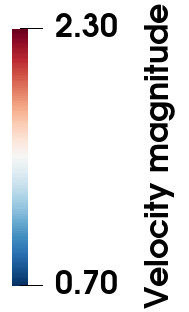}
  \end{subfigure}
  \begin{subfigure}[t]{0.2\textwidth}
    \centering
    \includegraphics[width=1.0\textwidth]{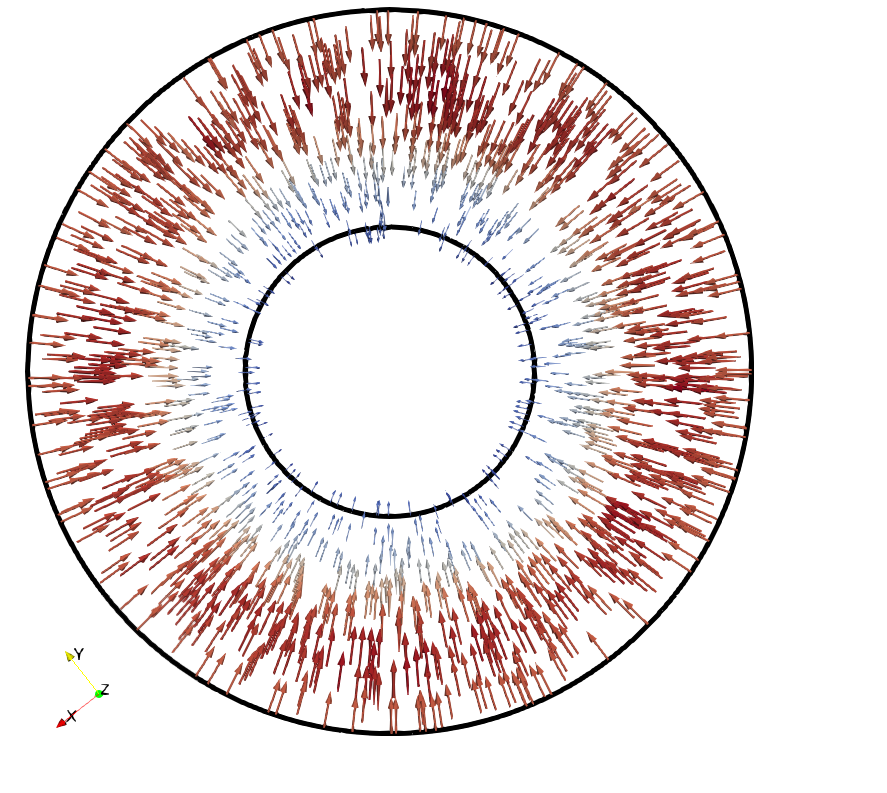}
    \caption{$\mathrm{t} = 0.0$}\label{fig:torus-relaxation-e}
  \end{subfigure}
  \begin{subfigure}[t]{0.2\textwidth}
    \centering
    \includegraphics[width=1.0\textwidth]{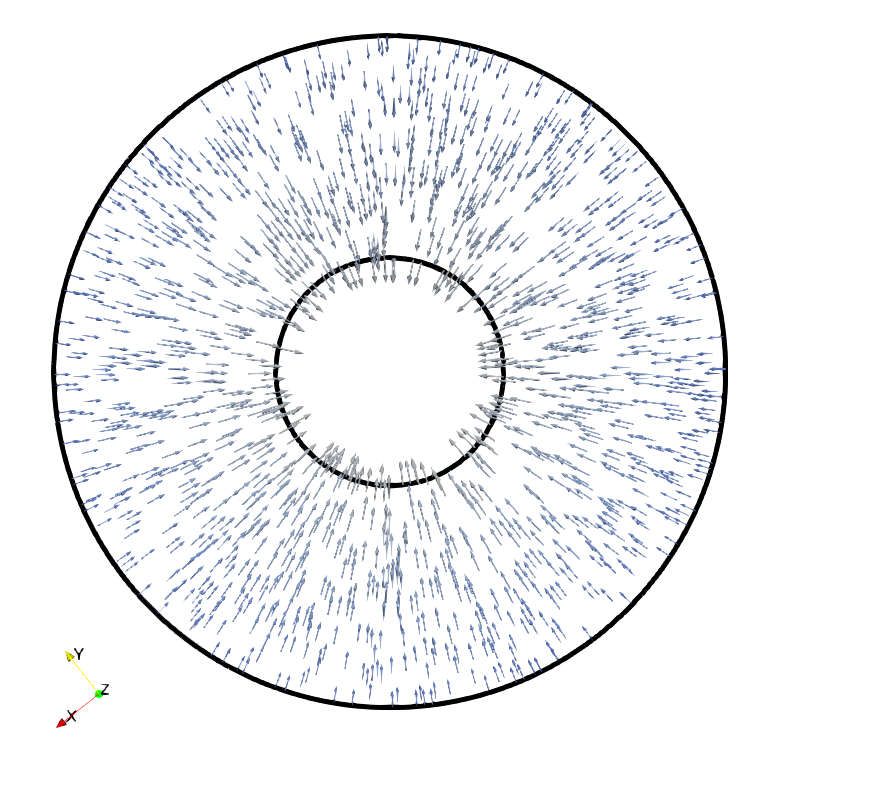}
    \caption{$\mathrm{t} = 0.06$}\label{fig:torus-relaxation-f}
  \end{subfigure}
  \begin{subfigure}[t]{0.2\textwidth}
    \centering
    \includegraphics[width=1.0\textwidth]{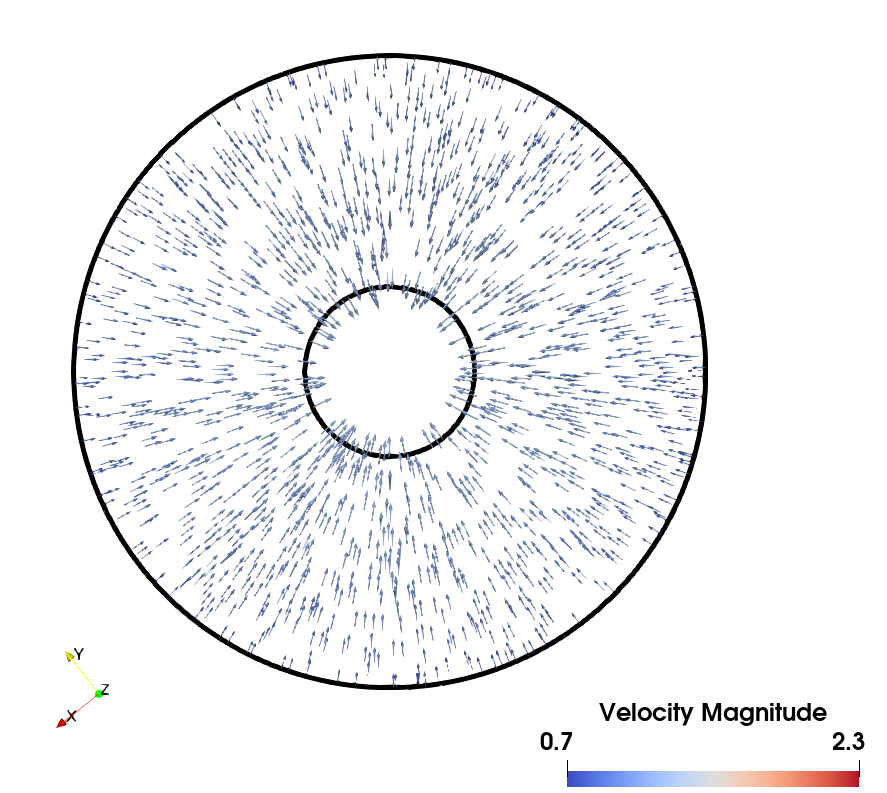}
    \caption{$\mathrm{t} = 0.12$}\label{fig:torus-relaxation-g}
  \end{subfigure}
  \begin{subfigure}[t]{0.2\textwidth}
    \centering
    \includegraphics[width=1.0\textwidth]{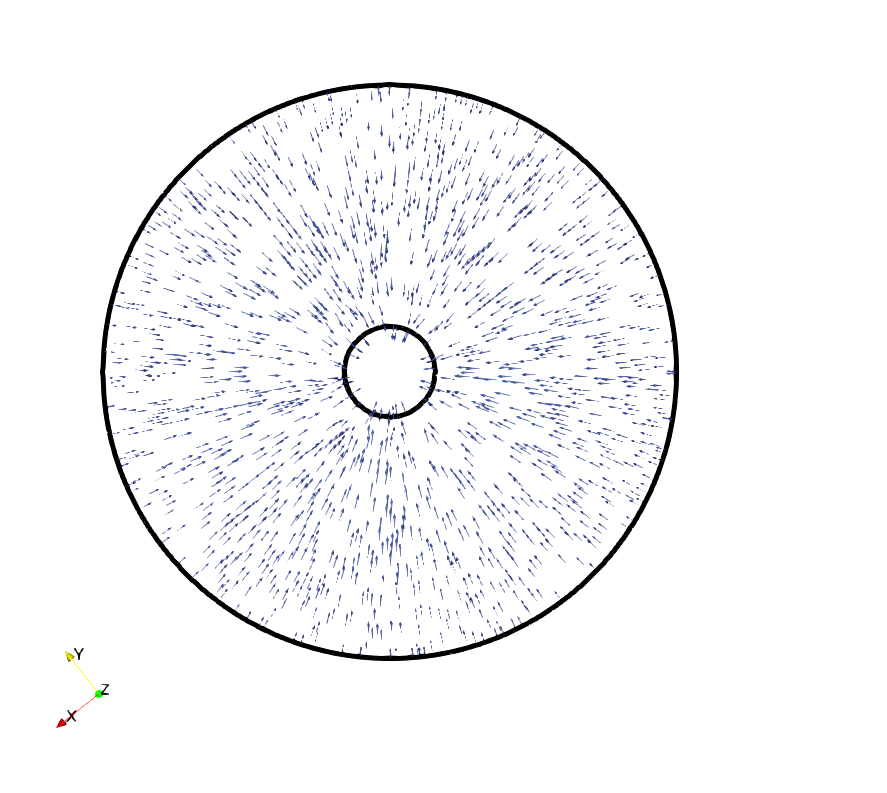}
    \caption{$\mathrm{t} = 0.23$}\label{fig:torus-relaxation-h}
  \end{subfigure}
  \caption{\textbf{Relaxation dynamics.} Marangoni effects generated by mean curvature gradients are balanced by evolving the shape towards a sphere. \emph{Relaxation of a popcorn, a pear and a torus.} \textbf{\textsf{(a)}}-\textbf{\textsf{(d)}}, \textbf{\textsf{(i)}}-\textbf{\textsf{(l)}} and \textbf{\textsf{(q)}}-\textbf{\textsf{(t)}} Surface concentration of stress-regulating molecular species. \textbf{\textsf{(e)}}-\textbf{\textsf{(h)}}, \textbf{\textsf{(m)}}-\textbf{\textsf{(p)}} and \textbf{\textsf{(u)}}-\textbf{\textsf{(x)}} Bulk flow field.}
  \label{fig:relaxation}
\end{figure}

Figure~\ref{fig:relaxation} represents the relaxation dynamics on a popcorn, pear and torus shapes. We start, in both cases, with a constant surface concentration of the molecular species. All simulation parameters are gathered in Table~\ref{tab:params-relaxation}. Setting $\mbox{\textit{Pe}} = 5.0$ and $\tau_D k_{\rm off} = 50.0$ to force a diffusion-dominated regime, we observe the shape evolving to suppress the mean curvature inhomogeneities; regions with higher curvature reduce their curvature, and viceversa. During the relaxation process, transient inhomogeneities in the surface concentration of the species appear and dissipate, due to the local expansion and contraction of the surface. This example underscores the role of active forces in sustaining a non-spherical equilibrium shape. Despite nontrivial 3D shape dynamics, errors in volume conservation remain low, as shown in Figure~\ref{fig:relaxation-vols}.


\begin{table}[h!]
  \centering
  \renewcommand{\arraystretch}{1.1}
  \scriptsize
  \begin{tabular}{|c|c|c|c|}
    \hline
    \textbf{Parameter} & \textbf{Popcorn} & \textbf{Pear}  & \textbf{Torus} \\ \hline
    Bounding box $\mathcal{B}$ & $[-2.01,2.01]^3$ & $[-2.25,1.25]\times[-1.75,1.75]^2$ & $[-1.1,1.1]^3$ \\ \hline
    Péclet number $\mbox{\textit{Pe}}$ & \multicolumn{3}{c|}{5.0} \\ \hline
    $\tau_D k_{\rm off}$ & \multicolumn{3}{c|}{50.0} \\ \hline
    Hydrodynamic length $L_\eta$ & \multicolumn{3}{c|}{$10^{-5}$} \\ \hline
    Characteristic length $R_0$ & 0.6 & 2.0 & 1.0 \\ \hline
    Number of cells per axis $n$ & 35 & 30 & 30 \\ \hline
    Time step $\Delta \mathrm{t}$ & \multicolumn{3}{c|}{$10^{-4}$} \\ \hline
  \end{tabular}
  \vspace{0.2cm}
  \caption{\textbf{Relaxation dynamics.} Simulation parameters.}
  \label{tab:params-relaxation}
\end{table}

\subsection{Bi- and uni-lateral cytokinesis}
\label{subsec:cleavage}

We now showcase the capabilities of the unfitted \ac{fe} framework to deal with constricted deformations. The goal is to simulate cytokinesis, i.e., the physical process at the end of cell division that partitions the cell into two daughter cells~\cite{green2012cytokinesis}. In animal cells, cytokinesis is achieved by an actomyosin furrow forming at the equator that constricts the cell until scission. Meanwhile, cortical tension at the poles resists the increasing cell pressure~\cite{sedzinski2011polar}. As reviewed in~\cite{da2022viscous}, both early hypotheses and recent experimental and modelling studies support the idea that cell division is driven by a gradient of surface tension directed toward the division axis. Moreover, the dominant mechanical contributor is cortical tension, rather than active or viscous torque contributions~\cite{turlier2014furrow,da2022viscous}. The mathematical model of Equations~\eqref{eq:full-non-dimensional-1}-\eqref{eq:full-non-dimensional-9} is thus adequate for this type of cellular process.

\begin{figure}[!ht]
  \centering
  \begin{subfigure}[t]{0.08\textwidth}
    \centering
    \includegraphics[width=0.9\textwidth]{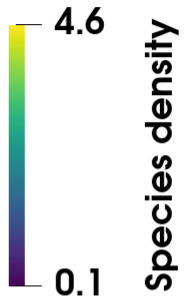}
  \end{subfigure}
  \begin{subfigure}[t]{0.24\textwidth}
    \centering
    \includegraphics[width=0.9\textwidth]{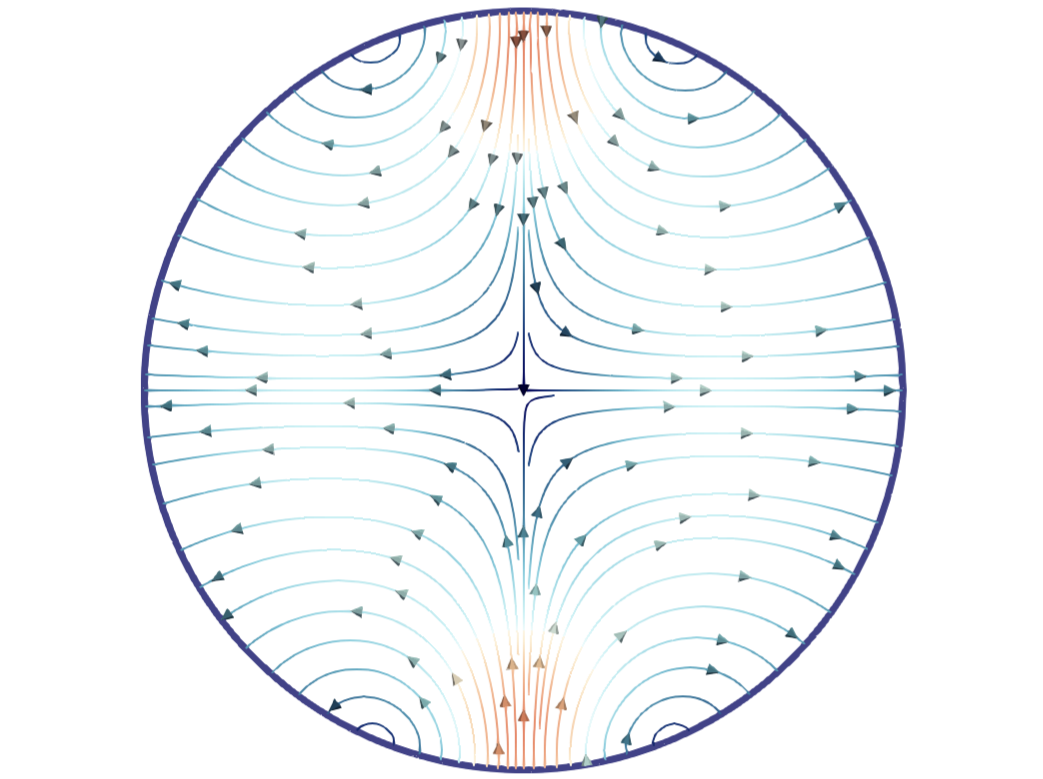}
    \caption{$\mathrm{t} = 0.00$}\label{fig:2D-division-a}
  \end{subfigure}
  \begin{subfigure}[t]{0.24\textwidth}
    \centering
    \includegraphics[width=0.9\textwidth]{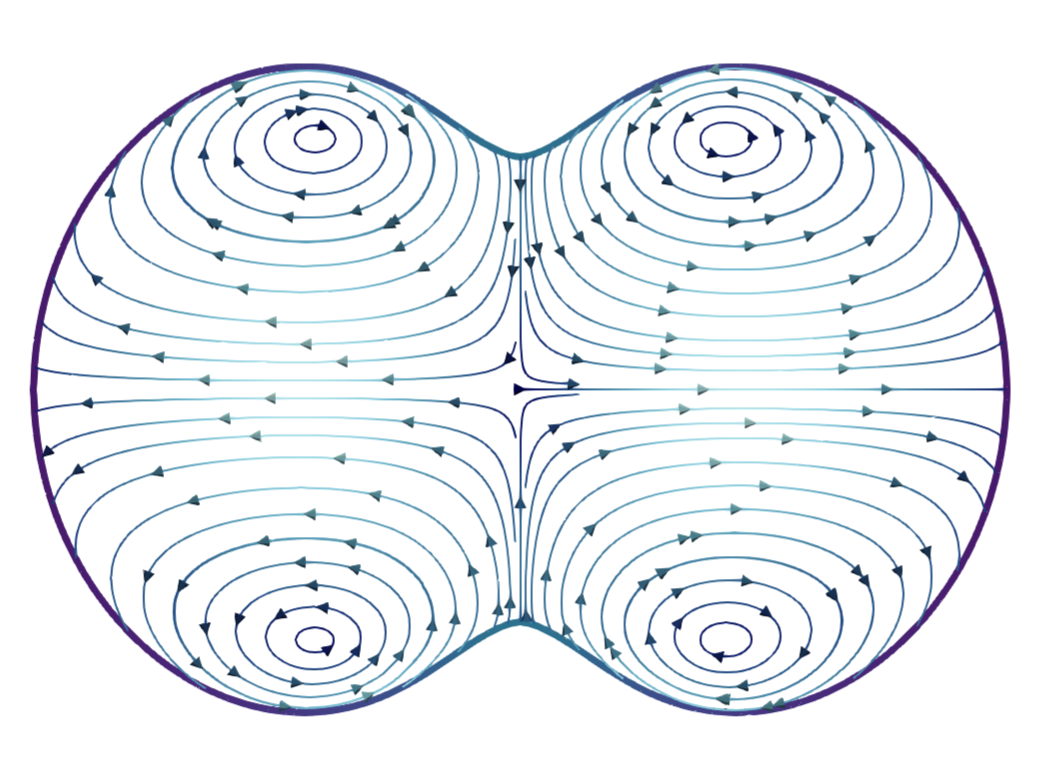}
    \caption{$\mathrm{t} = 0.08$}\label{fig:2D-division-b}
  \end{subfigure}
  \begin{subfigure}[t]{0.24\textwidth}
    \centering
    \includegraphics[width=0.9\textwidth]{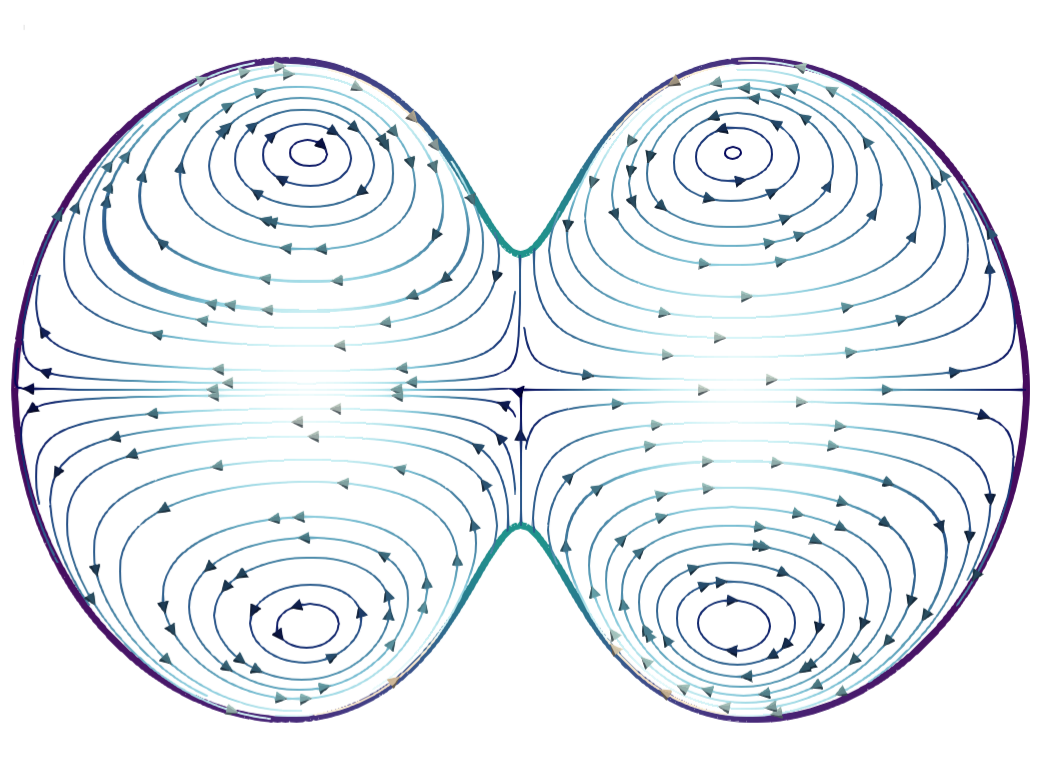}
    \caption{$\mathrm{t} = 0.16$}\label{fig:2D-division-c}
  \end{subfigure} \\
  \begin{subfigure}[t]{0.08\textwidth}
    \centering
    \includegraphics[width=0.9\textwidth]{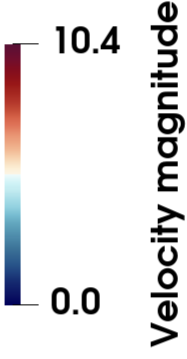}
  \end{subfigure}
  \begin{subfigure}[t]{0.24\textwidth}
    \centering
    \includegraphics[width=0.9\textwidth]{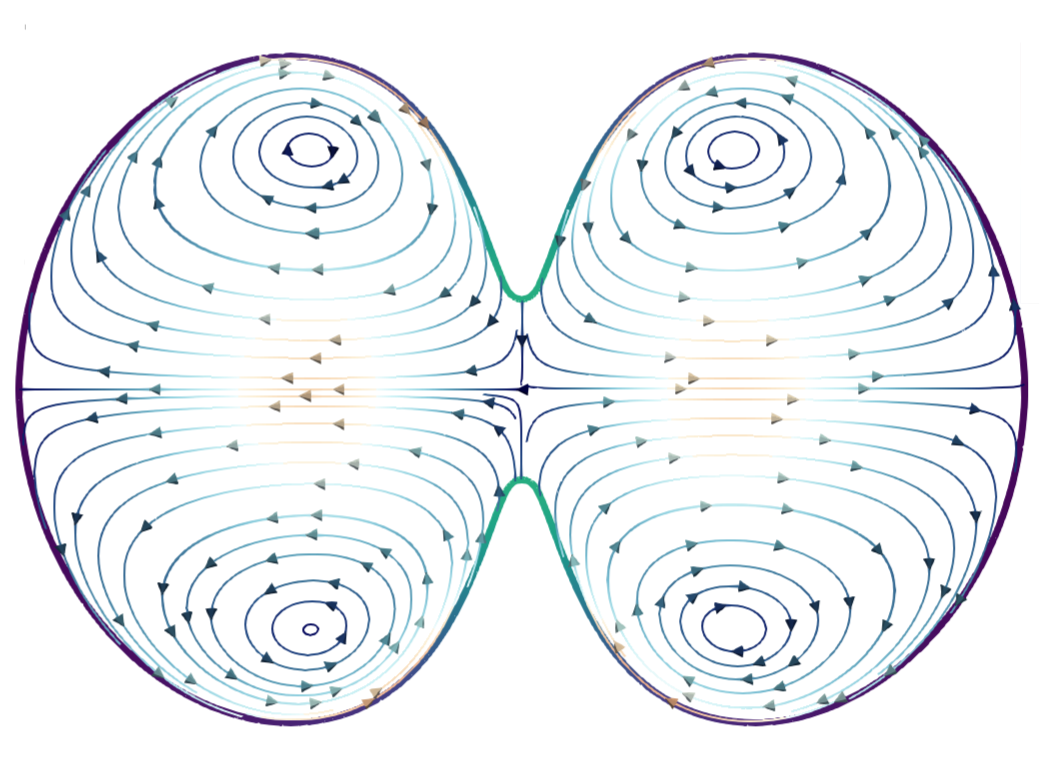}
    \caption{$\mathrm{t} = 0.20$}\label{fig:2D-division-d}
  \end{subfigure}
  \begin{subfigure}[t]{0.24\textwidth}
    \centering
    \includegraphics[width=0.9\textwidth]{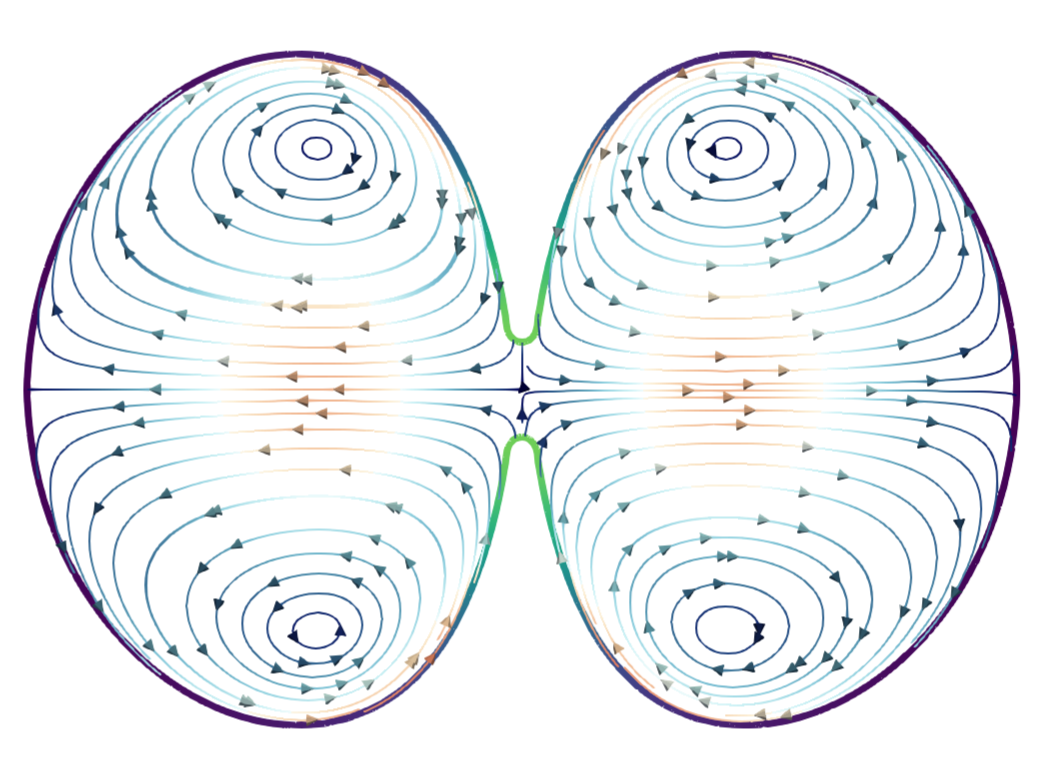}
    \caption{$\mathrm{t} = 0.24$}\label{fig:2D-division-e}
  \end{subfigure}
  \begin{subfigure}[t]{0.24\textwidth}
    \centering
    \includegraphics[width=0.9\textwidth]{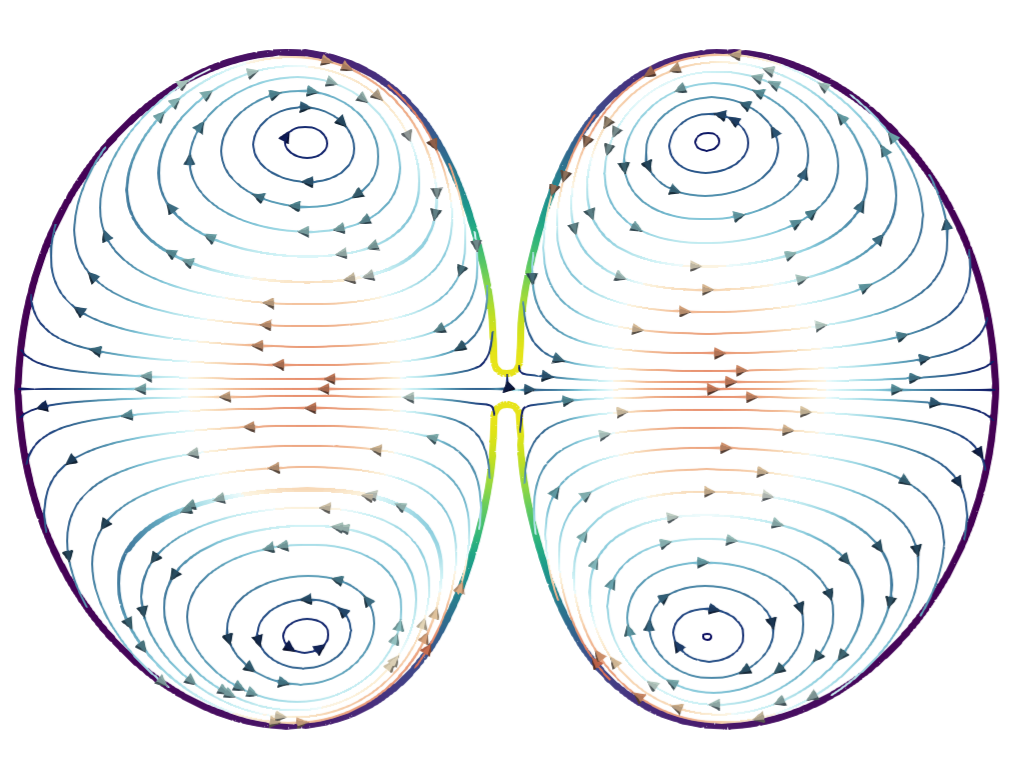}
    \caption{$\mathrm{t} = 0.28$}\label{fig:2D-division-f}
  \end{subfigure}
  \caption{\textbf{2D axisymmetric cleavage}. Evolution of bulk flow field and surface density of the molecular species.}
  \label{fig:2D-division}
\end{figure}

\begin{figure}[!ht]
  \centering
  \begin{subfigure}[t]{0.06\textwidth}
    \centering
    \includegraphics[width=0.9\textwidth]{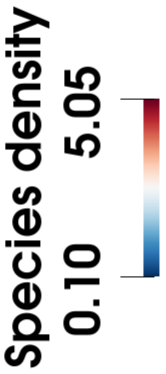}
  \end{subfigure}
  \begin{subfigure}[t]{0.22\textwidth}
    \centering
    \includegraphics[width=0.9\textwidth]{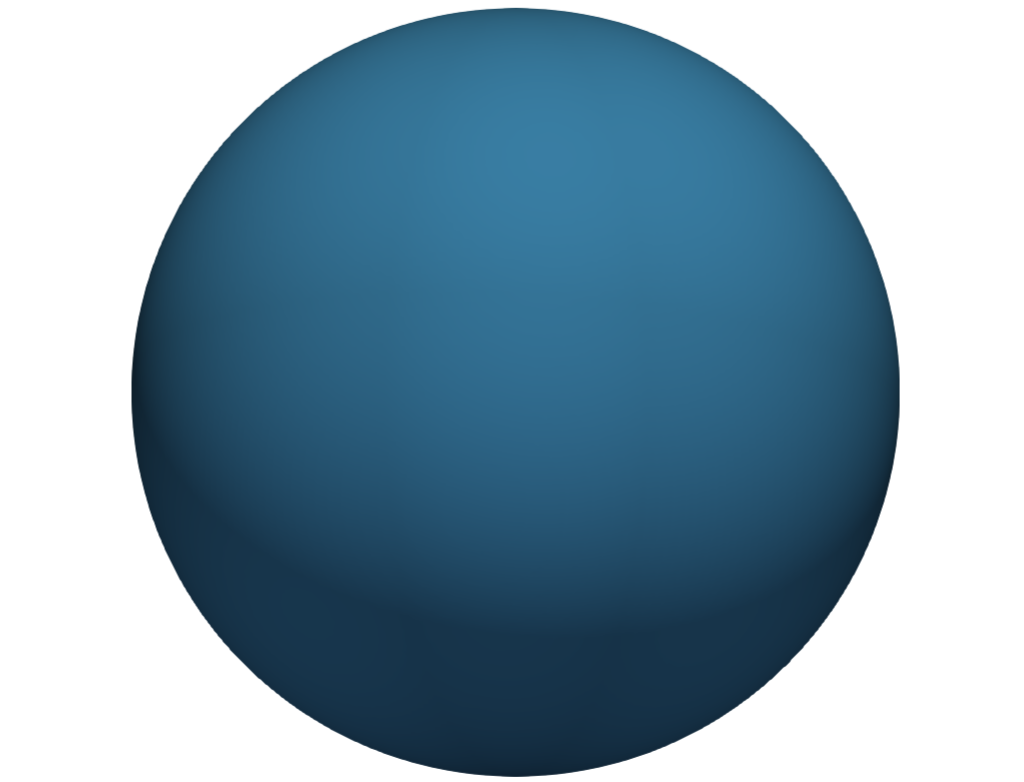}
    \caption{$\mathrm{t} = 0.00$}\label{fig:3D-division-a}
  \end{subfigure}
  \begin{subfigure}[t]{0.22\textwidth}
    \centering
    \includegraphics[width=0.9\textwidth]{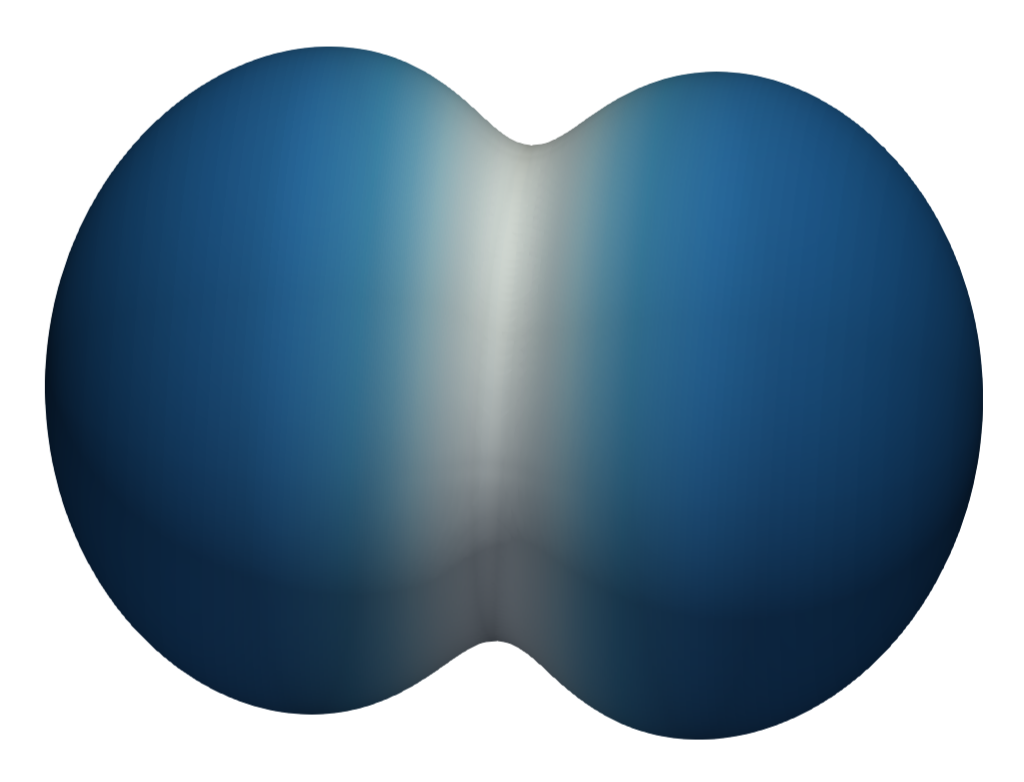}
    \caption{$\mathrm{t} = 0.03$}\label{fig:3D-division-b}
  \end{subfigure}
  \begin{subfigure}[t]{0.22\textwidth}
    \centering
    \includegraphics[width=0.9\textwidth]{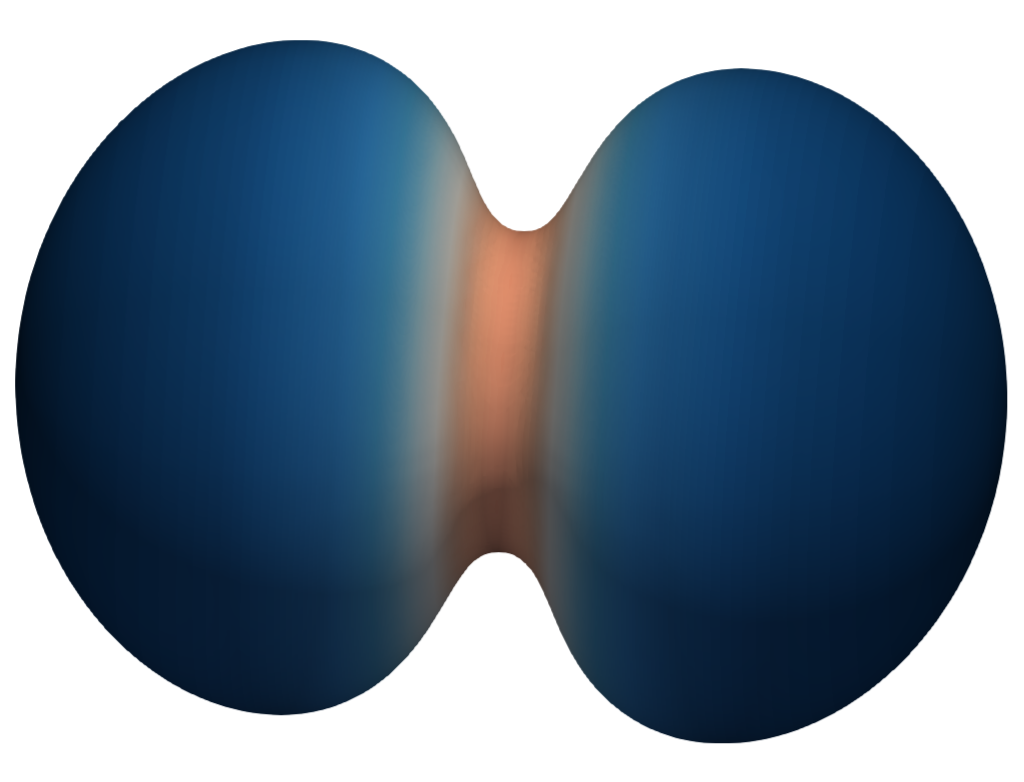}
    \caption{$\mathrm{t} = 0.06$}\label{fig:3D-division-c}
  \end{subfigure}
  \begin{subfigure}[t]{0.22\textwidth}
    \centering
    \includegraphics[width=0.9\textwidth]{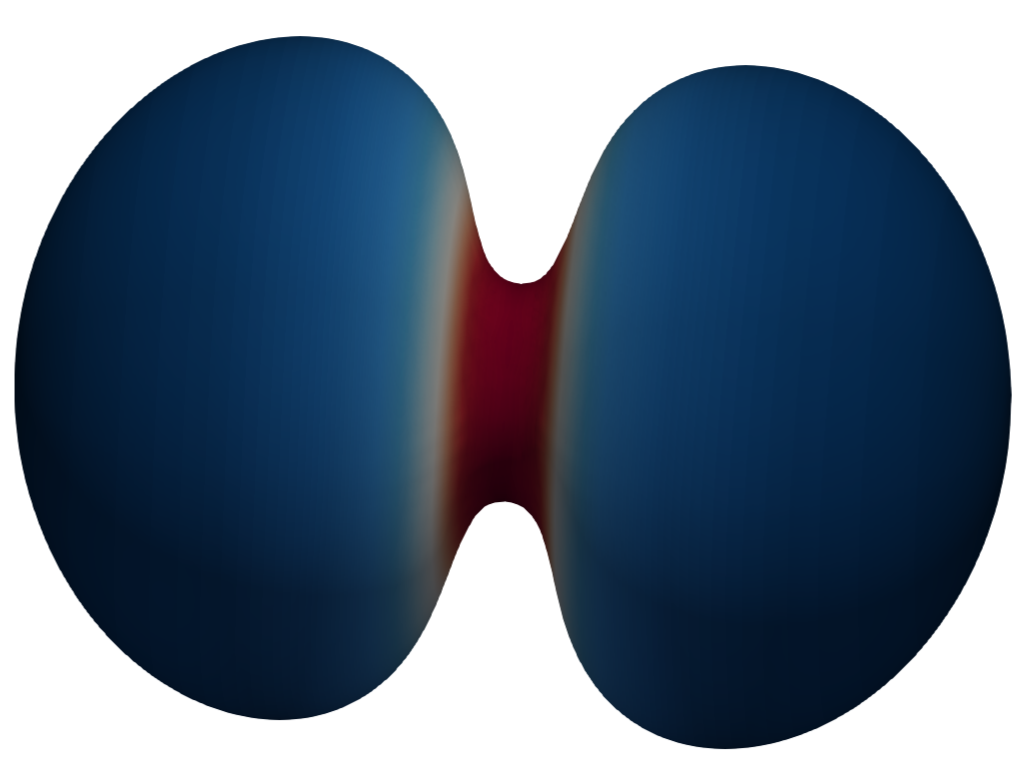}
    \caption{$\mathrm{t} = 0.08$}\label{fig:3D-division-d}
  \end{subfigure} \\
  \begin{subfigure}[t]{0.06\textwidth}
    \centering
    \includegraphics[width=0.9\textwidth]{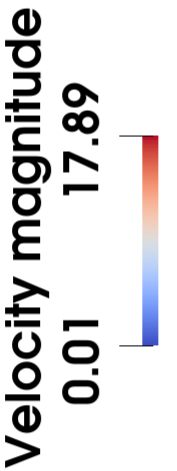}
  \end{subfigure}
  \begin{subfigure}[t]{0.22\textwidth}
    \centering
    \includegraphics[width=0.9\textwidth]{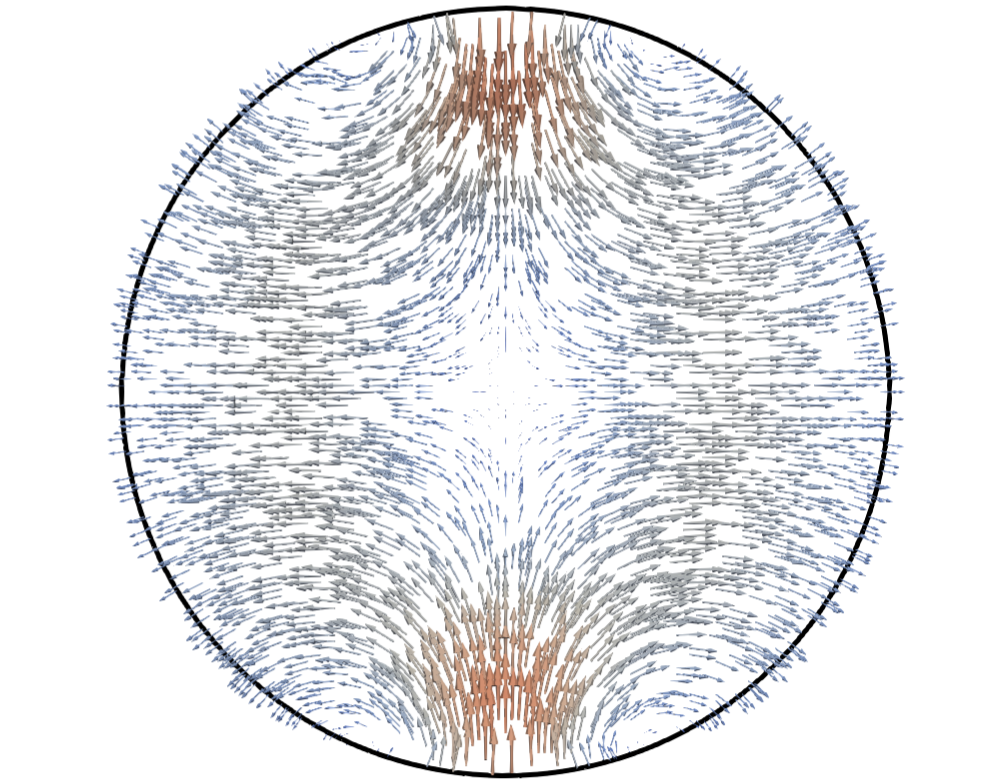}
    \caption{$\mathrm{t} = 0.00$}\label{fig:3D-division-e}
  \end{subfigure}
  \begin{subfigure}[t]{0.22\textwidth}
    \centering
    \includegraphics[width=0.9\textwidth]{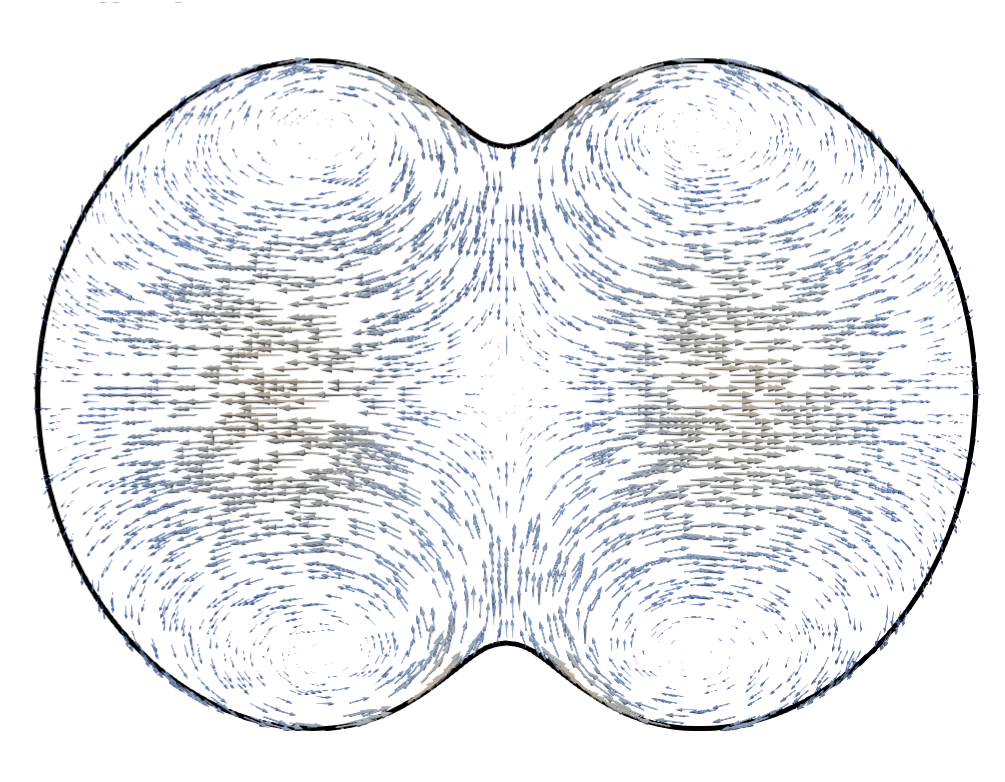}
    \caption{$\mathrm{t} = 0.03$}\label{fig:3D-division-f}
  \end{subfigure}
  \begin{subfigure}[t]{0.22\textwidth}
    \centering
    \includegraphics[width=0.9\textwidth]{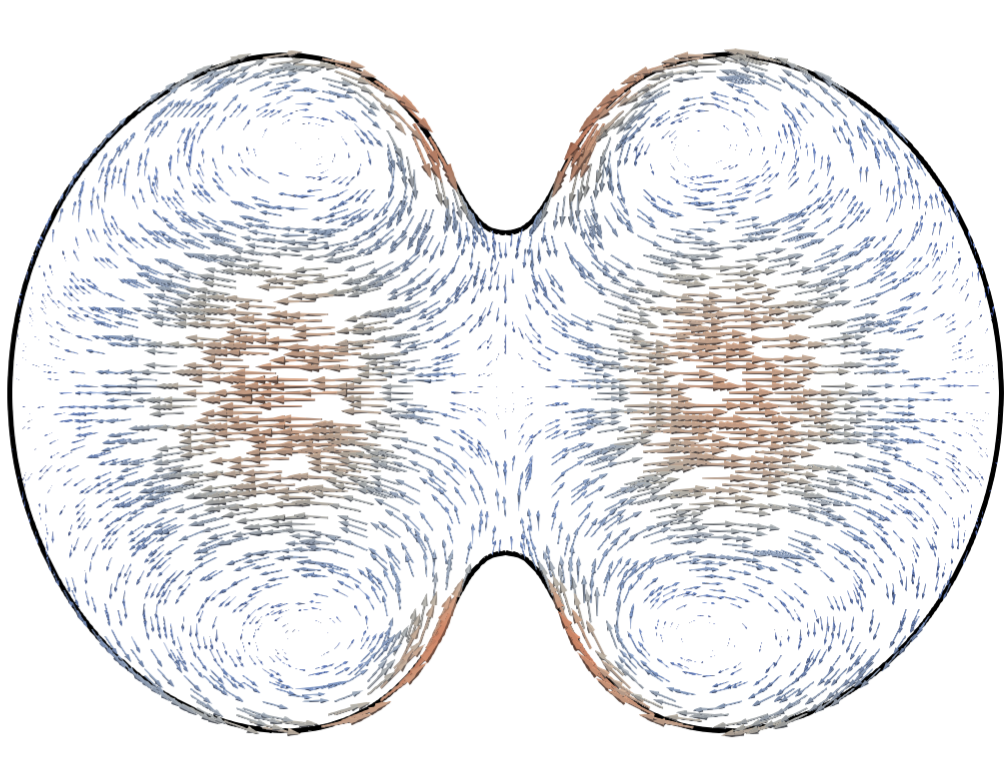}
    \caption{$\mathrm{t} = 0.06$}\label{fig:3D-division-g}
  \end{subfigure}
  \begin{subfigure}[t]{0.22\textwidth}
    \centering
    \includegraphics[width=0.9\textwidth]{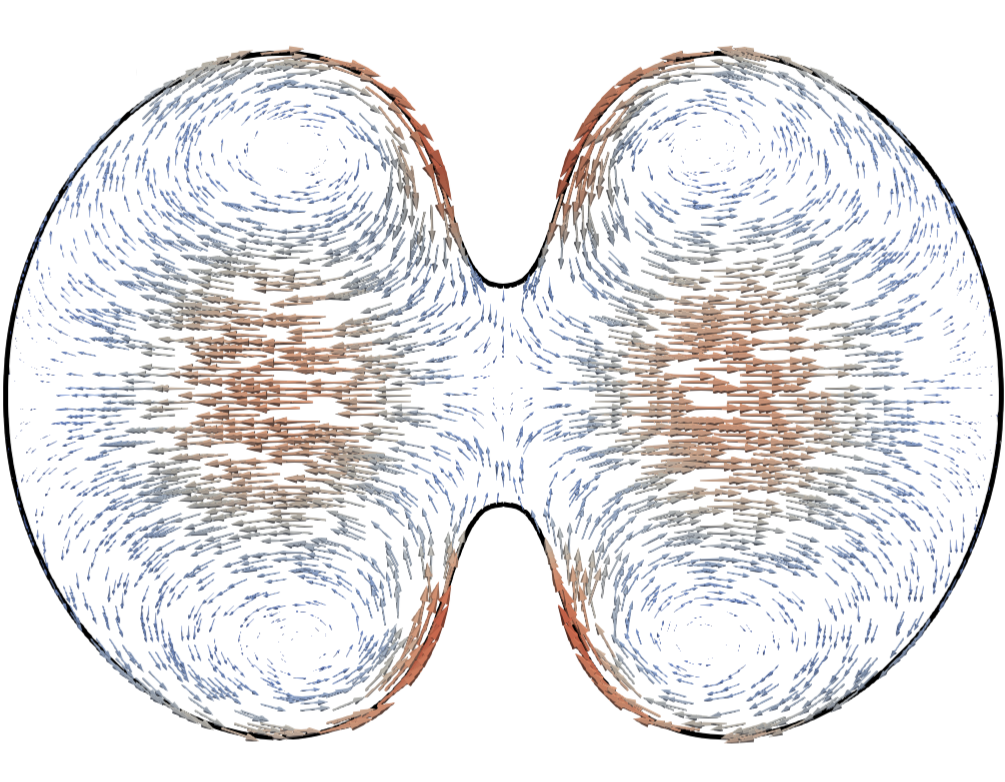}
    \caption{$\mathrm{t} = 0.08$}\label{fig:3D-division-h}
  \end{subfigure}
  \caption{\textbf{3D symmetric cleavage}. Evolution of bulk flow field and surface density of the molecular species.}
  \label{fig:3D-division}
\end{figure}

\begin{figure}[!ht]
  \centering
  \begin{subfigure}{0.49\textwidth}
    \centering
    \includegraphics[width=0.95\textwidth]{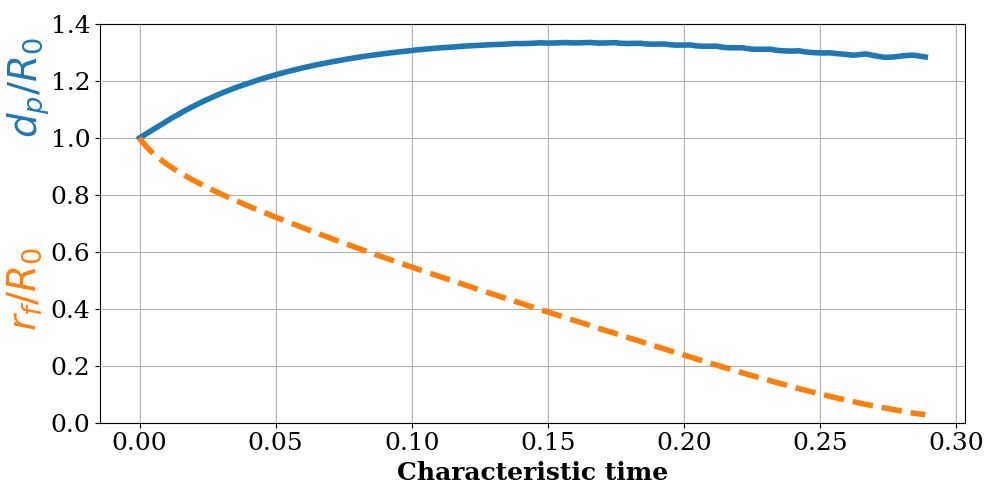}
    \caption{2D axisymmetric cleavage}\label{fig:2D-division-radii}
  \end{subfigure}
  \begin{subfigure}{0.49\textwidth}
    \centering
    \includegraphics[width=0.95\textwidth]{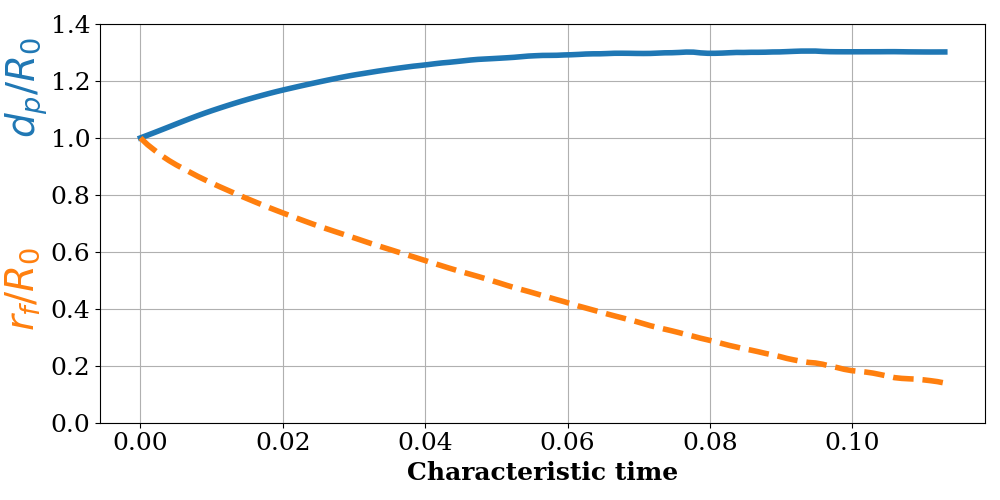}
    \caption{3D cleavage}\label{fig:3D-division-radii}
  \end{subfigure}
  \caption{\textbf{Symmetric cleavage}. Evolution of the radius of the contractile ring $r_f$ and the distance between poles $d_p$ relative to the initial radius of the sphere $R_0$.}
  \label{fig:division-radii}
\end{figure}

We first focus on the stereotypical mode of cell division, i.e., symmetric cytokinesis, which cuts the cell into two equal-sized daughter cells. This type of division is achieved by a circumferential cleavage furrow that forms and develops a ring constriction with two cortical poles under tensile stress at both sides. We will then refer to this configuration as \emph{bilateral} cytokinesis. The starting point is a spherical cell of radius $R_0$. Experimental measurements have shown that the mitotic apparatus of animal cells generates and sustains a Gaussian-like band of myosin overactivity to position the furrow and guide the constriction at the cell equator~\cite{rappaport1996cytokinesis,bement2005microtubule}. Following~\cite{turlier2014furrow}, we model the overactivity at the contractile ring by spatially modulating the active surface tension $\mathbf{N}_\Gamma^{\rm act}$ in Equation~\eqref{eq:active-surface-tension} through the myosin activity field $\xi$, defined as
\begin{equation}
  \label{eq:overactivity}
  \xi(x_\alpha) = \xi_0 + ( \delta \xi - \xi_0 ) e^{ - \frac{1}{2} \frac{x_\alpha^2}{\omega^2} },
\end{equation}
where $\xi_0$ represents a basal level of activity, controlling active tension at the poles, $\delta \xi$ is the overactivity, expanding along the axis of rotational symmetry $x_\alpha$ over a typical width $\omega$. Figure~\ref{fig:2D-division} and~\ref{fig:3D-division} show the evolution of myosin concentration at the surface and the flows in the cytoplasm for the 2D axisymmetric and 3D models. We prescribe $R_0 \! = \! 0.46$, $\omega \! = \! 0.1$, $\xi_0 \! = \! 1$ and $\delta \xi \! = \! 10$; the rest of simulation parameters are listed in Table~\ref{tab:params-division}. We observe the furrow constriction appearing due to overactivity and being reinforced by flows of myosin from the poles to the equator. Flow patterns in the cytoplasm, with two toroidal vortices forming at each side of the equator, agree with experimental observations~\cite{hiramoto1958quantitative}. Figure~\ref{fig:division-radii} represents the evolution of the radius of the contractile ring $r_f$ and the distance between the poles $d_p$, relative to the initial cell radius $R_0$. We observe tangent flows at the poles when $d_p$ plateaus.

\begin{table}[h!]
  \centering
  \renewcommand{\arraystretch}{1.1}
  \scriptsize
  \begin{tabular}{|c|c|c|}
    \hline
    \textbf{Parameter} & \textbf{2D axisymmetric} & \textbf{3D} \\ \hline
    Bounding box $\mathcal{B}$ & $[-0.75,0.75] \times [0.0,0.75]$ & $[-0.75,0.75]^3$ \\ \hline
    Péclet number $\mbox{\textit{Pe}}$ & 5.0 & 15.0 \\ \hline
    $\tau_D k_{\rm off}$ & \multicolumn{2}{c|}{1.0} \\ \hline
    Hydrodynamic length $L_\eta$ & \multicolumn{2}{c|}{$10^{5}$} \\ \hline
    Characteristic length $R_0$ & \multicolumn{2}{c|}{0.46} \\ \hline
    Number of cells per axis $n$ & 100 & 40 \\ \hline
    Time step $\Delta \mathrm{t}$ & \multicolumn{2}{c|}{$10^{-4}$} \\ \hline
  \end{tabular}
  \vspace{0.2cm}
  \caption{\textbf{Cytokinesis.} Simulation parameters.}
  \label{tab:params-division}
\end{table}

\begin{figure}[!ht]
  \centering
  \begin{subfigure}[t]{0.06\textwidth}
    \centering
    \includegraphics[width=0.9\textwidth]{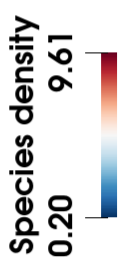}
  \end{subfigure}
  \begin{subfigure}[t]{0.22\textwidth}
    \centering
    \includegraphics[width=0.9\textwidth]{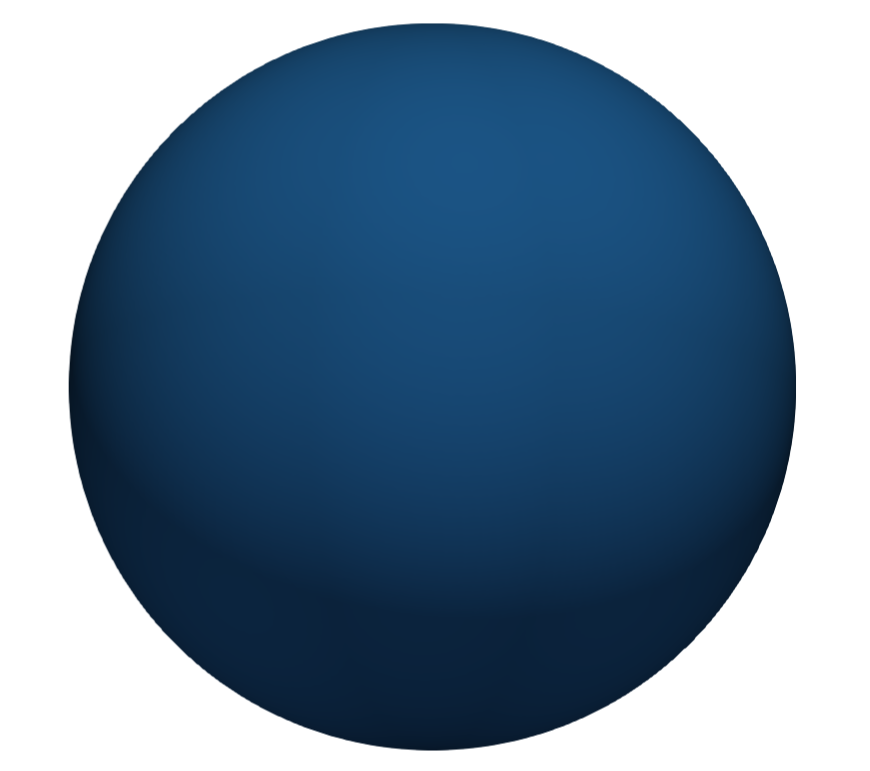}
    \caption{$\mathrm{t} = 0.000$}\label{fig:3D-division-unilateral-a}
  \end{subfigure}
  \begin{subfigure}[t]{0.22\textwidth}
    \centering
    \includegraphics[width=0.9\textwidth]{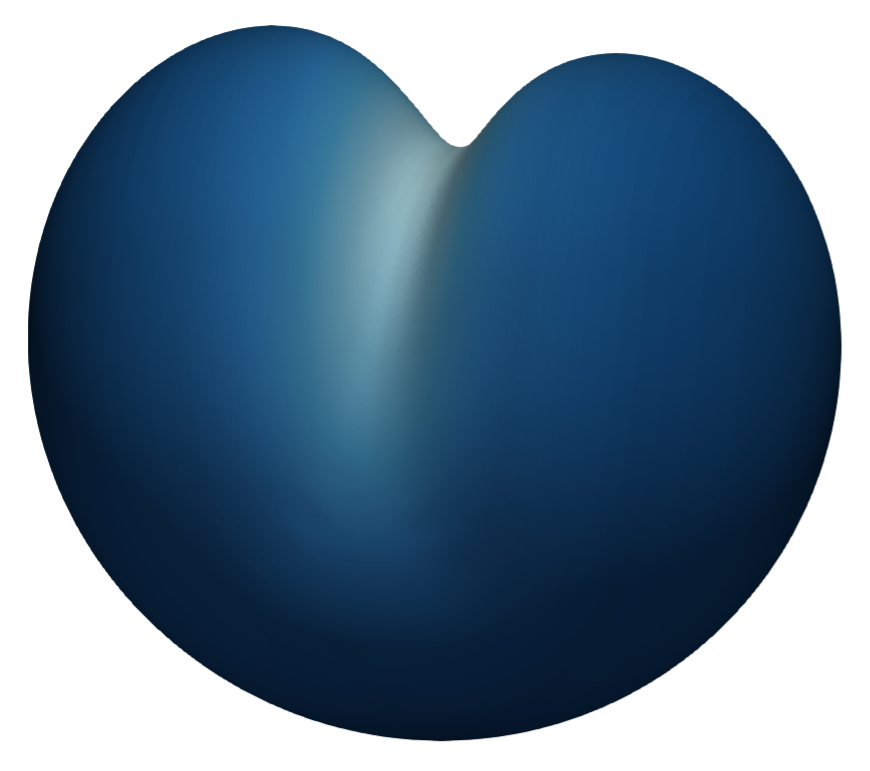}
    \caption{$\mathrm{t} = 0.027$}\label{fig:3D-division-unilateral-b}
  \end{subfigure}
  \begin{subfigure}[t]{0.22\textwidth}
    \centering
    \includegraphics[width=0.9\textwidth]{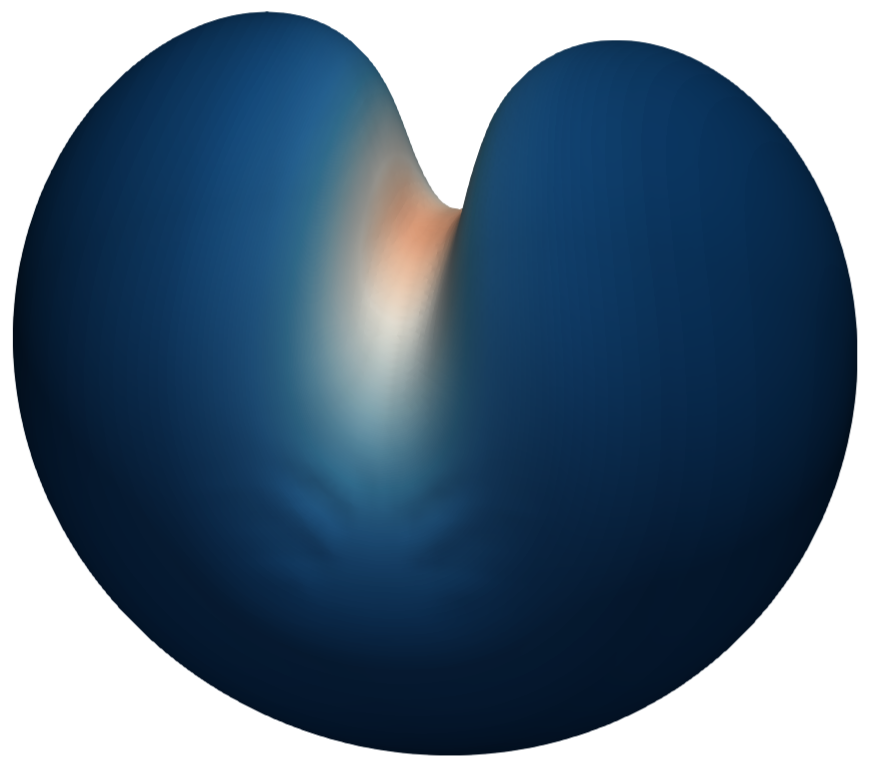}
    \caption{$\mathrm{t} = 0.054$}\label{fig:3D-division-unilateral-c}
  \end{subfigure}
  \begin{subfigure}[t]{0.22\textwidth}
    \centering
    \includegraphics[width=0.9\textwidth]{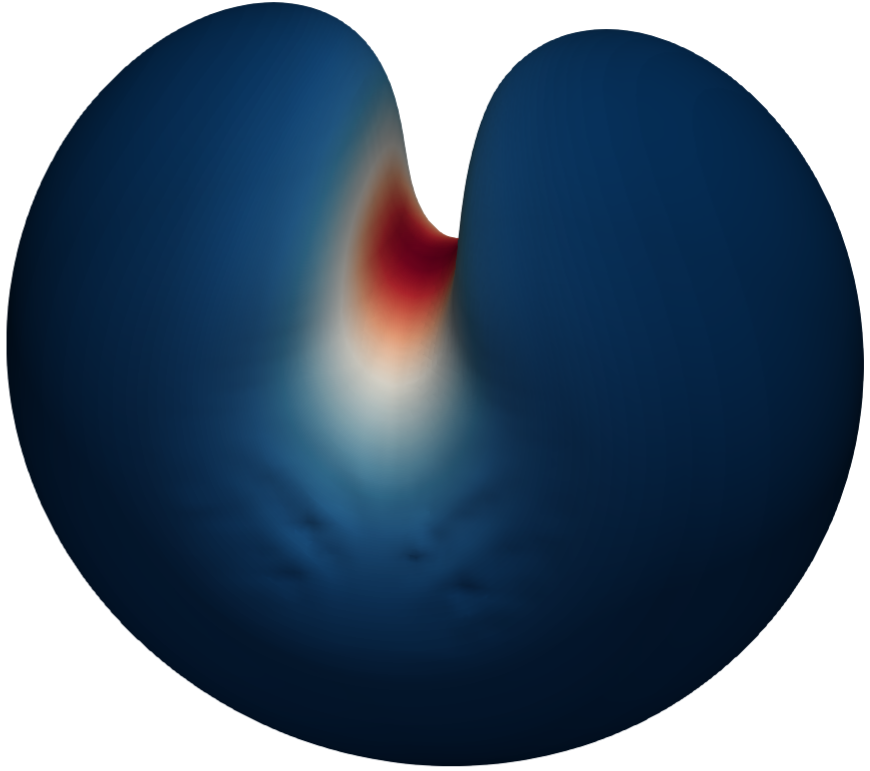}
    \caption{$\mathrm{t} = 0.081$}\label{fig:3D-division-unilateral-d}
  \end{subfigure} \\
  \begin{subfigure}[t]{0.06\textwidth}
    \centering
    \includegraphics[width=0.9\textwidth]{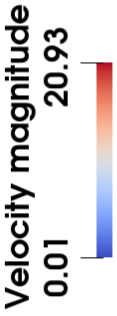}
  \end{subfigure}
  \begin{subfigure}[t]{0.22\textwidth}
    \centering
    \includegraphics[width=0.9\textwidth]{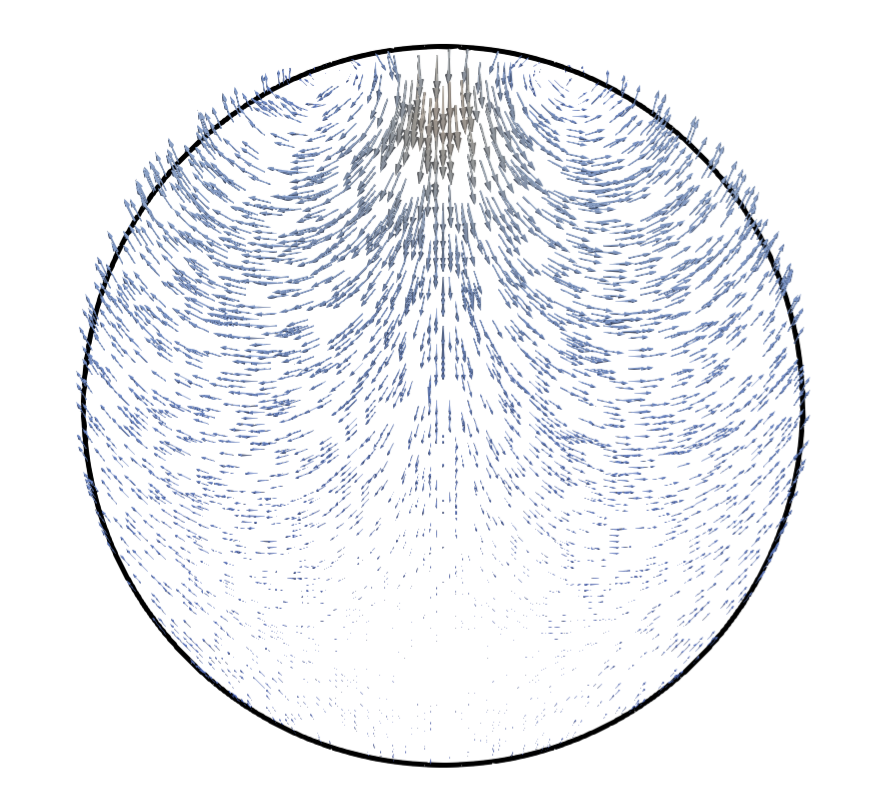}
    \caption{$\mathrm{t} = 0.000$}\label{fig:3D-division-unilateral-e}
  \end{subfigure}
  \begin{subfigure}[t]{0.22\textwidth}
    \centering
    \includegraphics[width=0.9\textwidth]{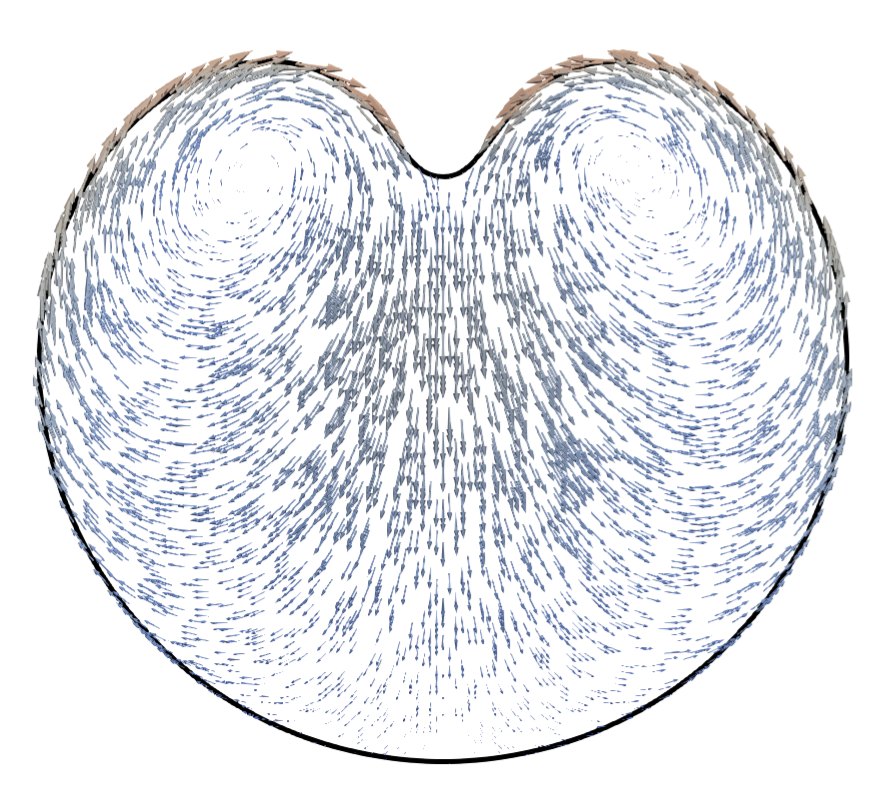}
    \caption{$\mathrm{t} = 0.027$}\label{fig:3D-division-unilateral-f}
  \end{subfigure}
  \begin{subfigure}[t]{0.22\textwidth}
    \centering
    \includegraphics[width=0.9\textwidth]{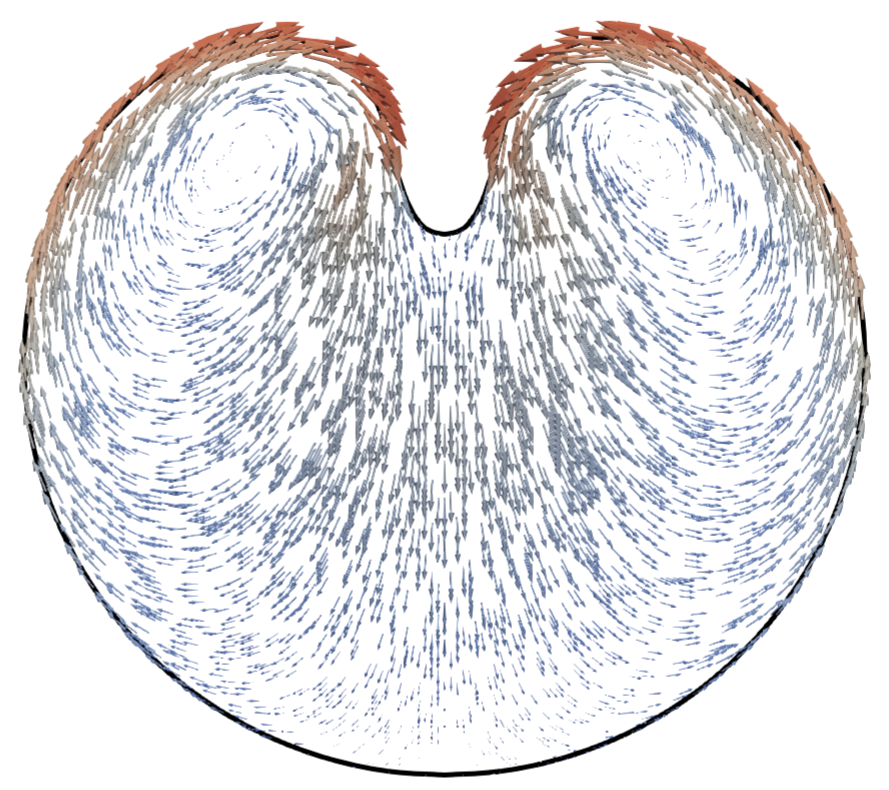}
    \caption{$\mathrm{t} = 0.054$}\label{fig:3D-division-unilateral-g}
  \end{subfigure}
  \begin{subfigure}[t]{0.22\textwidth}
    \centering
    \includegraphics[width=0.9\textwidth]{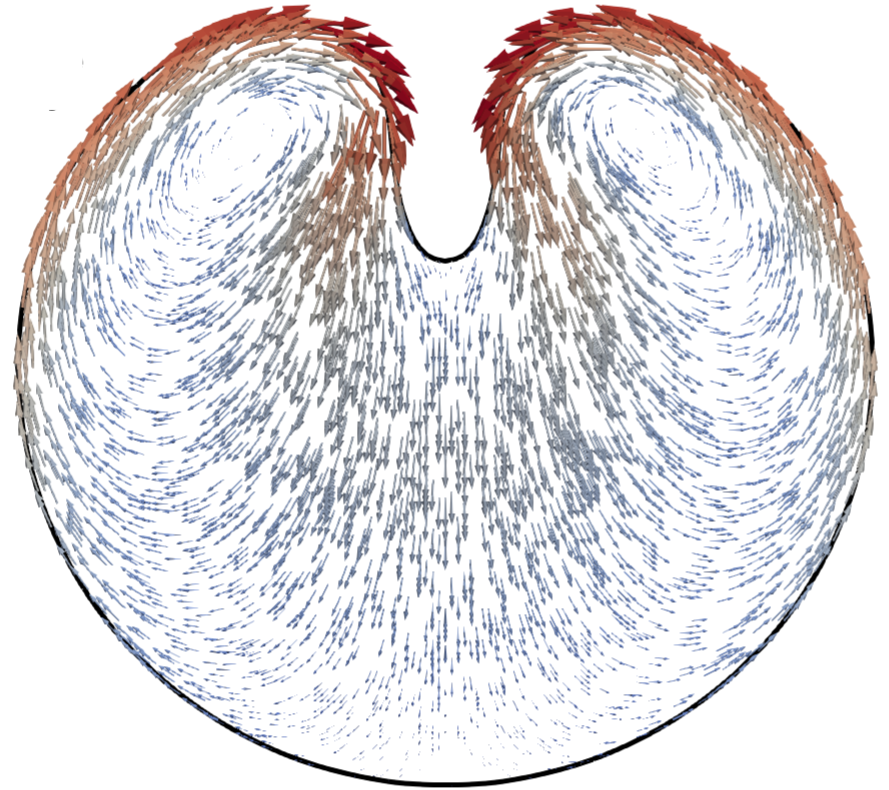}
    \caption{$\mathrm{t} = 0.081$}\label{fig:3D-division-unilateral-h}
  \end{subfigure}
  \caption{\textbf{3D unilateral cleavage}. Evolution of bulk flow field and surface density of the molecular species.}
  \label{fig:3D-division-unilateral}
\end{figure}

Our last experiment targets the non-rotationally symmetric mode of cell division known as unilateral cytokinesis~\cite{sugioka2022symmetry}. In the zygote of certain animal species, including jellyfish, corals and comb jellies (cnidaria and ctenophora, in general)~\cite{rappaport1963experimental}, but also in some epithelial tissues~\cite{herszterg2013interplay}, the cleavage furrow ingresses from only one side of the cell. This implies the formation of a single extensile pole at the opposite side of the furrow. Figure~\ref{fig:3D-division-unilateral} describes the dynamics of myosin concentration on the surface and bulk viscous flows during a unilateral cytokinesis that takes the same parameters as the bilateral case, except that $\xi$ in Equation~\eqref{eq:overactivity} only acts in half the equator. In contrast to the bilateral case, only one toroidal vortex forms around the ingression furrow.

We finally observe that both 2D axisymmetric and 3D models fail to complete cell pinching: In the 2D axisymmetric case, the axisymmetry boundary condition at the bulk phase (zero vertical displacement) is not compatible with the downwards displacement of the cleavage furrow. In the 3D cases, we observe shape instabilities arising when approximating pinch off. \added{On the other hand, we identified that alternative models for cytokinesis with stress-dependent myosin detachment rates or stress-dependent surface shear viscosities~\cite{bonati2022role} require to solve the transport problem with a mass-conserving scheme, which we have not included in the scope of this work. All these issues will be investigated in future work.}

\section{Conclusions}
\label{sec:conclusions}

Studying morphogenesis is a multidisciplinary effort that draws on various disciplines of science and technology, including developmental biologists, physicists, applied mathematicians or microscopy. Interdisciplinary collaboration between experts in these fields is often essential to gain a comprehensive understanding of morphogenesis. As experimental methods and theoretical models become more and more sophisticated, there is an increasing need of advanced computational methods to solve the complex mathematical models postulated by physicists and to interpret the experimental observations.

Morphogenetic processes are frequently characterised by very intricate three-dimensional and shape-evolving fluid dynamics, such as the cortex-cytoplasm interactions that concern this work. \ac{fe} analysis is one of the few computational approaches suitable for modelling this level of complexity, making it relatively easy to solve multiphysics \ac{pde} problems, with couplings between biochemical and mechanical signaling and between different topological dimensions. However, problems with moving boundaries and interfaces have been a major challenge in the \ac{fe} community. This is due --in part-- to the difficulty with \ac{fe} methods to strike a good balance between coping with large deformations and accurately resolving free surfaces and internal interfaces~\cite{Donea2004}.


Nowadays, a new generation of \emph{unfitted} \ac{fe} methods is challenging the state-of-the-art in \acp{fe} for moving boundaries and interfaces~\cite{de2023stability,Olshanskii2017,Hansbo2020}. They rely on a Eulerian description of motion and a sharp-interface representation of the geometry. By providing a robust and accurate way to track moving surfaces in a fixed computational grid, they are questioning the classical trade-off between Lagrangian and Eulerian \ac{fe} methods~\cite{Donea2004}. In this work, we specialise this emerging class of unfitted \ac{fe} technologies to model surface-bulk viscous flows in animal cells. To this end, we have formulated a novel partitioned \ac{fe} method that combines the aggregated \ac{fem}~\cite{Badia2018Mixed} for the bulk \acp{pde} with trace \ac{fem}~\cite{jankuhn2021trace} for the surface \acp{pde}. We have implemented the numerical model in \texttt{Gridap.jl}~\cite{Badia2020,Verdugo2022}, an advanced Julia \ac{fe} software ecosystem, leveraging high-order algorithms to compute quadrature rules and closest-point projections from the \texttt{algoim} library~\cite{saye2014high,saye2022high}. Our numerical experiments consider a minimal model of cellular symmetry breaking~\cite{mietke2019minimal}. They illustrate the capacity of the numerical framework to simulate nontrivial 3D dynamics, with very large distortions and topological changes, and the potential to address applications in animal morphogenesis that have been barely object of numerical modelling.

Our unfitted methodology offers several promising avenues for future development to enhance both its fidelity and applicability to address increasingly complex morphogenetic phenomena. A natural extension would involve a more faithful representation of the actomyosin cortex as a thin shell, incorporating membrane (e.g., tension) and torque stress resultants previously derived in~\cite{da2022viscous}. While membrane tension resultants can be readily integrated into the current model, capturing surface torques—whether of active~\cite{salbreux2017mechanics}, viscous~\cite{da2022viscous}, or elastic origin (such as plasma membrane bending~\cite{torres2019modelling})—requires second-order derivatives of the unknown fields. This necessity motivates an extension of our framework toward $\mathcal{H}^2$-conforming approximations~\cite{gfrerer2021ac}.

Beyond morphogenetic deformations, our approach could be extended to study 3D cell motility in viscous environments~\cite{poincloux2011contractility,hawkins2011spontaneous,farutin2019crawling}, including key additional mechanisms such as the membrane-to-cortex mechanical coupling~\cite{de2023cell}, and the mechanochemical interaction with regulatory proteins~\cite{gross2017active}. In particular, mutually inhibitory signaling circuits—often modeled by nonlinear reaction-diffusion systems~\cite{goehring2011polarization,gross2019guiding,de2024long}—play a key role in symmetry breaking and polarity establishment.

Mechanochemical coupling between cortical mechanics, surface-localized active proteins, and bulk-distributed passive species~\cite{howard2011turing} generally involves surface-bulk exchange of molecular components. Such coupling has been shown to underlie intracellular pattern formation~\cite{halatek2018self,brauns2021bulk,francis2024spatial}. Our unfitted framework \added{could address these scenarios with mass-conserving or, more generally, structure-preserving schemes~\cite{contri2025finite} for the robust solution of} advection-reaction-diffusion equations both on the cortex and in the cytoplasmic bulk, while preserving a sharp interface description. This feature provides a significant advantage over diffuse-interface or phase-field approaches~\cite{marth2014signaling,camley2017crawling,aland2023phase}, enabling a more accurate resolution of the interface, and avoids alternative approximations using projection methods~\cite{burkart2024dimensionality}.

Further developments could incorporate the polar and nematic ordering of cortical actin filaments~\cite{salbreux2009hydrodynamics,salbreux2022theory}, which are known to interact with gradients in cortical flow during cell division~\cite{reymann2016cortical,Mirza_2024}. Our unfitted framework could again prove advantageous in this context, simplifying the otherwise technical body-fitted FE formulation and implementation of tensorial fields on moving curved surfaces~\cite{nestler2019finite,torres2020approximation,nitschke2025active}.

As for modelling the bulk cytoplasm, we have treated the medium as a passive viscous fluid. However, the cytoplasm is a heterogeneous, crowded medium composed of a cytoskeletal meshwork permeated by cytosol and embedding various membrane-bound organelles such as the nucleus, endoplasmic reticulum, Golgi apparatus, and mitochondria. Capturing such complexity would require, at least, a two-phase poroviscous or poroelastic model~\cite{moeendarbary2013cytoplasm,mogilner2018intracellular}, potentially coupled to elastic-like inclusions~\cite{royer2019quasi,Liao2024}. \added{Stable and geometrically robust unfitted FE formulations for these type of problems are still rather limited and object of ongoing research~\cite{zhao2024discrete,berre2025cut}.}

Finally, a natural direction for generalization lies in the extension to multicellular systems, following surface-based approaches as in~\cite{maitre2016asymmetric,torres2022interacting,firmin2024mechanics}. This would pave the way for studying surface-bulk interactions in early animal embryogenesis, a domain in which computational modelling remains scarce—particularly using unfitted finite element methods. \added{Extending our \ac{fe} framework for evolving multicellular systems faces several challenges that include, but are not limited to, the capability of handling full cell division and pinch-off and an automatic approach to label cells and their free and internal surfaces.} Overall, the intersection between finite element technologies and animal morphogenesis offers a wide frontier of methodological and biological challenges, and we anticipate this cross-disciplinary interaction to remain fertile and active for years to come.


\section*{Declaration of competing interest}
\added{The authors declare that they have no known competing financial interests or personal relationships that could have appeared to influence the work reported in this paper}

\section*{CRediT authorship contribution statement}
\added{\textbf{Eric Neiva}: Conceptualization, Funding acquisition, Formal analysis, Investigation, Methodology, Software, Visualization, Writing - original draft preparation, review and editing. \textbf{Hervé Turlier}: Conceptualization, Funding acquisition, Methodology, Resources, Supervision, Validation, Writing - review and editing.}

\section*{Acknowledgments}

\newcommand{\thethanks}{We thank all members of the Turlier team for the fruitful discussions and Gridap developers for the many contributions that made this work possible. E.N. received funding from the European Union's Horizon EU research and innovation programme under grant agreement no.~101105565. H.T. received funding from the European Union's Horizon 2020 research and innovation programme under the European Research Council grant agreement no.~949267, from the Human Frontier Science Program research grant no.~RGP023/2025 and has been supported by the CNRS and the Collège de France.}

\thethanks

\appendix

\section{Axisymmetric formulation in 2D}
\label{app:axisymmetric}

\begin{figure}[h!]
  \centering
  \includegraphics[width=0.4\textwidth]{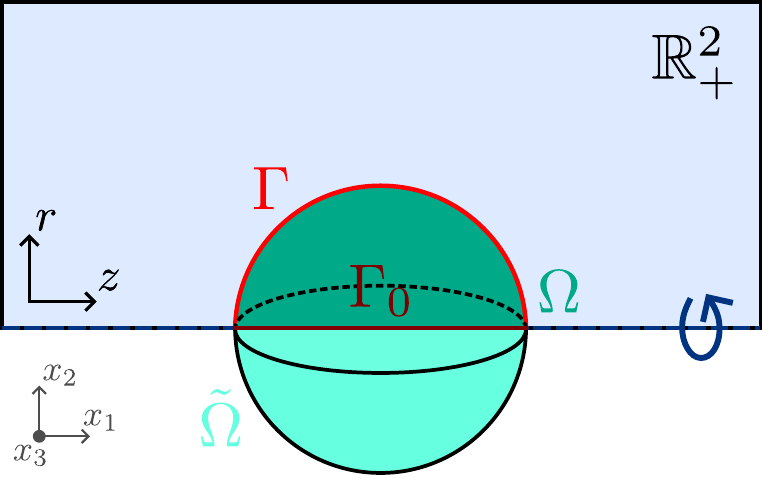}
  \caption{Geometry for the 2D axisymmetric analysis.}
  \label{fig:axisymmetric}
\end{figure}

This section details the 2D axisymmetric variational formulation of Equations~\eqref{eq:full-non-dimensional-1}-\eqref{eq:full-non-dimensional-9}. We denote with $(r,z)$ the Cartesian coordinates in $\mathbb{R}^2$ and assume the problem (solution and data) is rotationally invariant around the axis $r = 0$, see Figure~\ref{fig:axisymmetric}.

Let $\Omega$ denote a bounded domain contained in the half-space of positive $r$ coordinates $\mathbb{R}_{+}^2$. We use $\Gamma_0$ to refer to the portion of the boundary $\partial \Omega$ contained in the axis $r = 0$ and set $\Gamma = \partial \Omega \setminus \Gamma_0$. The axisymmetric domain $\tilde{\Omega}$ is the 3D set obtained by rotating $\Omega$ around the axis $r = 0$.

We now use the rotational invariance and the 3D transformation from Cartesian to cylindrical coordinates to formulate the 3D weak problem in $\tilde{\Omega}$ in the 2D domain $\Omega$ of the half-space $\mathbb{R}_{+}^2$, thus reducing the dimension of the problem by one order of magnitude. Note that, in this case, the symmetry boundary condition $u_h^r = 0$ on $\Gamma_0$ applies for the bulk flow problem. Meanwhile, in the surface flow problem, $U_h^r = 0$ on $\partial \Gamma$ and the rigid body modes reduce to the $z$-translational mode.  

Given the differential operators in cylindrical coordinates $(r,\theta,z)$~\cite{Bernardi1999}, in particular, for a scalar field $v$ and a vector field $\boldsymbol{u}$,

\[
  \tilde{\nabla} v = \nabla v, \enskip \tilde{\nabla} \cdot u = \frac{1}{r} \nabla \cdot ( r \boldsymbol{u} ), \enskip \text{and} \enskip \tilde{\boldsymbol{\varepsilon}}( \boldsymbol{u} )
  = \begin{bmatrix}
     \frac{\partial u^r}{\partial r} & 0 & \frac{1}{2} \left( \frac{\partial u^r}{\partial z} + \frac{\partial u^z}{\partial r} \right)\\
     0 & \frac{u^r}{r} & 0 \\
     \frac{1}{2} \left( \frac{\partial u^r}{\partial z} + \frac{\partial u^z}{\partial r} \right) & 0 & \frac{\partial u^z}{\partial z} \\
  \end{bmatrix},
\]
with analogous expressions holding for the surface operators; and the Jacobian of the coordinate transformation being $2 \pi r$, the bilinear forms and linear functionals of Section~\ref{subsec:space} in the 2D axisymmetric formulation read:

\subsection{Bulk viscous flows in Equation~\eqref{eq:discrete-bulk-problem}}

\[
  a_h(\boldsymbol{u}_h,\boldsymbol{v}_h) = \frac{2 R}{L_\eta} \int_\Omega \left\{ \, \boldsymbol{\varepsilon} (\boldsymbol{u}_h) : \boldsymbol{\varepsilon} (\boldsymbol{v}_h) + \frac{u_h^r}{r} \frac{v_h^r}{r} \, \right\} r \, \mathrm{d}\Omega \enskip \, \text{and} \enskip \,
  b_h(\boldsymbol{u}_h,p_h) = - \int_\Omega p_h \left( \nabla \cdot \boldsymbol{u}_h + \frac{u_h^r}{r} \right) \, r \, \mathrm{d}\Omega
\]

\[
  \begin{aligned}
    i_h(\boldsymbol{u}_h,p_h,\boldsymbol{v}_h,q_h) &= \int_\Gamma \left\{ \, \frac{\alpha}{h} \ \boldsymbol{u}_h \cdot \boldsymbol{v}_h - \left[ \frac{2 R}{L_\eta} \boldsymbol{\varepsilon} ( \boldsymbol{u}_h ) \boldsymbol{n}_\Gamma - p_h \boldsymbol{n}_\Gamma \right] \cdot \boldsymbol{v}_h - \left[ \frac{2 R}{L_\eta} \boldsymbol{\varepsilon} ( \boldsymbol{v}_h ) \boldsymbol{n}_\Gamma - q_h \boldsymbol{n}_\Gamma \right] \cdot \boldsymbol{u}_h \, \right\} r \, \mathrm{d}\Gamma \\
    \text{and} \; j_h(\boldsymbol{v}_h,q_h;\boldsymbol{U}_h) &= \int_\Gamma \left\{ \, \frac{\alpha}{h} \ \boldsymbol{U}_h \cdot \boldsymbol{v}_h - \left[ \frac{2 R}{L_\eta} \boldsymbol{\varepsilon} ( \boldsymbol{v}_h ) \boldsymbol{n}_\Gamma - q_h \boldsymbol{n}_\Gamma \right] \cdot \boldsymbol{U}_h \,
    \right\} r \, \mathrm{d}\Gamma
  \end{aligned}
\]

\subsection{Surface viscous flows in Equation~\eqref{eq:discrete-surface-flow-problem}}

\[
  A_h^U(\boldsymbol{U}_h,\boldsymbol{V}_h) = 2 \int_\Gamma \left\{ \, \boldsymbol{\varepsilon}_\Gamma (\boldsymbol{U}_h) : \boldsymbol{\varepsilon}_\Gamma ( \boldsymbol{V}_h ) + \frac{U_h^r}{r} \frac{V_h^r}{r} \, \right\} r \, \mathrm{d}\Gamma
\]

\[
  \begin{aligned}
    F_h^{\rm act}(\boldsymbol{V}_h;C_h) &= - \int_\Gamma \mbox{\textit{Pe}} f(C_h,1) \left( \mathrm{div}_\Gamma \boldsymbol{V}_h + \frac{V_h^r}{r} \right) r \, \mathrm{d}\Gamma \\
    F_h^{\rm cyt}(\boldsymbol{V}_h;\boldsymbol{u}_h,p_h) &= - \int_\Gamma \left[ \frac{2 R}{L_\eta} \boldsymbol{\varepsilon} ( \boldsymbol{u}_h ) \boldsymbol{n}_\Gamma - p_h \boldsymbol{n}_\Gamma \right] \cdot \boldsymbol{V}_h \ r \, \mathrm{d}\Gamma
  \end{aligned}
\]

\[
  S_h^U(\boldsymbol{U}_h,\boldsymbol{V}_h) = \int_{\mathcal{N}_h^\Gamma} \frac{\beta}{h} \ \boldsymbol{\varepsilon}_\Gamma (\boldsymbol{U}_h) \boldsymbol{n}_\Gamma \cdot \boldsymbol{\varepsilon}_\Gamma (\boldsymbol{V}_h) \boldsymbol{n}_\Gamma \ r \, \mathrm{d}\Omega
\]

\subsection{Surface molecular transport in Equation~\eqref{eq:discrete-surface-transport-problem}}

\[
  M_h(C_h^n,D_h^n) = \int_{\Gamma^n} C_h^n D_h^n \ r \, \mathrm{d}\Gamma,
  \quad A_h^C(C_h^n,D_h^n) = \int_{\Gamma^n} \nabla C_h^n \cdot \nabla D_h^n \ r \, \mathrm{d}\Gamma,
  \quad \text{and} \quad L_h(D_h^n) = \int_{\Gamma^n} D_h^n \ r \, \mathrm{d}\Gamma.
\]

\[
  S_h^C(C_h^n,D_h^n) = \int_{\mathcal{N}_h^{\Gamma^n}} \frac{\gamma}{h} \ ( \nabla C_h^n \cdot \boldsymbol{n}_\Gamma ) ( \nabla D_h^n \cdot \boldsymbol{n}_\Gamma ) \ r \, \mathrm{d}\Omega
\]

\[
  B_h(C_h^n,D_h^n;\boldsymbol{U}_h^n) = \int_{\Gamma^n} \left\{ \left( \boldsymbol{U}_h \cdot \nabla_\Gamma C_h^n \right) D_h^n + \left( \mathrm{div}_\Gamma \boldsymbol{U}_h^n + \frac{U_h^r}{r} \right) C_h^n D_h^n \right\} r \, \mathrm{d}\Gamma
\]


\renewcommand*{\bibfont}{\small}
\printbibliography
  
\end{document}